\documentclass[12pt]{article}
\usepackage{amsmath,amsthm}
\usepackage{eufrak}

\newcommand{\ar}{\renewcommand{\arraystretch}{1}} 
\DeclareMathAlphabet{\bb}{U}{msb}{m}{n} \gdef\C{\bb C}     \gdef\R{\bb R}
\gdef\K{\bb K} \gdef\BH{\bb H} \gdef\F{\bb F} 

 \DeclareMathOperator{\spin}{{\bf
Spin}}

\DeclareMathOperator{\Sym}{Sym} 
 
 \DeclareMathOperator{\Mat}{Mat}
 
 \DeclareMathOperator{\SO}{SO}
\DeclareMathOperator{\SL}{SL} \DeclareMathOperator{\SU}{SU}
\DeclareMathOperator{\Sp}{Sp}\DeclareMathOperator{\ISO}{ISO}

\newcommand{\scr}{\scriptstyle}

\newcommand{\cA}{\mathcal{A}}

\newcommand{\cE}{\mathcal{E}}
\newcommand{\cM}{{\cal M}}

\newcommand{\cK}{\mathcal{K}}

\newcommand{\bi}{{\bf i}}
\newcommand{\bj}{{\bf j}}
\newcommand{\bk}{{\bf k}}
\newcommand{\bx}{{\bf x}}
\newcommand{\by}{{\bf y}}

\newcommand{\fM}{\mathfrak{M}}

\newcommand{\fP}{\mathfrak{P}}
\newcommand{\fK}{\mathfrak{K}}

\newcommand{\fS}{\mathfrak{S}}
\newcommand{\fZ}{\mathfrak{Z}}
\newcommand{\fa}{\mathfrak{a}}
\newcommand{\fg}{\mathfrak{g}}

\newcommand{\fn}{\mathfrak{n}}
\newcommand{\fp}{\mathfrak{p}}
\newcommand{\fq}{\mathfrak{q}}

\newcommand{\biM}{\textbf{\emph{M}}}
\newcommand{\biP}{\textbf{\emph{P}}}
\newcommand{\biN}{\textbf{\emph{N}}}
\newcommand{\biV}{\textbf{\emph{V}}}

\newcommand{\cl}{C\kern -0.2em \ell}

\newcommand{\e}{\mbox{\bf e}}

\newcommand{\hypergeom}[5]{\mbox{$
_#1 F_#2\left. \!\! \left( \!\!\!\!
\begin{array}{c}
\multicolumn{1}{c}{\begin{array}{c} #3
\end{array}}\\[1mm]
\multicolumn{1}{c}{\begin{array}{c} #4
\end{array}}\end{array}
\!\!\!\! \right|\displaystyle{#5}\right) $} }

\newcommand{\Appell}[4]{\mbox{$
F_#1\left. \!\! \left[ \!\!\ar
\begin{array}{c}
\multicolumn{1}{c}{\begin{array}{c} #2
\end{array}}\\[1mm]
\multicolumn{1}{c}{\begin{array}{c} #3
\end{array}}\end{array}
\!\! \right|\displaystyle{#4}\right] $} }

\newcommand{\Lauricella}[4]{\mbox{$
\boldsymbol{\Psi}_#1\left. \!\! \left[ \!\!\ar
\begin{array}{c}
\multicolumn{1}{c}{\begin{array}{c} #2
\end{array}}\\[1mm]
\multicolumn{1}{c}{\begin{array}{c} #3
\end{array}}\end{array}
\!\! \right|\displaystyle{#4}\right] $} }

\newcommand{\ld}{\left[}
\newcommand{\rd}{\right]}

\newcommand{\tg}{\tan}
\newcommand{\ch}{\cosh}
\newcommand{\sh}{\sinh}
\newcommand{\tnh}{\tanh}

\newcommand{\ctg}{\cot}

\begin{document}
\title{Spherical functions on the de Sitter group}
\author{V. V. Varlamov\\
{\small\it Department of Mathematics, Siberia State University of Industry}\\
{\small\it Kirova 42, Novokuznetsk 654007, Russia}}
\date{}
\maketitle
\begin{abstract}
Matrix elements and spherical functions of irreducible
representations of the de Sitter group are studied on the various
homogeneous spaces of this group. It is shown that a universal
covering of the de Sitter group gives rise to quaternion Euler
angles. An explicit form of Casimir and Laplace-Beltrami operators
on the homogeneous spaces is given. Different expressions of the
matrix elements and spherical functions are given in terms of
multiple hypergeometric functions both for finite-dimensional and
unitary representations of the principal series of the de Sitter
group. Applications of the functions obtained to hydrogen atom
problem are considered.
\end{abstract}

\section{Introduction}
The representation theory of the de Sitter group, and also all the
questions concerning this group and the de Sitter spacetime, comes
in the forefront due the recent discoveries in modern cosmology. One
of the most important problem in this area is a construction of
quantum field theory in the de Sitter spacetime (see, for example,
\cite{All85,BM96,MRT05,Var05a}). As is known, in the standard
quantum field theory in Minkowski spacetime solutions (wave
functions) of relativistic wave equations are expressed via an
expansion in relativistic spherical functions (matrix elements of
the Lorentz group representations) \cite{AAV69,Var03,Var05,Var06}.
The analogous problem in five dimensions (solutions of wave
equations in de Sitter space) requires the most exact definition for
the matrix elements and spherical functions of irreducible
representations of the de Sitter group.

In the present work spherical functions are studied on the various
homogeneous spaces of the de Sitter group $\SO_0(1,4)$. A starting
point of this research is an analogue between universal coverings of
the Lorentz and de Sitter groups, which was first established by
Takahashi \cite{Tak63} (see also the work of Str\"{o}m
\cite{Str69}). Namely, the universal covering of $\SO_0(1,4)$ is
$\spin_+(1,4)\simeq\Sp(1,1)$ and the spinor group $\spin_+(1,4)$ is
described in terms of $2\times 2$ quaternionic matrices. On the
other hand, the universal covering of the Lorentz group $\SO_0(1,3)$
is $\spin_+(1,3)\simeq\SL(2,\C)$, where the spinor group
$\spin_+(1,3)$ is described in terms of $2\times 2$ complex
matrices. This analogue allows us to apply (with some restrictions)
Gel'fand-Naimark representation theory of the Lorentz group
\cite{GMS,Nai58} to $\SO_0(1,4)$. The section 2 contains a further
development of the Takahashi-Str\"{o}m analogue (quaternionic
description of $\SO_0(1,4)$). It is shown that for the group
$\spin_+(1,4)\simeq\Sp(1,1)$ there are quaternion Euler angles which
contain complex Euler angles of $\spin_+(1,3)\simeq\SL(2,\C)$ as a
particular case. Differential operators (Laplace-Beltrami and
Casimir operators) are defined on $\Sp(1,1)$ in terms of the
quaternion Euler angles. Spherical functions on the group
$\SO_0(1,4)$ are understood as functions of representations of the
class 1 realized on the homogeneous spaces of $\SO_0(1,4)$. A list
of homogeneous spaces of $\SO_0(1,4)$, including symmetric
Riemannian and non-Riemannian spaces, is given at the end of section
2. Spherical functions on the group $\SO(4)$ (maximal compact
subgroup of $\SO_0(1,4)$) are studied in the section 3. It is shown
that for a universal covering $\spin(4)\simeq\SU(2)\otimes\SU(2)$ of
$\SO(4)$ there are double Euler angles. It should be noted that all
the hypercomplex extensions (complex, double, quaternion) of usual
Euler angles of the group $\SU(2)$ follow directly from the
algebraic structure underlying the groups $\spin_+(p,q)$ and
describing within the framework of Clifford algebras $\cl_{p,q}$
\cite{Var04}. Matrix elements and spherical functions of $\SO(4)$
are expressed via the product of two hypergeometric functions.
Further, spherical functions of finite-dimensional representations
of $\SO_0(1,4)$ are studied in the section 4 on the various
homogeneous spaces of $\SO_0(1,4)$. It is shown that matrix elements
of $\SO_0(1,4)$ admit factorizations with respect to the matrix
elements of subgroups $\SO(4)$ and $\SO_0(1,3)$, since double and
complex angles are particular cases of the quaternion angles. In
turn, matrix elements and spherical functions of $\SO_0(1,4)$ are
expressed via multiple hypergeometric series (the product of three
hypergeometric functions). At the end of the section 4 we consider
applications of the spherical functions, defined on the
four-dimensional hyperboloid, to hydrogen and antihydrogen atom
problems. Spherical functions of the principal series
representations of $\SO_0(1,4)$ are considered in the section 5
within the Dixmier-Str\"{o}m representation basis of the de Sitter
group $\SO_0(1,4)$ \cite{Dix61,Str69}.

\section{The de Sitter group $\SO_0(1,4)$}
The homogeneous de Sitter group $\SO_0(1,4)$ consists of all real
matrices of fifth order with the unit determinant which leave
invariant the quadratic form
\[
Q(x)=x^2_0-x^2_1-x^2_2-x^2_3-x^2_4.
\]
The Lie algebra $\mathfrak{so}(1,4)$ of $\SO_0(1,4)$ consists of all
real matrices
\begin{equation}\label{Sit1}
\begin{bmatrix}
0 & a_{01} & a_{02} & a_{03} & a_{04}\\
a_{01} & 0 &-a_{12}&-a_{13} &-a_{14}\\
a_{02} & a_{12} & 0 & -a_{23} & -a_{24}\\
a_{03} & a_{13} & a_{23} & 0 & -a_{34}\\
a_{04} & a_{14} & a_{24} & a_{34} & 0
\end{bmatrix}.
\end{equation}
Thus, the algebra $\mathfrak{so}(1,4)$ has basis elements of the
form
\begin{equation}\label{Sit2}
L_{rs}=-e_{rs}+e_{sr},\quad s,r=1,2,3,4,\;\; s<r,
\end{equation}
\begin{equation}\label{Sit3}
L_{0r}=e_{0r}+e_{r0},\quad r=1,2,3,4,
\end{equation}
where $e_{rs}$ is a matrix with elements
$(e_{rs})_{pq}=\delta_{rp}\delta_{sq}$. The basis elements
(\ref{Sit2}) and (\ref{Sit3}) satisfy the following commutation
relations:
\begin{equation}
\ld
L_{\mu\nu},L_{\rho\sigma}\rd=g_{\nu\rho}L_{\mu\sigma}+g_{\mu\sigma}L_{\nu\rho}-
g_{\mu\rho}L_{\nu\sigma}-g_{\nu\sigma}L_{\mu\rho},\label{Sit4}
\end{equation}
\[
\rho,\,\mu,\,\nu,\,\sigma=0,1,2,3,4,
\]
where $g_{k0}=g_{0k}=\delta_{0k}$, $g_{ks}=-\delta_{ks}$;
$k,s=1,2,3,4$. $\SO_0(1,4)$ is a 10-parametric group.

The maximal compact subgroup $K$ of $\SO_0(1,4)$ is isomorphic to
the group $\SO(4)$ and consists of the matrices
\[
\begin{pmatrix}
1 & 0\\
0 & \SO(4)
\end{pmatrix}.
\]
Further, Cartan decomposition of the algebra $\mathfrak{so}(1,4)$
and Iwasawa decomposition of the group $\SO_0(1,4)$ have a great
importance at the construction of representations of the de Sitter
group $\SO_0(1,4)$. So, in the Cartan decomposition
$\mathfrak{so}(1,4)=\mathfrak{so}(4)+\fp$ a subspace $\fp$ consists
of the basis elements (\ref{Sit3}). The group $\SO_0(1,4)$ has a
real rank 1. For that reason the commutative subalgebra $\fa$ of
$\mathfrak{so}(1,4)$ is one dimensional. We can take the matrix
$L_{04}$ as a basis element of $\fa$. Therefore, the commutative
subgroup $A$ consists of the matrices
\begin{equation}\label{Sit5}
\begin{bmatrix}
\cosh\alpha & 0 & 0 & 0 & \sinh\alpha\\
0 & 1 & 0 & 0 & 0\\
0 & 0 & 1 & 0 & 0\\
0 & 0 & 0 & 1 & 0\\
\sinh\alpha & 0 & 0 & 0 & \cosh\alpha
\end{bmatrix},\quad 0\leq\alpha\leq\infty.
\end{equation}
Using the relations (\ref{Sit4}), we verify that a nilpotent
subalgebra $\fn$ of $\mathfrak{so}(1,4)$ is defined by the matrices
$L_{02}+L_{24}$, $L_{03}+L_{34}$ and $L_{01}+L_{14}$. Making an
exponential mapping of the subalgebra $\fn$ into the subgroup $N$,
we find that the nilpotent subgroup $N$ consists of the matrices
\begin{equation}\label{Sit6}
\begin{bmatrix}
1+(r^2+s^2+t^2)/2 & t & r & s & -(r^2+s^2+t^2)\\
t & 1 & 0 & 0 & -t\\
r & 0 & 1 & 0 & -r\\
s & 0 & 0 & 1 & -s\\
(r^2+s^2+t^2)/2 & t & r & s & 1-(r^2+s^2+t^2)
\end{bmatrix}.
\end{equation}
The subgroups $K$, $A$ and $N$ define the Iwasawa decomposition
$\SO_0(1,4)=\SO(4)\cdot NA$. In accordance with the definition of
the subgroup $M$ of $\SO_0(1,4)$ (see, for example, \cite{KK85}),
the subgroup $M$ is isomorphic to $\SO(3)$. Thus, a minimal
parabolic subgroup $P$ has a decomposition $P=\SO(3)\cdot NA$. Since
the rank of $\SO_0(1,4)$ is equal to 1, then there exist no other
parabolic subgroups containing $P$.

In the group $\SO_0(1,4)$ there are two independent Casimir
operators
\begin{equation}\label{Sit7}
F=L^2_{12}+L^2_{13}+L^2_{14}+L^2_{23}+L^2_{24}+L^2_{34}-L^2_{01}-L^2_{02}-L^2_{03}
-L^2_{04},
\end{equation}
\begin{multline}
W=(L_{12}L_{24}-L_{13}L_{24}+L_{14}L_{23})^2-(L_{12}L_{34}-L_{03}L_{24}+L_{04}L_{23})^2-\\
-(L_{01}L_{34}-L_{03}L_{14}+L_{04}L_{13})^2-(L_{01}L_{24}-L_{02}L_{14}+L_{04}L_{12})^2
-(L_{01}L_{23}-L_{02}L_{13}+L_{03}L_{12})^2.\label{Sit8}
\end{multline}
It is known that Casimir operator $W$ is equal to zero on the
representations $T^\sigma$ of the class 1 \cite{Boy71}. The Casimir
operator $F$ takes the values $\sigma(\sigma+3)$ on the
representations $T^\sigma$.

With the aim to obtain selfconjugated operators we will consider
generators $J_{\mu\nu}=\bi L_{\mu\nu}$ instead the elements
$L_{\mu\nu}$ of the algebra $\mathfrak{so}(1,4)$. In unitary
representations we have $J^\ast_{\mu\nu}=J_{\mu\nu}$. Let us
introduce the following designations for the ten generators
$J_{\mu\nu}$ of $\SO_0(1,4)$:
\begin{eqnarray}
\biM&=&(M_1\equiv J_{23},\;M_2\equiv J_{31},\;M_3\equiv
J_{12}),\nonumber\\
\biP&=&(P_1\equiv J_{14},\;P_2\equiv J_{24},\;P_3\equiv
J_{34}),\nonumber\\
\biN&=&(N_1\equiv J_{01},\;N_2\equiv J_{02},\;N_3\equiv
J_{03}),\label{Sit12}\\
P_0&=&J_{04}.\nonumber
\end{eqnarray}
Casimir operators of the group $\SO_0(1,4)$ in this designation have
the form
\begin{eqnarray}
F&=&(P^2_0+\biN^2)-(\biP^2+\biM^2),\nonumber\\
W&=&(\biM\cdot\biP)^2-(P_0\biM-\biP\times\biN)^2-(\biM\cdot\biN)^2.\nonumber
\end{eqnarray}
The generators (\ref{Sit12}) satisfy the following commutation
relations:
\begin{eqnarray}
&&\ld M_k,M_l\rd=\bi\varepsilon_{klm}M_m,\quad\ld
N_k,N_l\rd=-\bi\varepsilon_{klm}M_m,\nonumber\\
&&\ld P_k,P_l\rd=\bi\varepsilon_{klm}M_m,\nonumber\\
&&\ld M_k,N_l\rd=\bi\varepsilon_{klm}N_m,\quad\ld
M_k,P_l\rd=\bi\varepsilon_{klm}P_m,\nonumber\\
&&\ld M_k,N_k\rd=\ld M_k,P_k\rd=\ld M_k, P_0\rd=0,\label{Sit13}\\
&&\ld P_0,N_k\rd=\bi P_k,\;\ld P_0,P_k\rd=\bi N_k,\;\ld
P_k,N_l\rd=\bi\delta_{kl}P_0,\nonumber
\end{eqnarray}
where $\varepsilon_{klm}$ is an antisymmetric tensor of third rank,
which takes the values 0 or $\pm 1$ ($k,l,m=1,2,3$).

\subsection{Quaternionic description of $\SO_0(1,4)$}
Universal covering of the de Sitter group $\SO_0(1,4)$ is a spinor
group $\spin_+(1,4)\simeq\Sp(1,1)$ \cite{Hel78,Var04}. In its turn,
$\spin_+(1,4)\in\cl^+_{1,4}$, where $\cl^+_{1,4}$ is an even
subalgebra of the Clifford algebra $\cl_{1,4}$ associated with the
de Sitter space $\R^{1,4}$. Further, there is an isomorphism
$\cl^+_{1,4}\simeq\cl_{1,3}$, where $\cl_{1,3}$ is a space-time
algebra associated with the Minkowski space $\R^{1,3}$.

In virtue of the Karoubi theorem \cite{Kar78}, the space-time
algebra $\cl_{1,3}$ admits the following decomposition\footnote{This
decomposition is a particular case of the most general formula
$\cl(V\oplus V^\prime,Q\oplus
Q^\prime)\simeq\cl(V,Q)\otimes\cl(V^\prime,-Q^\prime)$, where $V$
and $V^\prime$ are vector spaces endowed with quadratic forms $Q$
and $Q^\prime$ over the field $\F$, $\dim V$ is even \cite[prop.
3.16]{Kar78}.}:
\[
\cl_{1,3}\simeq\cl_{1,1}\otimes\cl_{0,2}.
\]
The decomposition $\cl_{1,3}\simeq\cl_{1,1}\otimes\cl_{0,2}$ means
that for the algebra $\cl_{1,3}$ there exists a transition from the
real coordinates to quaternion coordinates of the form
$a+b\zeta_1+c\zeta_2+d\zeta_1\zeta_2$, where $\zeta_1=\e_{123}$,
$\zeta_2=\e_{124}$. At this point,
$\zeta^2_1=\zeta^2_2=(\zeta_1\zeta_2)^2=-1$, $\e^2_1=1$,
$\e^2_2=\e^2_3=\e^2_4=-1$. It is easy to see that the units
$\zeta_1$ and $\zeta_2$ form a basis of the quaternion algebra,
since $\zeta_1\sim\bi$, $\zeta_2\sim\bj$, $\zeta_1\zeta_2\sim\bk$.
Therefore, a general element
\[
\cA_{\cl_{1,3}}=a^0\e_0+\sum^4_{i=1}a^i\e_i+\sum^4_{i=1}\sum^4_{j=1}a^{ij}\e_i\e_j
+\sum^4_{i=1}\sum^4_{j=1}\sum^4_{k=1}a^{ijk}\e_i\e_j\e_k+
a^{1234}\e_1\e_2\e_3\e_4
\]
of the space-time algebra $\cl_{1,3}$ can be written in the form
\[
\cA_{\cl_{1,3}}=\cl^0_{1,1}+\cl^1_{1,1}\zeta_1+\cl^2_{1,1}\zeta_2+
\cl^3_{1,1}\zeta_1\zeta_2,
\]
where the each coefficient $\cl^i_{1,1}$ $(i=0,1,2,3)$ is isomorphic
to the anti-quaternion algebra $\cl_{1,1}$\footnote{$\cl_{1,1}$ is a
real Clifford algebra of the type $p-q\equiv 0\pmod{8}$ with a
division ring $\K\simeq\R$. This algebra is called the
anti-quaternion algebra by Rozenfel'd \cite{Roz55}.}:
\begin{eqnarray}
\cl^0_{1,1}&=&a^0+a^1\e_1+a^2\e_2+a^{12}\e_{12},\nonumber\\
\cl^1_{1,1}&=&a^{123}-a^{23}\e_1-a^{13}\e_2-a^3\e_{12},\nonumber\\
\cl^2_{1,1}&=&a^{124}-a^{24}\e_1+a^{14}\e_2+a^4\e_{12},\nonumber\\
\cl^3_{1,1}&=&-a^{34}-a^{134}\e_1-a^{234}\e_2+a^{1234}\e_{12}.\nonumber
\end{eqnarray}
It is easy to verify that the units $\zeta_1$ and $\zeta_2$ commute
with all the basis elements of $\cl_{1,1}$.

Further, let us define matrix representations of the quaternion
units $\zeta_1$ and $\zeta_2$ as follows:
\[
\zeta_1\longmapsto\begin{pmatrix} 0 & -1\\
1 & 0\end{pmatrix},\quad\zeta_2\longmapsto\begin{pmatrix} 0 & \bi\\
\bi & 0\end{pmatrix}.
\]
Thus, in virtue of the Karoubi theorem we have
\[
\cl_{1,3}\simeq\Mat_2(\cl_{1,1})=\begin{bmatrix}\cl^0_{1,1}-\bi\cl^3_{1,1}
& -\cl^1_{1,1}+\bi\cl^2_{1,1}\\
\cl^1_{1,1}+\bi\cl^2_{1,1} & \cl^0_{1,1}+\bi\cl^3_{1,1}
\end{bmatrix}.
\]
Or,
\begin{gather}
\cl_{1,3}\simeq\Mat_2(\cl_{1,1})=\begin{bmatrix} a & b\\
c & d\end{bmatrix}=\nonumber\\
\begin{bmatrix}\scr
a^0-a^{134}+(a^1+a^{34})\bi-(a^{13}+a^4)\bj+(a^{14}-a^3)\bk &\scr
a^{24}-a^{123}+(a^{23}+a^{124})\bi+(a^{13}-a^4)\bj+(a^{14}+a^3)\bk\\
\scr
a^{24}+a^{123}+(a^{124}-a^{23})\bi-(a^{13}+a^4)\bj+(a^{14}-a^3)\bk &
\scr
a^0+a^{134}+(a^1-a^{34})\bi+(a^2-a^{1234})\bj+(a^{12}+a^{234})\bk
\end{bmatrix},\nonumber
\end{gather}
where $\bi=\e_1$, $\bj=\e_2$, $\bk=\e_{12}$ are anti-quaternion
units, which satisfy the relations
\begin{gather}
\bi^2=-1,\quad\bj^2=1,\quad\bk^2=1,\nonumber\\
\bi\bj=-\bj\bi=\bk,\quad\bk\bi=-\bi\bk=\bj,\quad\bk\bj=-\bj\bk=\bi.
\nonumber
\end{gather}

In such a way, the universal covering of the de Sitter group
$\SO_0(1,4)$ is
\[\ar
\spin_+(1,4)\simeq\left\{\begin{bmatrix} a & b \\ c & c
\end{bmatrix}\in\BH(2):\;\;\det\begin{bmatrix}a & b \\ c & d
\end{bmatrix}=1\right\}=\Sp(1,1),
\]
where $\det\begin{bmatrix}a & b \\ c & d
\end{bmatrix}=1$ means that
\[
\overline{a}b=\overline{c}d,\quad |a|^2-|c|^2=1,\quad |d|^2-|b|^2=1,
\]
or,
\[
a\overline{c}=b\overline{d},\quad |a|^2-|b|^2=1,\quad |d|^2-|c|^2=1,
\]
here $\overline{a}$ means a quaternion conjugation.

The ten-parameter group $\spin_+(1,4)\simeq\Sp(1,1)$ has the
following one-parameter subgroups:
\[
m_{12}(\psi)=\begin{pmatrix} e^{\bi\frac{\psi}{2}} & 0\\
0 & e^{-\bi\frac{\psi}{2}}\end{pmatrix},\quad
m_{13}(\varphi)=\begin{pmatrix}
\cos\frac{\varphi}{2} & -\sin\frac{\varphi}{2}\\
\sin\frac{\varphi}{2} & \cos\frac{\varphi}{2}\end{pmatrix},\quad
m_{23}(\theta)=\begin{pmatrix} \cos\frac{\theta}{2} &
\bi\sin\frac{\theta}{2}\\
\bi\sin\frac{\theta}{2} & \cos\frac{\theta}{2}\end{pmatrix},
\]
\[
p_{14}(\phi)=\begin{pmatrix}\cos\frac{\phi}{2} & \bi\sin\frac{\phi}{2}\\
\bi\sin\frac{\phi}{2} & \cos\frac{\phi}{2}\end{pmatrix},\quad
p_{24}(\varsigma)=\begin{pmatrix}\cos\frac{\varsigma}{2} &
-\bj\sin\frac{\varsigma}{2}\\
\bj\sin\frac{\varsigma}{2} &
\cos\frac{\varsigma}{2}\end{pmatrix},\quad
p_{34}(\chi)=\begin{pmatrix}e^{\bk\frac{\chi}{2}} & 0\\
0 & e^{-\bk\frac{\chi}{2}}\end{pmatrix},
\]
\[
n_{01}(\tau)=\begin{pmatrix}\cosh\frac{\tau}{2} & \sinh\frac{\tau}{2}\\
\sinh\frac{\tau}{2} & \cosh\frac{\tau}{2}\end{pmatrix},\quad
n_{02}(\epsilon)=\begin{pmatrix}\cosh\frac{\epsilon}{2} &
\bi\sinh\frac{\epsilon}{2}\\
-\bi\sinh\frac{\epsilon}{2} &
\cosh\frac{\epsilon}{2}\end{pmatrix},\quad
n_{03}(\varepsilon)=\begin{pmatrix}e^{\frac{\varepsilon}{2}} & 0\\
0 & e^{-\frac{\varepsilon}{2}}\end{pmatrix},
\]
\[
p_{04}(\omega)=\begin{pmatrix}e^{\frac{\omega}{2}} & 0\\
0 & e^{-\frac{\omega}{2}}\end{pmatrix},
\]
where the ranges of parameters (Euler angles) are
\begin{equation}\label{QEA}
{\renewcommand{\arraystretch}{1.05}
\begin{array}{ccccc}
0 &\leq&\theta& \leq& \pi,\\
0 &\leq&\varphi& <&2\pi,\\
-2\pi&\leq&\psi&<&2\pi,
\end{array}\quad\quad
\begin{array}{ccccc}
0 &\leq&\phi& \leq& \pi,\\
0 &\leq&\varsigma& <&2\pi,\\
-2\pi&\leq&\chi&<&2\pi,
\end{array}}
\end{equation}
\begin{equation}\label{QEA2}
{\renewcommand{\arraystretch}{1.05}
\begin{array}{ccccc}
-\infty &<&\tau&<&+\infty,\\
-\infty&<&\epsilon&<&+\infty,\\
-\infty&<&\varepsilon&<&+\infty,\\
-\infty&<&\omega&<&+\infty.
\end{array}}
\end{equation}

Let us find a general transformation $\fq$ of $\spin_+(1,4)$ in the
space of representation with the smallest weight (a so-called
fundamental representation). In general, this form of the element
$g\in G$ is related closely with the Cartan decomposition $G=KAK$,
where $G$ is a connected Lie group, $K$ is a maximal compact
subgroup of $G$ and $A$ is a maximal commutative subgroup of $G$.
For example, the 3-parameter group $\SU(2)$ (a universal covering of
$\SO(3)$) has the following subgroups:
\begin{equation}\label{Sub1}
K=\left\{\begin{pmatrix}\cos\frac{\theta}{2} &
\bi\sin\frac{\theta}{2}\\
\bi\sin\frac{\theta}{2} &
\cos\frac{\theta}{2}\end{pmatrix}\right\},\quad
A=\left\{\begin{pmatrix}e^{\bi\frac{t}{2}} & 0\\
0 & e^{-\bi\frac{t}{2}}\end{pmatrix}\right\},
\end{equation}
where $t=\{\varphi,\psi\}$. Therefore, the Cartan decomposition
$\SU(2)=KAK$ of the element $u\in\SU(2)$ is (see, for example,
\cite{VK90})
\begin{equation}\label{Elem1}
g\equiv u(\varphi,\theta,\psi)=\begin{pmatrix}
e^{\bi\frac{\varphi}{2}} & 0\\
0 &
e^{-\bi\frac{\varphi}{2}}\end{pmatrix}\!\!\begin{pmatrix}\cos\frac{\theta}{2}
&
\bi\sin\frac{\theta}{2}\\
\bi\sin\frac{\theta}{2} &
\cos\frac{\theta}{2}\end{pmatrix}\!\!\begin{pmatrix}e^{\bi\frac{\psi}{2}}
& 0\\
0 & e^{-\bi\frac{\psi}{2}}\end{pmatrix},
\end{equation}
where $\varphi$, $\theta$, $\psi$ are Euler angles.

In its turn, the 6-parameter group $\spin_+(1,3)\simeq\SL(2,\C)$ (a
universal covering of the Lorentz group $\SO_0(1,3)$) is a complex
extension of the group $\SU(2)$, that is,
$\SL(2,\C)=[\SU(2)]^c=K^cA^cK^c$, where $K^c$ and $A^c$ are complex
extensions of the groups (\ref{Sub1}):
\begin{eqnarray}
K^c&=&\left\{\begin{pmatrix}\cos\frac{\theta^c}{2} &
\bi\sin\frac{\theta^c}{2}\\
\bi\sin\frac{\theta^c}{2} &
\cos\frac{\theta^c}{2}\end{pmatrix}=\begin{pmatrix}\cos\frac{\theta}{2}
&
\bi\sin\frac{\theta}{2}\\
\bi\sin\frac{\theta}{2} & \cos\frac{\theta}{2}\end{pmatrix}
\begin{pmatrix}
\cosh\frac{\tau}{2} & \sinh\frac{\tau}{2}\\
\sinh\frac{\tau}{2} &
\cosh\frac{\tau}{2}\end{pmatrix}\right\},\nonumber\\
A^c&=&\left\{\begin{pmatrix}e^{\bi\frac{t^c}{2}} & 0\\
0 & e^{-\bi\frac{t^c}{2}}\end{pmatrix}=\begin{pmatrix}e^{\bi\frac{p}{2}} & 0\\
0 & e^{-\bi\frac{p}{2}}\end{pmatrix}\begin{pmatrix} e^{\frac{q}{2}}
&
0\\
0 & e^{-\frac{q}{2}}\end{pmatrix}\right\},\nonumber
\end{eqnarray}
where $p=\{\varphi,\psi\}$, $q=\{\epsilon,\varepsilon\}$. Thus, the
Cartan decomposition $\SL(2,\C)=K^cA^cK^c$ of the element
$\fg\in\spin_+(1,3)\simeq\SL(2,\C)$ is
\begin{gather}g\equiv\fg(\varphi^c,\theta^c,\psi^c)=
\mathfrak{g}(\varphi,\,\epsilon,\,\theta,\,\tau,\,\psi,\,\varepsilon)=
\nonumber\\[0.2cm]
\begin{pmatrix}
e^{\bi\frac{\varphi}{2}} & 0\\
0 & e^{-\bi\frac{\varphi}{2}}
\end{pmatrix}{\renewcommand{\arraystretch}{1.1}\!\!\!\begin{pmatrix}
e^{\frac{\epsilon}{2}} & 0\\
0 & e^{-\frac{\epsilon}{2}}
\end{pmatrix}}\!\!\!{\renewcommand{\arraystretch}{1.3}\begin{pmatrix}
\cos\frac{\theta}{2} & \bi\sin\frac{\theta}{2}\\
\bi\sin\frac{\theta}{2} & \cos\frac{\theta}{2}
\end{pmatrix}\!\!\!\!
\begin{pmatrix}
\ch\frac{\tau}{2} & \sh\frac{\tau}{2}\\
\sh\frac{\tau}{2} & \ch\frac{\tau}{2}
\end{pmatrix}}\!\!\!{\renewcommand{\arraystretch}{1.1}\begin{pmatrix}
e^{\bi\frac{\psi}{2}} & 0\\
0 & e^{-\bi\frac{\psi}{2}}
\end{pmatrix}}\!\!\!
\begin{pmatrix}
e^{\frac{\varepsilon}{2}} & 0\\
0 & e^{-\frac{\varepsilon}{2}}
\end{pmatrix}=\nonumber\\[0.2cm]
=\begin{pmatrix}
e^{\bi\frac{\varphi^c}{2}} & 0\\
0 &
e^{-\bi\frac{\varphi^c}{2}}\end{pmatrix}\!\!\!\begin{pmatrix}\cos\frac{\theta^c}{2}
&
\bi\sin\frac{\theta^c}{2}\\
\bi\sin\frac{\theta^c}{2} &
\cos\frac{\theta^c}{2}\end{pmatrix}\!\!\!\begin{pmatrix}e^{\bi\frac{\psi^c}{2}}
& 0\\
0 & e^{-\bi\frac{\psi^c}{2}}\end{pmatrix},\label{Elem2}
\end{gather}
where
\begin{eqnarray}
\varphi^c&=&\varphi-\bi\epsilon,\nonumber\\
\theta^c&=&\theta-\bi\tau,\nonumber\\
\psi^c&=&\psi-\bi\varepsilon\nonumber
\end{eqnarray}
are {\it complex Euler angles}. Hence it follows that the element
(\ref{Elem2}) is a complex extension of (\ref{Elem1}).

Further, the 6-parameter spinor group $\spin(4)$ (a universal
covering of $\SO(4)$) due to an isomorphism
$\spin(4)\simeq\SU(2)\otimes\SU(2)$ admits the decomposition
$\spin(4)=K^eA^eK^e$, where $K^e$ and $A^e$ are double extensions of
the subgroups (\ref{Sub1}):
\begin{eqnarray}
K^e&=&\left\{\begin{pmatrix}\cos\frac{\theta^e}{2} &
\bi\sin\frac{\theta^e}{2}\\
\bi\sin\frac{\theta^e}{2} &
\cos\frac{\theta^e}{2}\end{pmatrix}=\begin{pmatrix}\cos\frac{\theta}{2}
&
\bi\sin\frac{\theta}{2}\\
\bi\sin\frac{\theta}{2} & \cos\frac{\theta}{2}\end{pmatrix}
\begin{pmatrix}
\cos\frac{\phi}{2} & \bi\sin\frac{\phi}{2}\\
\bi\sin\frac{\phi}{2} &
\cos\frac{\phi}{2}\end{pmatrix}\right\},\nonumber\\
A^e&=&\left\{\begin{pmatrix}e^{\bi\frac{t^e}{2}} & 0\\
0 & e^{-\bi\frac{t^e}{2}}\end{pmatrix}=\begin{pmatrix}e^{\bi\frac{p}{2}} & 0\\
0 & e^{-\bi\frac{p}{2}}\end{pmatrix}\begin{pmatrix}
e^{\bi\frac{q}{2}} &
0\\
0 & e^{-\bi\frac{q}{2}}\end{pmatrix}\right\},\nonumber
\end{eqnarray}
where $p=\{\varphi,\psi\}$, $q=\{\varsigma,\chi\}$. In this case,
the Cartan decomposition $\spin(4)=K^eA^eK^e$ of the element
$g\in\SU(2)\otimes\SU(2)$ is
\begin{gather}g\equiv g(\varphi^e,\theta^e,\psi^e)=
g(\varphi,\,\varsigma,\,\theta,\,\phi,\,\psi,\,\chi)=\nonumber\\[0.2cm]
\begin{pmatrix}
e^{\bi\frac{\varphi}{2}} & 0\\
0 & e^{-\bi\frac{\varphi}{2}}
\end{pmatrix}{\renewcommand{\arraystretch}{1.1}\!\!\!\begin{pmatrix}
e^{\bi\frac{\varsigma}{2}} & 0\\
0 & e^{-\bi\frac{\varsigma}{2}}
\end{pmatrix}}\!\!\!{\renewcommand{\arraystretch}{1.3}\begin{pmatrix}
\cos\frac{\theta}{2} & \bi\sin\frac{\theta}{2}\\
\bi\sin\frac{\theta}{2} & \cos\frac{\theta}{2}
\end{pmatrix}\!\!\!\!
\begin{pmatrix}
\cos\frac{\phi}{2} & \bi\sin\frac{\phi}{2}\\
\bi\sin\frac{\phi}{2} & \cos\frac{\phi}{2}
\end{pmatrix}}\!\!\!{\renewcommand{\arraystretch}{1.1}\begin{pmatrix}
e^{\bi\frac{\psi}{2}} & 0\\
0 & e^{-\bi\frac{\psi}{2}}
\end{pmatrix}}\!\!\!\!
\begin{pmatrix}
e^{\bi\frac{\chi}{2}} & 0\\
0 & e^{-\bi\frac{\chi}{2}}
\end{pmatrix}=\nonumber\\[0.2cm]
=\begin{pmatrix}
e^{\bi\frac{\varphi^e}{2}} & 0\\
0 &
e^{-\bi\frac{\varphi^e}{2}}\end{pmatrix}\!\!\!\begin{pmatrix}\cos\frac{\theta^e}{2}
&
\bi\sin\frac{\theta^e}{2}\\
\bi\sin\frac{\theta^e}{2} &
\cos\frac{\theta^e}{2}\end{pmatrix}\!\!\!\begin{pmatrix}e^{\bi\frac{\psi^e}{2}}
& 0\\
0 & e^{-\bi\frac{\psi^e}{2}}\end{pmatrix},\label{Elem3}
\end{gather}
where
\begin{equation}\label{DEA}
{\renewcommand{\arraystretch}{1.3} \left.\begin{array}{ccc}
\theta^e&=&\theta+\phi,\\
\varphi^e&=&\varphi+\varsigma,\\
\psi^e&=&\psi+\chi
\end{array}\right\}}
\end{equation}
are {\it double Euler angles}. It is easy to see that the element
(\ref{Elem3}) is a double extension of (\ref{Elem1}).

Finally, the 10-parameter spinor group $\spin_+(1,4)\simeq\Sp(1,1)$
(a universal covering of the de Sitter group $\SO_0(1,4)$) is
defined in terms of $2\times 2$ quaternionic matrices. This fact
allows us to introduce a decomposition $\Sp(1,1)=K^qA^qK^q$, where
$K^q$ and $A^q$ are quaternionic extensions of the groups
(\ref{Sub1}):
\begin{eqnarray}
K^q&=&\left\{\begin{pmatrix}\cos\frac{\theta^q}{2} &
\bi\sin\frac{\theta^q}{2}\\
\bi\sin\frac{\theta^q}{2} &
\cos\frac{\theta^q}{2}\end{pmatrix}=\begin{pmatrix}\cos\frac{\theta}{2}
&
\bi\sin\frac{\theta}{2}\\
\bi\sin\frac{\theta}{2} & \cos\frac{\theta}{2}\end{pmatrix}
\begin{pmatrix}
\cosh\frac{\tau}{2} & \sinh\frac{\tau}{2}\\
\sinh\frac{\tau}{2} &
\cosh\frac{\tau}{2}\end{pmatrix}\begin{pmatrix}
\cos\frac{\phi}{2} & \bi\sin\frac{\phi}{2}\\
\bi\sin\frac{\phi}{2} &
\cos\frac{\phi}{2}\end{pmatrix}\right\},\nonumber\\
A^q&=&\left\{\begin{pmatrix}e^{\bi\frac{\varphi^q}{2}} & 0\\
0 & e^{-\bi\frac{\varphi^q}{2}}\end{pmatrix}=\begin{pmatrix}e^{\bi\frac{\varphi}{2}} & 0\\
0 & e^{-\bi\frac{\varphi}{2}}\end{pmatrix}\begin{pmatrix}
e^{\frac{\epsilon}{2}} &
0\\
0 & e^{-\frac{\epsilon}{2}}\end{pmatrix}\begin{pmatrix}
e^{\bk\frac{\varsigma}{2}} & 0 \\
0 & e^{-\bk\frac{\varsigma}{2}}\end{pmatrix},\right.\nonumber\\
&&\left.\begin{pmatrix}e^{\bi\frac{\psi^q}{2}} & 0\\
0 & e^{-\bi\frac{\psi^q}{2}}\end{pmatrix}=\begin{pmatrix}e^{\bi\frac{\psi}{2}} & 0\\
0 & e^{-\bi\frac{\psi}{2}}\end{pmatrix}\begin{pmatrix}
e^{\frac{\varepsilon}{2}} &
0\\
0 & e^{-\frac{\varepsilon}{2}}\end{pmatrix}\begin{pmatrix}
e^{\frac{\omega}{2}} & 0 \\
0 &
e^{-\frac{\omega}{2}}\end{pmatrix}\begin{pmatrix}e^{\bj\frac{\chi}{2}}
& 0\\
0 & e^{-\bj\frac{\chi}{2}}\end{pmatrix}\right\}.\nonumber
\end{eqnarray}
Therefore, the Cartan decomposition $\Sp(1,1)=K^qA^qK^q$ of the
element $\fq\in\Sp(1,1)$ is
\begin{gather}
g\equiv\fq(\varphi^q,\theta^q,\psi^q)=\fq(\varphi,\epsilon,\varsigma,\theta,
\tau,\phi,\psi,\varepsilon,\omega,\chi)=\nonumber\\[0.2cm]
=\begin{pmatrix}e^{\bi\frac{\varphi}{2}} & 0\\
0 & e^{-\bi\frac{\varphi}{2}}\end{pmatrix}\!\!\!\begin{pmatrix}
e^{\frac{\epsilon}{2}} &
0\\
0 & e^{-\frac{\epsilon}{2}}\end{pmatrix}\!\!\!\begin{pmatrix}
e^{\bk\frac{\varsigma}{2}} & 0 \\
0 & e^{-\bk\frac{\varsigma}{2}}\end{pmatrix}\!\!\!
\begin{pmatrix}\cos\frac{\theta}{2}
&
\bi\sin\frac{\theta}{2}\\
\bi\sin\frac{\theta}{2} & \cos\frac{\theta}{2}\end{pmatrix}\!\!\!
\begin{pmatrix}
\cosh\frac{\tau}{2} & \sinh\frac{\tau}{2}\\
\sinh\frac{\tau}{2} &
\cosh\frac{\tau}{2}\end{pmatrix}\!\!\!\begin{pmatrix}
\cos\frac{\phi}{2} & \bi\sin\frac{\phi}{2}\\
\bi\sin\frac{\phi}{2} &
\cos\frac{\phi}{2}\end{pmatrix}\times\nonumber\\[0.2cm]
\times\begin{pmatrix}e^{\bi\frac{\psi}{2}} & 0\\
0 & e^{-\bi\frac{\psi}{2}}\end{pmatrix}\!\!\!\begin{pmatrix}
e^{\frac{\varepsilon}{2}} &
0\\
0 & e^{-\frac{\varepsilon}{2}}\end{pmatrix}\!\!\!\begin{pmatrix}
e^{\frac{\omega}{2}} & 0 \\
0 &
e^{-\frac{\omega}{2}}\end{pmatrix}\!\!\!\begin{pmatrix}e^{\bj\frac{\chi}{2}}
& 0\\
0 & e^{-\bj\frac{\chi}{2}}\end{pmatrix}=\nonumber\\[0.2cm]
=\begin{pmatrix}
e^{\bi\frac{\varphi^q}{2}} & 0\\
0 &
e^{-\bi\frac{\varphi^q}{2}}\end{pmatrix}\!\!\!\begin{pmatrix}\cos\frac{\theta^q}{2}
&
\bi\sin\frac{\theta^q}{2}\\
\bi\sin\frac{\theta^q}{2} &
\cos\frac{\theta^q}{2}\end{pmatrix}\!\!\!\begin{pmatrix}e^{\bi\frac{\psi^q}{2}}
& 0\\
0 & e^{-\bi\frac{\psi^q}{2}}\end{pmatrix},\label{Elem4}
\end{gather}
where
\begin{equation}\label{QEuler}
{\renewcommand{\arraystretch}{1.3} \left.\begin{array}{ccc}
\theta^q&=&\theta+\phi-\bi\tau,\\
\varphi^q&=&\varphi-\bi\epsilon+\bj\varsigma,\\
\psi^q&=&\psi-\bi\varepsilon-\bi\omega+\bk\chi
\end{array}\right\}}
\end{equation}
are {\em quaternion Euler angles}\footnote{Quaternion Euler angles
of $\spin_+(1,4)\simeq\Sp(1,1)$ contain complex Euler angles
$\theta^c=\theta-\bi\tau$, $\varphi^c=\varphi-\bi\epsilon$,
$\psi^c=\psi-\bi\varepsilon$ of the group
$\spin_+(1,3)\simeq\SL(2,\C)$ as a particular case (for more details
see \cite{Var06}).}. Hence it immediately follows that the element
(\ref{Elem4}) is a quaternionic extension of (\ref{Elem1}).

\subsection{Differential operators on the group $\Sp(1,1)$}
\begin{sloppypar}\noindent
Let $\Omega(t)$ be the one-parameter subgroup of $\Sp(1,1)$ and let
$\omega(t)$ be a matrix from the group $\Omega(t)$. The operators of
the right regular representation of $\Sp(1,1)$, corresponding to the
elements of the subgroup $\Omega(t)$, transfer quaternion functions
$f(\mathfrak{q})$ into
$R(\omega(t))f(\mathfrak{q})=f(\mathfrak{q}\omega(t))$. For that
reason the infinitesimal operator of the right regular
representation\index{representation!right regular}
$R(\mathfrak{q})$, associated with one--parameter subgroup
$\Omega(t)$, transfers the function $f(\mathfrak{q})$ into
$\frac{df(\mathfrak{q}\omega(t))}{dt}$ at $t=0$.\end{sloppypar}

Let us denote quaternion Euler angles of the element
$\mathfrak{q}\omega(t)$ via $\varphi^q(t),\theta^q(t),\psi^q(t)$.
Then there is an equality
\[
\left.\frac{df(\mathfrak{q}\omega(t))}{dt}\right|_{t=0}=
\frac{\partial
f}{\partial\varphi^q}\left(\varphi^q(0)\right)^\prime+
\frac{\partial f}{\partial\theta^q}\left(\theta^q(0)\right)^\prime+
\frac{\partial f}{\partial\psi^q}\left(\psi^q(0)\right)^\prime.
\]
The infinitesimal operator\index{operator!infinitesimal} $J_\omega$,
corresponding to the subgroup $\Omega(t)$, has a form
\[
J_\omega=
\left(\varphi^q(0)\right)^\prime\frac{\partial}{\partial\varphi^q}+
\left(\theta^q(0)\right)^\prime\frac{\partial}{\partial\theta^q}+
\left(\psi^q(0)\right)^\prime\frac{\partial}{\partial\psi^q}.
\]

Let us calculate infinitesimal operators $J^q_{\omega_1}$,
$J^q_{\omega_2}$, $J^q_{\omega_3}$ corresponding to the quaternion
subgroups $\Omega^q_1$, $\Omega^q_2$, $\Omega^q_3$. The quaternion
subgroups $\Omega^q_i$ ($i=1,2,3$) arise from the fact that all the
ten parameters of $\Sp(1,1)$ can be divided in three groups
according the Cartan decomposition (\ref{Elem4}) for the element
$\fq\in\Sp(1,1)$. The subgroup $\Omega^q_3$ consists of the matrices
\[
\omega_3(t^q)=
\begin{pmatrix}
e^{\bi\frac{t^q}{2}} & 0\\
0 & e^{-\bi\frac{t^q}{2}}
\end{pmatrix},
\]
where the variable $t^q$ has the form of quaternionic angles. Let
$\mathfrak{q}=\mathfrak{q}(\varphi^q,\theta^q,\psi^q)$ be a matrix
with quaternion Euler angles (the matrix (\ref{Elem4}))
$\varphi^q=\varphi-\bi\epsilon+\bj\varsigma$,
$\theta^q=\theta+\phi-\bi\tau$,
$\psi^q=\psi-\bi\varepsilon-\bi\omega+\bk\chi$. Therefore, Euler
angles of the matrix $\mathfrak{q}\omega_3(t^q)$ equal to
$\varphi^q$, $\theta^q$, $\psi^q=t-\bi t-\bi t+\bk t$. Hence it
follows that
\begin{gather}
\varphi^\prime(0)=0,\;\;
\epsilon^\prime(0)=0,\;\;\omega^\prime(0)=-\bi,\;\;
\theta^\prime(0)=0,\;\;\phi^\prime(0)=0,\;\; \tau^\prime(0)=0,\;\;
\psi^\prime(0)=1,\nonumber\\
\varepsilon^\prime(0)=-\bi,\;\;\varsigma^\prime(0)=\bj,\;\;\chi^\prime(0)=\bk.
\nonumber
\end{gather}
So, the operator $J^q_{\omega_3}$, corresponding to the subgroup
$\Omega^c_3$, has the form
\begin{equation}\label{J3}
J^q_{\omega_3}=\frac{\partial}{\partial\psi}-
\bi\frac{\partial}{\partial\varepsilon}-\bi\frac{\partial}{\partial\omega}
+\bk\frac{\partial}{\partial\chi}.
\end{equation}
Whence
\begin{equation}\label{IO3}
M_3=\frac{\partial}{\partial\psi},\quad
N_3=\frac{\partial}{\partial\varepsilon},\quad
P_3=\frac{\partial}{\partial\chi},\quad
P_0=\frac{\partial}{\partial\omega}.
\end{equation}

Let us calculate the infinitesimal operator $J^q_{\omega_1}$
corresponding to the quaternion subgroup $\Omega^q_1$. The subgroup
$\Omega^q_1$ consists of the matrices
\[
\omega_1(t^q)= {\renewcommand{\arraystretch}{1.3}
\begin{pmatrix}
\cos\frac{t^q}{2} & \bi\sin\frac{t^q}{2}\\
\bi\sin\frac{t^q}{2} & \cos\frac{t^q}{2}
\end{pmatrix}}.
\]
The Euler angles of these matrices equal to $0,\,t^q=t+et-\bi
t,\,0$, $e$ is the double unit. Let us represent the matrix
$\fq\omega_1(t^q)$ by the following product:
\[
{\renewcommand{\arraystretch}{1.3} \mathfrak{q}\omega_1(t^q)=
\begin{pmatrix}
\cos\frac{\theta^q}{2}e^{\bi\frac{(\varphi^q+\psi^q)}{2}} &
\bi\sin\frac{\theta^q}{2}e^{\bi\frac{(\varphi^q-\psi^q)}{2}}\\
\bi\sin\frac{\theta^q}{2}e^{\bi\frac{(\psi^q-\varphi^q)}{2}} &
\cos\frac{\theta^q}{2}e^{-\bi\frac{(\varphi^q+\psi^q)}{2}}
\end{pmatrix}
\begin{pmatrix}
\cos\frac{t^q}{2} & \bi\sin\frac{t^q}{2}\\
\bi\sin\frac{t^q}{2} & \cos\frac{t^q}{2}
\end{pmatrix}.
}
\]
Multiplying the matrices on the right-side of the latter expression,
we obtain
\begin{eqnarray}
\cos\theta^q(t)&=&\cos\theta^q\cos t^q-\sin\theta^q\sin
t^q\cos\psi^q,
\label{SL9}\\[0.2cm]
e^{\bi\varphi^q(t)}&=&e^{\bi\varphi^q}\frac{\sin\theta^q\cos t^q+
\cos\theta^q\sin t^q\cos\psi^q+\bi\sin t^q\sin\psi^q}
{\sin\theta^q(t)},\label{SL10}\\[0.2cm]
e^{\bi\frac{[\varphi^q(t)+\psi^q(t)]}{2}}&=&e^{\bi\frac{\varphi^q}{2}}
\frac{\cos\frac{\theta^q}{2}\cos\frac{t^q}{2}e^{\frac{\bi\psi^q}{2}}-
\sin\frac{\theta^q}{2}\sin\frac{t^q}{2}e^{-\bi\frac{\psi^q}{2}}}
{\cos\frac{\theta^q(t)}{2}}.\label{SL11}
\end{eqnarray}
For calculation of derivatives $\varphi^\prime(t)$,
$\epsilon^\prime(t)$, $\omega^\prime(t)$, $\theta^\prime(t)$,
$\phi^\prime(t)$, $\tau^\prime(t)$, $\psi^\prime(t)$,
$\varepsilon^\prime(t)$, $\varsigma^\prime(t)$, $\chi^\prime(t)$ at
$t=0$ we must differentiate on $t$ the both parts of the each
equality from (\ref{SL9})--(\ref{SL11}). At this point, we have
$\varphi(0)=\varphi$, $\epsilon(0)=\epsilon$, $\ldots$,
$\chi(0)=\chi$.

So, let us differentiate the both parts of (\ref{SL9}). As a result,
we obtain
\[
-\sin\theta^q(t)\left[\theta^\prime(t)+e\phi^\prime(t)-\bi\tau^\prime(t)
\right]=-\cos\theta^q\sin t^q(1+e-\bi)-\sin\theta^q\cos
t^q\cos\psi^q(1+e-\bi).
\]
Taking $t=0$, we find that
\[
\theta^\prime(0)+e\phi^\prime(0)-\bi\tau^\prime(0)=\cos\psi^q(1+e-\bi).
\]
Whence
\[
\theta^\prime(0)=\cos\psi^q,\quad\phi^\prime(0)=\cos\psi^q,\quad
\tau^\prime(0)=\cos\psi^q.
\]
Differentiating now the both parts of (\ref{SL10}) and taking $t=0$,
we obtain
\[
\varphi^\prime(0)-\bi\epsilon^\prime(0)+\bj\varsigma^\prime(0)=
\frac{\sin\psi^q(1+e-\bi)}{\sin\theta^q}.
\]
Therefore,
\[
\varphi^\prime(0)=\frac{\sin\psi^q}{\sin\theta^q},\quad
\epsilon^\prime(0)=\frac{\sin\psi^q}{\sin\theta^q},\quad
\varsigma^\prime(0)=\frac{\sin\psi^q}{\sin\theta^q}.
\]
Further, differentiating the both parts of (\ref{SL11}) and taking
$t=0$, we find that
\[
\psi^\prime(0)-\bi\varepsilon^\prime(0)-\bi\omega^\prime(0)+
\bj\chi^\prime(0)=(-1-e+\bi)\cot\theta^q\sin\psi^q
\]
and
\[
\psi^\prime(0)=\varepsilon^\prime(0)=\chi^\prime(0)=
-\cot\theta^q\sin\psi^q,\quad\omega^\prime(0)=0.
\]
In such a way, we have
\begin{equation}\label{J1}
J^q_{\omega_1}=M_1+P_1-\bi N_1,
\end{equation}
where
\begin{eqnarray}
M_1&=&\cos\psi^q\frac{\partial}{\partial\theta}+
\frac{\sin\psi^q}{\sin\theta^q}\frac{\partial}{\partial\varphi}-
\ctg\theta^q\sin\psi^q\frac{\partial}{\partial\psi},\label{SL12}\\
N_1&=&\cos\psi^q\frac{\partial}{\partial\tau}+
\frac{\sin\psi^q}{\sin\theta^q}\frac{\partial}{\partial\epsilon}-
\ctg\theta^q\sin\psi^q\frac{\partial}{\partial\varepsilon},\label{SL13}\\
P_1&=&\cos\psi^q\frac{\partial}{\partial\phi}+
\frac{\sin\psi^q}{\sin\theta^q}\frac{\partial}{\partial\varsigma}-
\ctg\theta^q\sin\psi^q\frac{\partial}{\partial\chi}.\label{SL14}
\end{eqnarray}

Let us calculate now an infinitesimal operator $J^q_{\omega_2}$
corresponding to the quaternion subgroup $\Omega^q_2$. The subgroup
$\Omega^q_2$ consists of the matrices
\[
\omega_2(t^q)= {\renewcommand{\arraystretch}{1.3}
\begin{pmatrix}
\cos\frac{t^q}{2} & -\sin\frac{t^q}{2}\\
\sin\frac{t^q}{2} & \cos\frac{t^q}{2}
\end{pmatrix}},
\]
where the Euler angles equal correspondingly to $0,\,t^c=t-\bi t+\bj
t,\,0$. It is obvious that the matrix $\mathfrak{q}\omega_2(t^q)$
can be represented by the product
\[
{\renewcommand{\arraystretch}{1.3} \mathfrak{q}\omega_1(t^q)=
\begin{pmatrix}
\cos\frac{\theta^q}{2}e^{\bi\frac{(\varphi^q+\psi^q)}{2}} &
\bi\sin\frac{\theta^q}{2}e^{\bi\frac{(\varphi^q-\psi^q)}{2}}\\
\bi\sin\frac{\theta^q}{2}e^{\bi\frac{(\psi^q-\varphi^q)}{2}} &
\cos\frac{\theta^q}{2}e^{-\bi\frac{(\varphi^q+\psi^q)}{2}}
\end{pmatrix}
\begin{pmatrix}
\cos\frac{t^q}{2} & -\sin\frac{t^q}{2}\\
\sin\frac{t^q}{2} & \cos\frac{t^q}{2}
\end{pmatrix}.
}
\]
Multiplying the matrices on the right-side of this equality, we see
that Euler angles of the product $\fq\omega_2(t^q)$ are related by
the formulae
\begin{eqnarray}
\cos\theta^q(t)&=&\cos\theta^q\cos t^q+\sin\theta^q\sin
t^q\sin\psi^q,
\label{SL15}\\[0.2cm]
e^{\bi\varphi^q(t)}&=&e^{\bi\varphi^q}\frac{\sin\theta^q\cos t^q-
\cos\theta^q\sin t^q\sin\psi^q+\bi\sin t^q\cos\psi^q}
{\sin\theta^q(t)},\label{SL16}\\[0.2cm]
e^{\bi\frac{[\varphi^q(t)+\psi^q(t)]}{2}}&=&e^{\bi\frac{\varphi^q}{2}}
\frac{\cos\frac{\theta^q}{2}\cos\frac{t^q}{2}e^{\frac{\bi\psi^q}{2}}+
\sin\frac{\theta^q}{2}\sin\frac{t^q}{2}e^{-\bi\frac{\psi^q}{2}}}
{\cos\frac{\theta^q(t)}{2}}.\label{SL17}
\end{eqnarray}
Differentiating on $t$ the both parts of the each equalities
(\ref{SL15})--(\ref{SL17}) and taking $t=0$, we obtain
\begin{eqnarray}
&&\theta^\prime(0)=\tau^\prime(0)=\phi^\prime(0)=-\sin\psi^q,\nonumber\\
&&\varphi^\prime(0)=\epsilon^\prime(0)=\varsigma^\prime(0)=
\frac{\cos\psi^q}{\sin\theta^q},\nonumber\\
&&\psi^\prime(0)=\varepsilon^\prime(0)=\chi^\prime(0)=
-\cot\theta^q\cos\psi^q,\quad\omega^\prime(0)=0.\nonumber
\end{eqnarray}
Therefore, for the subgroup $\Omega^q_2$ we have
\begin{equation}\label{J2}
J^q_{\omega_2}=M_2-\bi N_2+\bj P_2,
\end{equation}
where
\begin{eqnarray}
M_2&=&-\sin\psi^q\frac{\partial}{\partial\theta}+
\frac{\cos\psi^q}{\cos\theta^q}\frac{\partial}{\partial\varphi}-
\ctg\theta^q\sin\psi^q\frac{\partial}{\partial\psi},\label{SL18}\\
N_2&=&-\sin\psi^q\frac{\partial}{\partial\tau}+
\frac{\cos\psi^q}{\sin\theta^q}\frac{\partial}{\partial\epsilon}-
\ctg\theta^q\cos\psi^q\frac{\partial}{\partial\varepsilon},\label{SL19}\\
P_2&=&-\sin\psi^q\frac{\partial}{\partial\phi}+
\frac{\cos\psi^q}{\sin\theta^q}\frac{\partial}{\partial\varsigma}-
\ctg\theta^q\cos\psi^q\frac{\partial}{\partial\chi}.\label{SL20}
\end{eqnarray}
Let us introduce an auxiliary quaternion angle
$\psi^q_1=\psi-\bi\varepsilon+\bk\chi$. It is easy to see that
$\psi^q=\psi^q_1-\bi\omega$; therefore, $\psi^q_1$ is the part of
$\psi^q$. Further, taking into account expressions (\ref{IO3}),
(\ref{SL12})--(\ref{SL14}) and (\ref{SL18})--(\ref{SL20}), we can
rewrite the operators (\ref{J3}), (\ref{J1}), (\ref{J2}) in the
form\footnote{These operators look like as $\SU(2)$ type (or
$\SU(2)\otimes\SU(2)$ type) infinitesimal operators. However, it is
easy to verify that they do not form a group, since
$\psi^q\ne\psi^q_1$.}
\begin{eqnarray}
J^q_{\omega_1}&=&\cos\psi^q\frac{\partial}{\partial\theta^q}+
\frac{\sin\psi^q}{\sin\theta^q}\frac{\partial}{\partial\varphi^q}-
\ctg\theta^q\sin\psi^q\frac{\partial}{\partial\psi^q_1},\label{qJ1}\\
J^q_{\omega_2}&=&-\sin\psi^q\frac{\partial}{\partial\theta^q}+
\frac{\cos\psi^q}{\sin\theta^q}\frac{\partial}{\partial\varphi^q}-
\ctg\theta^q\cos\psi^q\frac{\partial}{\partial\psi^q_1},\label{qJ2}\\
J^q_{\omega_3}&=&\frac{\partial}{\partial\psi^q},\label{qJ3}\\
\dot{J}^q_{\omega_1}&=&\cos\dot{\psi}^q\frac{\partial}{\partial\dot{\theta}^q}+
\frac{\sin\dot{\psi}^q}{\sin\dot{\theta}^q}
\frac{\partial}{\partial\dot{\varphi}^q}-
\ctg\dot{\theta}^q\sin\dot{\psi}^q\frac{\partial}{\partial\dot{\psi}^q_1},
\label{dqJ1}\\
\dot{J}^q_{\omega_2}&=&-\sin\dot{\psi}^q\frac{\partial}{\partial\dot{\theta}^q}+
\frac{\cos\dot{\psi}^q}{\sin\dot{\theta}^q}
\frac{\partial}{\partial\dot{\varphi}^q}-
\ctg\dot{\theta}^q\cos\dot{\psi}^q\frac{\partial}{\partial\dot{\psi}^q_1},
\label{dqJ2}\\
\dot{J}^q_{\omega_3}&=&\frac{\partial}{\partial\dot{\psi}^q},\label{dqJ3}
\end{eqnarray}
where
\[
{\renewcommand{\arraystretch}{1.55}
\begin{array}{ccl}
\dfrac{\partial}{\partial\theta^q}&=&\dfrac{\partial}{\partial\theta}+
\dfrac{\partial}{\partial\phi}+\bi\dfrac{\partial}{\partial\tau},\\
\dfrac{\partial}{\partial\varphi^q}&=&\dfrac{\partial}{\partial\varphi}+
\bi\dfrac{\partial}{\partial\epsilon}+\bj\dfrac{\partial}{\partial\varsigma},\\
\dfrac{\partial}{\partial\psi^q}&=&\dfrac{\partial}{\partial\psi}+
\bi\dfrac{\partial}{\partial\varepsilon}+\bi\dfrac{\partial}{\partial\omega}+
\bk\dfrac{\partial}{\partial\chi},\\
\dfrac{\partial}{\partial\psi^q_1}&=&\dfrac{\partial}{\partial\psi}+
\bi\dfrac{\partial}{\partial\varepsilon}+
\bk\dfrac{\partial}{\partial\chi}.
\end{array}
\quad
\begin{array}{ccl}
\dfrac{\partial}{\partial\dot{\theta}^q}&=&\dfrac{\partial}{\partial\theta}-
\dfrac{\partial}{\partial\phi}-\bi\dfrac{\partial}{\partial\tau},\\
\dfrac{\partial}{\partial\dot{\varphi}^q}&=&\dfrac{\partial}{\partial\varphi}-
\bi\dfrac{\partial}{\partial\epsilon}-\bj\dfrac{\partial}{\partial\varsigma},\\
\dfrac{\partial}{\partial\dot{\psi}^q}&=&\dfrac{\partial}{\partial\psi}-
\bi\dfrac{\partial}{\partial\varepsilon}-\bi\dfrac{\partial}{\partial\omega}-
\bk\dfrac{\partial}{\partial\chi},\\
\dfrac{\partial}{\partial\dot{\psi}^q_1}&=&\dfrac{\partial}{\partial\psi}-
\bi\dfrac{\partial}{\partial\varepsilon}-
\bk\dfrac{\partial}{\partial\chi}.
\end{array}}
\]
Using the expressions (\ref{qJ1})--(\ref{qJ3}), we see that for the
first Casimir operator $F$ of the group $\SO_0(1,4)$ there exists
the following equality:
\[
-F=-P^2_0-\biN^2+\biP^2+\biM^2=\left(J^q_{\omega_1}\right)^2+
\left(J^q_{\omega_2}\right)^2+\left(J^q_{\omega_3}\right)^2.
\]
Or,
\begin{equation}\label{FKO}
-F=\frac{\partial^2}{\partial{\theta^q}^2}+
\cot\theta^q\frac{\partial}{\partial\theta^q}+
\frac{1}{\sin^2\theta^q}\frac{\partial^2}{\partial{\varphi^q}^2}-
\frac{2\cos\theta^q}{\sin^2\theta^q}
\frac{\partial^2}{\partial\varphi^q\partial\psi^q_1}+
\cot^2\theta^q\frac{\partial^2}{\partial{\psi^q_1}^2}+
\frac{\partial^2}{\partial{\psi^q}^2}.
\end{equation}
Matrix elements $t^{\sigma}_{mn}(\fq)=
\fM^{\sigma}_{mn}(\varphi^q,\theta^q,\psi^q)$ of irreducible
representations of the group $\SO_0(1,4)$ are eigenfunctions of the
operator (\ref{FKO}):
\begin{equation}\label{FKO2}
\left[-F+\sigma(\sigma+3)\right]\fM^{\sigma}_{mn}(\fq)=0,
\end{equation}
where
\begin{equation}\label{MF}
\fM^{\sigma}_{mn}(\fq)=e^{-\bi(m\varphi^q+n(\psi^q_1-\bi\omega))}\fZ^{\sigma}_{mn}
(\cos\theta^q),
\end{equation}
since $\psi^q=\psi^q_1-\bi\omega$. Here, $\fM^\sigma_{mn}(\fq)$ are
general matrix elements of the representations of $\SO_0(1,4)$, and
$\fZ^\sigma_{mn}(\cos\theta^q)$ are {\it hyperspherical functions}.
Substituting the functions (\ref{MF}) into (\ref{FKO2}) and taking
into account the operator (\ref{FKO}), we arrive at the following
differential equation:
\begin{multline}
\frac{d^2\fZ^\sigma_{mn}(\cos\theta^q)}{d{\theta^q}^2}+\cot\theta^q
\frac{d
\fZ^\sigma_{mn}(\cos\theta^q)}{d\theta^q}-\frac{m^2}{\sin^2\theta^q}
\fZ^\sigma_{mn}(\cos\theta^q)+\frac{2mn\cos\theta^q}{\sin^2\theta^q}
\fZ^\sigma_{mn}(\cos\theta^q)-\\
-n^2\cot^2\theta^q\fZ^\sigma_{mn}(\cos\theta^q)-
n^2\fZ^\sigma_{mn}(\cos\theta^q)+\sigma(\sigma+3)\fZ^\sigma_{mn}(\cos\theta^q)=0,
\nonumber
\end{multline}
or
\[
\left[\frac{d^2}{d{\theta^q}^2}+\cot\theta^q\frac{d}{d\theta^q}-
\frac{m^2+n^2-2mn\cos\theta^q}{\sin^2\theta^q}+\sigma(\sigma+3)\right]
\fZ^\sigma_{mn}(\cos\theta^q)=0.
\]
After substitution $z=\cos\theta^q$ this equation can be rewritten
as
\begin{equation}\label{FKO3}
\left[(1-z^2)\frac{d^2}{dz^2}-2z\frac{d}{dz}-
\frac{m^2+n^2-2mnz}{1-z^2}+\sigma(\sigma+3)\right]
\fZ^{\sigma}_{mn}(z)=0.
\end{equation}
The latter equation has three singular points $-1$, $+1$, $\infty$.
It is a Fuchsian equation. Indeed, denoting
$w(z)=\fZ^\sigma_{mn}(z)$, we write the equation (\ref{FKO3}) in the
form
\begin{equation}\label{Fux1}
\frac{d^2w(z)}{dz^2}-p(z)\frac{dw(z)}{dz}+q(z)w(z)=0,
\end{equation}
where
\[
p(z)=\frac{2z}{(1-z)(1+z)},\quad
q(z)=\frac{\sigma(\sigma+3)(1-z^2)-m^2-n^2+2mnz}{(1-z)^2(1+z)^2}.
\]
Let us find solutions of (\ref{Fux1}). Applying the substitution
\[
t=\frac{1-z}{2},\quad
w(z)=t^{\frac{|m-n|}{2}}(1-t)^{\frac{|m+n|}{2}}v(t),\nonumber
\]
we arrive at hypergeometric equation
\begin{equation}\label{Hyper}
t(1-t)\frac{d^2v}{dt^2}+[c-(a+b+1)t]\frac{dv}{dt}-abv(t)=0,
\end{equation}
where
\begin{eqnarray}
a&=&\sigma+3+\frac{1}{2}(|m-n|+|m+n|),\nonumber\\
b&=&-\sigma+\frac{1}{2}(|m-n|+|m+n|),\nonumber\\
c&=&|m-n|+1.\nonumber
\end{eqnarray}
Therefore, a solution of (\ref{Hyper}) is
\[
v(t)=C_1\hypergeom{2}{1}{a,b}{c}{t}+C_2t^{1-c}
\hypergeom{2}{1}{b-c+1,a-c+1}{2-c}{t}.
\]
Coming back to initial variable, we obtain
\begin{multline}
w(z)=C_1\left(\frac{1-z}{2}\right)^{\frac{|m-n|}{2}}
\left(\frac{1+z}{2}\right)^{\frac{|m+n|}{2}}\times\\
\times\hypergeom{2}{1}{\sigma+3+\frac{1}{2}(|m-n|+|m+n|),-\sigma+\frac{1}{2}(|m-n|+|m+n|)}
{|m-n|+1}{\frac{1-z}{2}}+\\
+C_2\left(\frac{1-z}{2}\right)^{-\frac{|m-n|}{2}}
\left(\frac{1+z}{2}\right)^{\frac{|m+n|}{2}}\times\\
\times\hypergeom{2}{1}{-\sigma+\frac{1}{2}(|m+n|-|m-n|),\sigma+3+\frac{1}{2}(|m+n|-|m-n|)}
{1-|m-n|}{\frac{1-z}{2}}. \label{Sol1'}
\end{multline}
Thus, from (\ref{Sol1'}) it follows that the function
$\fZ^\sigma_{mn}$ can be represented by the following particular
solution:
\begin{multline}
\fZ^\sigma_{mn}(\cos\theta^q)=C_1\sin^{|m-n|}\frac{\theta^q}{2}
\cos^{|m+n|}\frac{\theta^q}{2}\times\\
\times\hypergeom{2}{1}{\sigma+3+\frac{1}{2}(|m-n|+|m+n|),-\sigma+\frac{1}{2}(|m-n|+|m+n|)}
{|m-n|+1}{\sin^2\frac{\theta^q}{2}}.\label{Hyper2}
\end{multline}
In section 4 and 5 we will give more explicit expressions for the
functions $\fZ^\sigma_{mn}(\cos\theta^q)$ via the multiple
hypergeometric series.

Finally, using the formulae (\ref{dqJ1})--(\ref{dqJ3}), we can
obtain the same differential equation for the function
$\fZ^{\dot{\sigma}}_{\dot{m}\dot{n}}(\cos\dot{\theta}^q)$. All the
calculations in this case are analogous to the previous calculations
for $\fZ^\sigma_{mn}(\cos\theta^q)$.

\subsection{Homogeneous spaces of $\SO_0(1,4)$}
Before introducing the spherical functions on the group $\SO_0(1,4)$
it is useful to give a general definition for spherical functions on
the group $G$. Let $T(g)$ be an irreducible representation of the
group $G$ in the space $L$ and let $H$ be a subgroup of $G$. The
vector $\boldsymbol{\xi}$ in the space $L$ is called {\it an
invariant with respect to the subgroup} $H$ if for all $h\in H$ the
equality $T(h)\boldsymbol{\xi}=\boldsymbol{\xi}$ holds. The
representation $T(g)$ is called {\it a representation of the class
one with respect to the subgroup} $H$ if in its space there are
non-null vectors which are invariant with respect to $H$. At this
point, a contraction of $T(g)$ onto its subgroup $H$ is unitary:
\[
(T(h)\boldsymbol{\xi}_1,T(h)\boldsymbol{\xi}_2)=(\boldsymbol{\xi}_1,
\boldsymbol{\xi}_2).
\]
Hence it follows that a function
\[
f(g)=(T(g)\boldsymbol{\eta},\boldsymbol{\xi})
\]
corresponds the each vector $\boldsymbol{\eta}\in L$. $f(g)$ are
called {\it spherical functions of the representation $T(g)$ with
respect to $H$}.

Spherical functions can be considered as functions on homogeneous
spaces ${\cal M}=G/H$. In its turn,
a homogeneous space $\cM$ of the group $G$ has the following properties:\\
a) It is a topological space on which the group $G$ acts
continuously, that is, let $y$ be a point in $\cM$, then $gy$ is
defined and is again
a point in $\cM$ ($g\in G$).\\
b) This action is transitive, that is, for any two points $y_1$ and
$y_2$ in $\cM$ it is always possible to find a group element $g\in
G$
such that $y_2=gy_1$.\\
There is a one-to-one correspondence between the homogeneous spaces
of $G$ and the coset spaces of $G$. Let $H_0$ be a maximal subgroup
of $G$ which leaves the point $y_0$ invariant, $hy_0=y_0$, $h\in
H_0$, then $H_0$ is called {\it the stabilizer of} $y_0$.
Representing now any group element of $G$ in the form $g=g_ch$,
where $h\in H_0$ and $g_c\in G/H_0$, we see that, by virtue of the
transitivity property, any point $y\in\cM$ can be given by
$y=g_chy_0=g_cy$. Hence it follows that the elements $g_c$ of the
coset space give a parametrization of $\cM$. The mapping
$\cM\leftrightarrow G/H_0$ is continuous since the group
multiplication is continuous and the action on $\cM$ is continuous
by definition. The stabilizers $H$ and $H_0$ of two different points
$y$ and $y_0$ are conjugate, since from $H_0g_0=g_0$, $y_0=g^{-1}y$,
it follows that $gH_0g^{-1}y=y$, that is, $H=gH_0g^{-1}$.

Coming back to the de Sitter group $G=\SO_0(1,4)$, we see that there
are the following homogeneous spaces of $\SO_0(1,4)$ depending on
the stabilizer $H$. First of all, when $H=0$ the homogeneous space
$\cM_{10}$ coincides with {\it a group manifold} $\mathfrak{S}_{10}$
of $\SO_0(1,4)$. Therefore, $\mathfrak{S}_{10}$ is a maximal
homogeneous space of the de Sitter group. Further, when
$H=\Omega^q_\psi$, where $\Omega^q_\psi$ is a group of diagonal
matrices
\[
\begin{pmatrix}
e^{\frac{\bi\psi^q}{2}} & 0\\
0 & e^{-\frac{\bi\psi^q}{2}}
\end{pmatrix},
\]
the homogeneous space $\cM_6$ coincides with a {\it two-dimensional
quaternion sphere} $S^q_2$,
$\cM_6=S^q_2\sim\Sp(1,1)/\Omega^q_\psi$\footnote{When the stabilizer
$H$ is a compact group, the homogeneous space $\cM=G/H$ is called
{\it a Riemannian symmetric space} \cite{Hel78}. When $H$ is a
non-compact group, we arrive at the non-Riemannian spaces. The
homogeneous space $\cM_6=S^q_2\sim\Sp(1,1)/\Omega^q_\psi$ is the
non-Riemannian space, since the stabilizer $H=\Omega^q_\psi$ is
non-compact subgroup of $\Sp(1,1)$. Quaternion and anti-quaternion
spheres were studied by Rozenfel'd \cite{Roz55}.}.

We obtain the following homogeneous space $\cM_4$ when the
stabilizer $H$ coincides with a maximal compact subgroup $K=\SO(4)$
of $\SO_0(1,4)$. In this case we arrive at the upper sheet of a
four-dimensional hyperboloid $\cM_4=H^4\sim\SO_0(1,4)/\SO(4)$. The
upper sheet $H^4_+$ of the two-sheeted hyperboloid $H^4$ can be
understood as a quotient space $\SO_0(1,4)/\SO(4)$. Indeed, let us
consider the upper sheet $H^4_+$ of $H^4$:
\begin{equation}\label{Sit9}
H^4_+:\; x^2_0-x^2_1-x^2_2-x^2_3-x^2_4=1,\quad x_0>0
\end{equation}
and the point $x^0=(1,0,0,0,0)$ on $H^4_+$. The group $\SO_0(1,4)$
transfers the hyperboloid $H^4_+$ into itself. Besides, for any two
points $x^\prime$ and $x^{\prime\prime}$ of $H^4_+$ there is such an
element $g\in\SO_0(1,4)$ that $gx^\prime=x^{\prime\prime}$, that is,
$\SO_0(1,4)$ is a transitive transformation group of the homogeneous
space. The set of elements from $\SO_0(1,4)$, leaving the point
$x^0$ invariant, coincides with the subgroup $\SO(4)$. Therefore,
$H^4_+$ is homeomorphic to the quotient space $\SO_0(1,4)/\SO(4)$.
It should be noted that {\it a four-dimensional Lobatchevski space}
$\mathcal{L}^4$, called also {\it a de Sitter space}, is realized on
the hyperboloid $H^4_+$\footnote{It is obvious that among the all
homogeneous spaces of $\SO_0(1,4)$ the space $H^4_+$ is the most
important for physics. In accordance with modern cosmology, $H^4_+$
is understood as a space-time endowed with a global topology of
constant negative curvature (the de Sitter universe).}.

In the case $x^2_0-x^2_1-x^2_2-x^2_3-x^4=0$ we arrive at a cone
$C^4$ which can also be considered as a homogeneous space of
$\SO_0(1,4)$. Usually, only the upper sheets $H^4_+$ and $C^4_+$ are
considered in applications.

The following homogeneous space $\cM_3$ of $\SO_0(1,4)$ is a
three-dimensional real sphere $S^3\sim\SO(4)/\SO(3)$. In contrast to
the previous homogeneous spaces, the sphere $S^3$ coincides with a
quotient space $\SO_0(1,4)/P$, where $P$ is a minimal parabolic
subgroup of $\SO_0(1,4)$. From the Iwasawa decompositions
$\SO_0(1,4)=KNA$ and $P=MNA$, where $M=\SO(3)$, $N$ and $A$ are
nilpotent and commutative subgroups of $\SO_0(1,4)$, it follows that
$\SO_0(1,4)/P=KNA/MNA\sim K/M\sim\SO(4)/\SO(3)$.
\begin{sloppypar}
A minimal homogeneous space $\cM_2$ of $\SO_0(1,4)$ is a
two-dimensional real sphere $S^2\sim\SO(3)/\SO(2)$.
\end{sloppypar}
Taking into account the list of homogeneous spaces of $\SO_0(1,4)$,
we now introduce the following types of spherical functions $f(\fq)$
on the de Sitter group:
\begin{itemize}
\item $f(\fq)=\fM^\sigma_{mn}(\fq)=
e^{-\bi m\varphi^q}\fZ^\sigma_{mn}(\cos\theta^q)e^{-\bi n\psi^q}$.
This function is defined on the group manifold $\fS_{10}$ of
$\SO_0(1,4)$. It is the most general spherical function on the group
$\SO_0(1,4)$. In this case $f(\fq)$ depends on all the ten
parameters of $\SO_0(1,4)$ and for that reason it should be called
as {\it a function on the de Sitter group}. An explicit form of
$\fM^\sigma_{mn}(\fq)$ (respectively
$\fM^{\dot{\sigma}}_{\dot{m}\dot{n}}(\fq)$) for finite-dimensional
representations and of $\fM^{-\frac{3}{2}+\bi\rho,l_0}_{mn}(\fq)$
(resp. $\fM^{-\frac{3}{2}-\bi\rho,l_0}_{\dot{m}\dot{n}}(\fq)$) for
infinite-dimensional representations of $\SO_0(1,4)$ will be given
in the sections 4 and 5, respectively.
\item $f(\varphi^q,\theta^q)=\fM^m_\sigma(\varphi^q,\theta^q,0)=e^{-\bi\varphi^q}
\fZ^m_\sigma(\cos\theta^q)$. This function is defined on the
homogeneous space $\cM_6=S^2_q\sim\Sp(1,1)/\Omega^q_\psi$, that is,
on the surface of the two-dimensional quaternion sphere $S^q_2$. The
function $\fM^m_\sigma(\varphi^q,\theta^q,0)$ is a five-dimensional
analogue of the usual spherical function $Y^m_l(\varphi,\theta)$
defined on the surface of the real two-sphere $S_2$. In its turn,
the function $f(\dot{\varphi}^q,\dot{\theta}^q)=
\fM^{\dot{m}}_{\dot{\sigma}}(\dot{\varphi}^q,\dot{\theta}^q,0)$ is
defined on the surface of the dual quaternion sphere $\dot{S}^2_q$.
An explicit form of the functions
$\fM^m_\sigma(\varphi^q,\theta^q,0)$
($\fM^{\dot{m}}_{\dot{\sigma}}(\dot{\varphi}^q,\dot{\theta}^q,0)$)
and $\fM^m_{-\frac{3}{2}+\bi\rho,l_0}(\varphi^q,\theta^q,0)$
($\fM^{\dot{m}}_{-\frac{3}{2}-\bi\rho,l_0}(\dot{\varphi}^q,\dot{\theta}^q,0)$)
will be given in the section 4 and 5.
\item
$f(\epsilon,\tau,\varepsilon,\omega)=\fM^\sigma_{mn}(\epsilon,\tau,\varepsilon,\omega)=
e^{\bi m\epsilon}\fP^\sigma_{mn}(\cosh\tau)e^{\bi
n(\varepsilon+\omega)}$. This function is defined on the homogeneous
space $\cM_4=H^4_+\sim\SO_0(1,4)/\SO(4)$, that is, on the upper
sheet of the hyperboloid $x^2_0-x^2_1-x^2_2-x^2_3-x^2_4=1$. An
explicit form of the functions
$\fM^\sigma_{mn}(\epsilon,\tau,\varepsilon,\omega)$
($\fM_{\dot{m}\dot{n}}^{\dot{l}}(\epsilon,\tau,\varepsilon,\omega)$)
and
$\fM_{mn}^{-\frac{3}{2}+\bi\rho}(\epsilon,\tau,\varepsilon,\omega)$
($\fM_{\dot{m}\dot{n}}^{-\frac{3}{2}-\bi\rho}(\epsilon,\tau,\varepsilon,\omega)$)
will be given in the section 4 and 5.
\item\begin{sloppypar}\noindent
$f(\varphi,\theta,\psi)=\fM^\sigma_{mn}(\varphi,\theta,\psi)=
e^{-\bi m\varphi}P^\sigma_{mn}(\cos\theta)e^{-\bi n\psi}$ (or
$f(\varsigma,\phi,\chi)=\fM^\sigma_{mn}(\varsigma,\phi,\chi)=
e^{-\bi m\varsigma}P^\sigma_{mn}(\cos\phi) e^{-\bi n\chi}$). This
function is defined on the homogeneous space $\cM_3\sim
S^3=\SO(4)/\SO(3)$, that is, on the surface of the real 3-sphere
$x^2_0+x^2_1+x^2_2+x^2_3=1$. In essence, we come here to
representations of $\SO_0(1,4)$ restricted to the subgroup $\SO(4)$.
\end{sloppypar}
\item\begin{sloppypar}\noindent $f(\varphi,\theta)=\fM^m_l(\varphi,\theta,0)=
e^{-\bi m\varphi}P^m_{\sigma}(\cos\theta)\sim
Y^m_\sigma(\varphi,\theta)$ (or
$f(\varsigma,\phi)=\fM^m_\sigma(\varsigma,\phi,0)= e^{-\bi
m\varsigma}P^m_{\sigma}(\cos\phi)\sim Y^m_\sigma(\varsigma,\phi)$).
This function is defined on the homogeneous space
$\cM_2=S^2\sim\SO(3)/\SO(2)$, that is, on the surface of the real
2-sphere $S^2$. We come here to the most degenerate representations
of $\SO_0(1,4)$ restricted to the subgroup $\SO(3)$.\end{sloppypar}
\end{itemize}

\section{Spherical functions on the group $\SO(4)$}
As is known, the group $\SO(4)$ is a maximal compact subgroup of
$\SO_0(1,4)$. $\SO(4)$ corresponds to basis elements
$\biM=(M_1,M_2,M_3)$ and $\biP=(P_1,P_2,P_3)$ of the algebra
$\mathfrak{so}(1,4)$:
\begin{equation}\label{Sit15}
\ld M_k,M_l\rd=\bi\varepsilon_{klm}M_m,\quad \ld
M_k,P_l\rd=\bi\varepsilon_{klm}P_m,\quad\ld
P_k,P_l\rd=\bi\varepsilon_{klm}M_m.
\end{equation}
Introducing linear combinations $\biV=(\biM+\biP)/2$ and
$\biV^\prime=(\biM-\biP)/2$, we obtain
\begin{equation}\label{Sit16}
\ld V_k,V_l\rd=\bi\varepsilon_{klm}V_m,\quad \ld
V^\prime_k,V^\prime_l\rd=\bi\varepsilon_{klm}V^\prime_m.
\end{equation}
The operators $\biV$ and $\biV^\prime$ form bases of the two
independent algebras $\mathfrak{so}(3)$. It means that $\SO(4)$ is
isomorphic to a direct product $\SO(3)\otimes\SO(3)$.

A universal covering of $\SO(4)$ is
$\spin(4)\simeq\SU(2)\otimes\SU(2)$. The one-parameter subgroups of
$\spin(4)$ are
\[
m_{12}(\psi)=\begin{pmatrix} e^{\bi\frac{\psi}{2}} & 0 \\
0 & e^{-\bi\frac{\psi}{2}}\end{pmatrix},\quad m_{13}(\varphi)=
\begin{pmatrix}
e^{\bi\frac{\varphi}{2}} & 0\\
0 & e^{-\bi\frac{\varphi}{2}}\end{pmatrix},\quad m_{23}(\theta)=
\begin{pmatrix}
\cos\frac{\theta}{2} & \bi\sin\frac{\theta}{2}\\
\bi\sin\frac{\theta}{2} & \cos\frac{\theta}{2}
\end{pmatrix},
\]
\[
p_{14}(\chi)=\begin{pmatrix} e^{\bi\frac{\chi}{2}} & 0 \\
0 & e^{-\bi\frac{\chi}{2}}\end{pmatrix},\quad p_{24}(\varsigma)=
\begin{pmatrix}
e^{\bi\frac{\varsigma}{2}} & 0\\
0 & e^{-\bi\frac{\varsigma}{2}}\end{pmatrix},\quad p_{34}(\phi)=
\begin{pmatrix}
\cos\frac{\phi}{2} & \bi\sin\frac{\phi}{2}\\
\bi\sin\frac{\phi}{2} & \cos\frac{\phi}{2}
\end{pmatrix},
\]
where
\[
{\renewcommand{\arraystretch}{1.05}
\begin{array}{ccccc}
0 &\leq&\theta& \leq& \pi,\\
0 &\leq&\varphi& <&2\pi,\\
-2\pi&\leq&\psi&<&2\pi,
\end{array}\quad\quad
\begin{array}{ccccc}
0 &\leq&\phi& \leq& \pi,\\
0 &\leq&\varsigma& <&2\pi,\\
-2\pi&\leq&\chi&<&2\pi.
\end{array}}
\]
A fundamental representation of the group
$\spin(4)\simeq\SU(2)\otimes\SU(2)$ is defined by the matrix
(\ref{Elem3}).

On the group $\SO(4)$ there exist the following Laplace-Beltrami
operators:
\begin{eqnarray}
\biV^2&=&V^2_1+V^2_2+V^2_3=\frac{1}{4}(\biM^2+\biP^2+2\biM\biP),\label{BLO1}\\
{\biV^\prime}^2&=&{V^\prime_1}^2+{V^\prime_2}^2+{V^\prime_3}^2=
\frac{1}{4}(\biM^2+\biP^2-2\biM\biP).\label{BLO2}
\end{eqnarray}
At this point, we see that operators (\ref{BLO1}), (\ref{BLO2})
contain Casimir operators $\biM^2+\biP^2$, $\biM\biP$ of the group
$\SO(4)$. Using expressions (\ref{DEA}), we obtain a Euler
parametrization of the Laplace-Beltrami operators,
\begin{eqnarray}
\biV^2&=&\frac{\partial^2}{\partial\theta^e{}^2}+
\ctg\theta^e\frac{\partial}{\partial\theta^e}+\frac{1}{\sin^2\theta^e}\left[
\frac{\partial^2}{\partial\varphi^e{}^2}-
2\cos\theta^e\frac{\partial}{\partial\varphi^e}
\frac{\partial}{\partial\psi^e}+
\frac{\partial^2}{\partial\psi^e{}^2}\right],\nonumber\\
{\biV^\prime}^2&=&\frac{\partial^2}{\partial\dot{\theta}^e{}^2}+
\ctg\dot{\theta}^e\frac{\partial}{\partial\dot{\theta}^e}+
\frac{1}{\sin^2\dot{\theta}^e}\left[
\frac{\partial^2}{\partial\dot{\varphi}^e{}^2}-
2\cos\dot{\theta}^e\frac{\partial}{\partial\dot{\varphi}^e}
\frac{\partial}{\partial\dot{\psi}^e}+
\frac{\partial^2}{\partial\dot{\psi}^e{}^2}\right].\label{KO2}
\end{eqnarray}
Here, $\dot{\theta}^e=\theta-\phi$,
$\dot{\varphi}^e=\varphi-\varsigma$, $\dot{\psi}^e=\psi-\chi$ are
conjugate double angles.

Matrix elements $t^{l}_{mn}(g)=
\fM^{l}_{mn}(\varphi^e,\theta^e,\psi^e)$ of irreducible
representations of the group $\SO(4)$ are eigenfunctions of the
operators (\ref{KO2}),
\begin{eqnarray}
\left[\biV^2+l(l+1)\right]\fM^{l}_{mn}(\varphi^e,\theta^e,\psi^e)&=&0,\nonumber\\
\left[{\biV^\prime}^2+\dot{l}(\dot{l}+1)\right]\fM^{\dot{l}}_{\dot{m}\dot{n}}
(\dot{\varphi}^e,\dot{\theta}^e,\dot{\psi}^e)&=&0,\label{EQ}
\end{eqnarray}
where
\begin{eqnarray}
\fM^{l}_{mn}(g)&=& e^{-\bi(m\varphi^e+n\psi^e)}\fZ^{l}_{mn}
(\cos\theta^e),\nonumber\\
\fM^{\dot{l}}_{\dot{m}\dot{n}}(g)&=&e^{\bi(\dot{m}\dot{\varphi}^e+
\dot{n}\dot{\psi}^e)}\fZ^{\dot{l}}_{\dot{m}\dot{n}}(\cos\dot{\theta}^e).
\label{HF3'}
\end{eqnarray}
Here, $\fM^l_{mn}(g)$ are general matrix elements of the
representations of $\SO(4)$, and $\fZ^l_{mn}(\cos\theta^e)$ are {\it
hyperspherical functions} of $\SO(4)$. Substituting the functions
(\ref{HF3'}) into (\ref{EQ}) and taking into account the operators
(\ref{KO2}) and substitutions $z=\cos\theta^e$,
$\overset{\ast}{z}=\cos\dot{\theta}^e$, we come to the following
differential equations:
\begin{eqnarray}
\left[(1-z^2)\frac{d^2}{dz^2}-2z\frac{d}{dz}-
\frac{m^2+n^2-2mnz}{1-z^2}+l(l+1)\right]
\fZ^{l}_{mn}(z)&=&0,\label{Leg1}\\
\left[(1-\overset{\ast}{z}{}^2)\frac{d^2}{d\overset{\ast}{z}{}^2}-
2\overset{\ast}{z}\frac{d}{d\overset{\ast}{z}}-
\frac{\dot{m}^2+\dot{n}^2-2\dot{m}\dot{n}\overset{\ast}{z}}
{1-\overset{\ast}{z}{}^2}+\dot{l}(\dot{l}+1)\right]
\fZ^{\dot{l}}_{\dot{m}\dot{n}}(\overset{\ast}{z})&=&0.\label{Leg2}
\end{eqnarray}
The latter equations have three singular points $-1$, $+1$,
$\infty$. The equations (\ref{Leg1}), (\ref{Leg2}) are Fuchsian
equations. Indeed, denoting $w(z)=\fZ^l_{mn}(z)$, we write the
equation (\ref{Leg1}) in the form
\begin{equation}\label{Fux}
\frac{d^2w(z)}{dz^2}-p(z)\frac{dw(z)}{dz}+q(z)w(z)=0,
\end{equation}
where
\[
p(z)=\frac{2z}{(1-z)(1+z)},\quad
q(z)=\frac{l(l+1)(1-z^2)-m^2-n^2+2mnz}{(1-z)^2(1+z)^2}.
\]
The solution of (\ref{Fux}) is
\begin{multline}
w(z)=C_1\left(\frac{1-z}{2}\right)^{\frac{|m-n|}{2}}
\left(\frac{1+z}{2}\right)^{\frac{|m+n|}{2}}\times\\
\times\hypergeom{2}{1}{l+1+\frac{1}{2}(|m-n|+|m+n|),-l+\frac{1}{2}(|m-n|+|m+n|)}
{|m-n|+1}{\frac{1-z}{2}}+\\
+C_2\left(\frac{1-z}{2}\right)^{-\frac{|m-n|}{2}}
\left(\frac{1+z}{2}\right)^{\frac{|m+n|}{2}}\times\\
\times\hypergeom{2}{1}{-l+\frac{1}{2}(|m+n|-|m-n|),l+1+\frac{1}{2}(|m+n|-|m-n|)}
{1-|m-n|}{\frac{1-z}{2}}. \label{Sol1}
\end{multline}
It is obvious that a solution of (\ref{Leg2}) has the analogous
structure.

Let us now consider spherical functions $f(g)$ and homogeneous
spaces $\cM=\SO(4)/H$ of the group $\SO(4)$ depending on the
stabilizer $H$. First of all, when $H=0$ the homogeneous space
$\cM_6$ coincides with {\it a group manifold} $\fK_6$ of $\SO(4)$.
Therefore, $\fK_6$ is a maximal homogeneous space of the group
$\SO(4)$. Further, when $H=\Omega^e_\psi$, where $\Omega^e_\psi$ is
a group of diagonal matrices
\[
\begin{pmatrix}
e^{\frac{\bi\psi^e}{2}} & 0\\
0 & e^{-\frac{\bi\psi^e}{2}}
\end{pmatrix},
\]
the homogeneous space $\cM_4$ coincides with a {\it two-dimensional
double sphere} $S^e_2$, $\cM_4=S^e_2\sim\spin(4)/\Omega^e_\psi$. The
sphere $S^e_2$ can be constructed from the quantities
$z_k=x_k+ey_k$, $\overset{\ast}{z}_k=x_k-ey_k$ $(k=1,2,3)$ as
follows:
\begin{equation}\label{DBS}
S^e_2:\;z^2_1+z^2_2+z^2_3=\bx^2+\by^2+2e\bx\by=r^2,
\end{equation}
where $e$ is {\it a double unit}, $e^2=1$. The conjugate (dual)
sphere $\dot{S}^e_2$ is
\begin{equation}\label{DDS}
\dot{S}^e_2:\;\overset{\ast}{z}_1{}^2+\overset{\ast}{z}_2{}^2+
\overset{\ast}{z}_3{}^2=\bx^2+\by^2-2e\bx\by=\overset{\ast}{r}{}^2.
\end{equation}
We obtain the following homogeneous space $\cM_3$  when the
stabilizer $H$ coincides with a subgroup $\SO(3)$. In this case we
have a three-dimensional sphere $\cM_3=S^3\sim\SO(4)/\SO(3)$ in the
space $\R^4$.

Finally, a minimal homogeneous space $\cM_2$ of $\SO(4)$ is a
two-dimensional real sphere $S_2\sim\SO(3)/\SO(2)$. All the
homogeneous spaces of $\SO(4)$ are symmetric Riemannian spaces.

Taking into account the list of homogeneous spaces of $\SO(4)$, we
now introduce the following types of spherical functions $f(g)$ on
the group $\SO(4)$.
\begin{itemize}
\item $f(g)=\fM^l_{mn}(g)$. This function is defined on the
group manifold $\fK_6$ of $\SO(4)$. It is the most general spherical
function on the group $\SO(4)$. In this case $f(g)$ depends on all
the six parameters of $\SO(4)$ and for that reason it should be
called as {\it a function on the group $\SO(4)$}.
\item $f(\varphi^e,\theta^e)=\fM^m_l(\varphi^e,\theta^e,0)$.
This function is defined on the homogeneous space
$\cM_4=S^e_2\sim\SO(4)/\Omega^e_\psi$, that is, on the surface of
the two-dimensional double sphere $S^e_2$. The function
$\fM^m_l(\varphi^e,\theta^e,0)$ is a four-dimensional analogue of
the usual spherical function $Y^m_l(\varphi,\theta)$ defined on the
surface of the real two-sphere $S^2$. In its turn, the function
$f(\dot{\varphi}^e,\dot{\theta}^e)=
\fM^{\dot{m}}_{\dot{l}}(\dot{\varphi}^e,\dot{\theta}^e,0)$ is
defined on the surface of the dual sphere $\dot{S}^e_2$.
\item
$f(\varphi,\theta,\psi)=e^{-\bi m\varphi}P^l_{mn}(\cos\theta)e^{-\bi
n\psi}$ (or $f(\varsigma,\phi,\chi)=e^{-\bi
m\varsigma}P^l_{mn}(\cos\phi) e^{-\bi n\chi}$). This function is
defined on the homogeneous space $\cM_3\sim S^3=\SO(4)/\SO(3)$, that
is, on the surface of the real 3-sphere $x^2_0+x^2_1+x^2_2+x^2_3=1$.
\item $f(\varphi,\theta)=e^{-\bi m\varphi}P^m_{l}(\cos\theta)\sim
Y^m_l(\varphi,\theta)$ (or $f(\varsigma,\phi)=e^{-\bi
m\varsigma}P^m_{l}(\cos\phi)\sim Y^m_l(\varsigma,\phi)$). This
function is defined on the homogeneous space
$\cM_2=S^2\sim\SO(3)/\SO(2)$, that is, on the surface of the real
2-sphere $S^2$. We come here to the most degenerate representations
of $\SO(4)$ restricted to the subgroup $\SU(2)$.
\end{itemize}

First, let us consider spherical functions
$f(g)=\fM^l_{mn}(g)=e^{-\bi
m\varphi^e}\fZ^l_{mn}(\cos\theta^e)e^{-\bi n\psi^e}$ on the group
manifold $\fK_6$ of $\SO(4)$. The Laplace-Beltrami operators
$\bigtriangleup_L(\fK_6)$ and $\overline{\bigtriangleup}_L(\fK_6)$
are coincide with (\ref{BLO1}) and (\ref{BLO2}). Spherical functions
of the first type $f(g)=\fM^l_{mn}(g)$
($f(\dot{g})=\fM^{\dot{l}}_{\dot{m}\dot{n}}(\dot{g})$) are
eigenfunctions of the operator $\bigtriangleup_L(\fK_6)$
($\overline{\bigtriangleup}_L(\fK_6)$). With the aim to find an
explicit form of hyperspherical functions on
$\fZ^l_{mn}(\cos\theta^e)$, we will use an addition theorem for
generalized spherical functions $P^l_{mn}(\cos\theta)$ of the group
$\SU(2)$ \cite{Vil65}:
\begin{equation}\label{Add1}
e^{-\bi(m\varphi+n\psi)}P^l_{mn}(\cos\theta)=\sum_{k=-l}^le^{-\bi
k\varphi_2} P^l_{mk}(\cos\theta_1)P^l_{kn}(\cos\theta_2),
\end{equation}
where the angles $\varphi$, $\psi$, $\theta$, $\theta_1$,
$\varphi_2$, $\theta_2$ are related by the formulae
\begin{eqnarray}
\cos\theta&=&\cos\theta_1\cos\theta_2-\sin\theta_1\sin\theta_2\cos\varphi_2,\label{Add2}\\
e^{\bi\varphi}&=&\frac{\sin\theta_1\cos\theta_2+\cos\theta_1\sin\theta_2\cos\varphi_2+
\bi\sin\theta_2\sin\varphi_2}{\sin\theta},\label{Add3}\\
e^{\frac{\bi(\varphi+\psi)}{2}}&=&\frac{\cos\frac{\theta_1}{2}\cos\frac{\theta_2}{2}
e^{\bi\frac{\varphi_2}{2}}-\sin\frac{\theta_1}{2}\sin\frac{\theta_2}{2}
e^{-\bi\frac{\varphi_2}{2}}}{\cos\frac{\theta}{2}}.\label{Add4}
\end{eqnarray}
Let $\cos(\theta+\phi)=\cos\theta^e$ and $\varphi_2=0$, then the
formulae (\ref{Add2})--(\ref{Add4}) take the form
\begin{eqnarray}
\cos\theta^e&=&\cos\theta\cos\phi-\sin\theta\sin\phi,\nonumber\\
e^{\bi\varphi}&=&\frac{\sin\theta\cos\phi+\cos\theta\sin\phi}{\sin\theta^e}=1,\nonumber\\
e^{\frac{\bi(\varphi+\psi)}{2}}&=&\frac{\cos\frac{\theta}{2}\cos\frac{\phi}{2}-
\sin\frac{\theta}{2}\sin\frac{\phi}{2}}{\cos\frac{\theta^e}{2}}=1.
\nonumber
\end{eqnarray}
Hence it follows that $\varphi=\psi=0$ and the formula (\ref{Add1})
can be written as
\begin{equation}\label{HFSO4}
\fZ^l_{mn}(\cos\theta^e)=\sum^l_{k=-l}P^l_{mk}(\cos\theta)P^l_{kn}(\cos\phi).
\end{equation}
$\fZ^l_{mn}(\cos\theta^e)$ are {\it hyperspherical functions of the
group $\SO(4)$}\footnote{The functions $\fZ^l_{mn}(\cos\theta^e)$
and $\fZ^{\dot{l}}_{\dot{m}\dot{n}}(\cos\dot{\theta}^e)$ form a
representation of the type $(l,0)\oplus(0,\dot{l})$, that is, when
$l=\dot{l}$. In the case of tensor representations, when
$l\ne\dot{l}$, we arrive at the functions
$\fZ^{l\dot{l}}_{mn;\dot{m}\dot{n}}(\cos\theta^e,\cos\dot{\theta}^e)=
\fZ^l_{mn}(\cos\theta^e)\fZ^{\dot{l}}_{\dot{m}\dot{n}}(\cos\dot{\theta}^e)$
({\it generalized hyperspherical functions of} $\SO(4)$), which can
be expressed via the product of the two generalized hypergeometric
functions
$\hypergeom{3}{2}{\alpha,\beta,\gamma}{\delta,\epsilon}{x}$. In the
case of Lorentz group, general solutions of relativistic wave
equations for arbitrary spin chains (tensor representations) are
defined via an expansion in generalized hyperspherical functions
$\fZ^{l\dot{l}}_{mn;\dot{m}\dot{n}}(\cos\theta^c,\cos\dot{\theta}^c)$
of $\SO_0(1,3)$, where $\theta^c$, $\dot{\theta}^c$ are complex
Euler angles of $\spin_+(1,3)\simeq\SL(2,\C)$ \cite{Var05}.}. Using
an explicit expression for the function $P^l_{mn}$
\cite{Vil65,Var06}, we obtain
\begin{multline}
\fZ^l_{mn}(\cos\theta^e)= \sum^l_{k=-l}\bi^{m+n-2k}
\sqrt{\Gamma(l-m+1)\Gamma(l+m+1)\Gamma(l-k+1)\Gamma(l+k+1)}\times\\
\cos^{2l}\frac{\theta}{2}\tg^{m-k}\frac{\theta}{2}\times\\[0.2cm]
\sum^{\min(l-m,l+k)}_{j=\max(0,k-m)}
\frac{\bi^{2j}\tg^{2j}\dfrac{\theta}{2}}
{\Gamma(j+1)\Gamma(l-m-j+1)\Gamma(l+k-j+1)\Gamma(m-k+j+1)}\times\\[0.2cm]
\sqrt{\Gamma(l-n+1)\Gamma(l+n+1)\Gamma(l-k+1)\Gamma(l+k+1)}
\cos^{2l}\frac{\phi}{2}\tan^{n-k}\frac{\phi}{2}\times\\[0.2cm]
\sum^{\min(l-n,l+k)}_{s=\max(0,k-n)}
\frac{\bi^{2s}\tan^{2s}\dfrac{\phi}{2}}
{\Gamma(s+1)\Gamma(l-n-s+1)\Gamma(l+k-s+1)\Gamma(n-k+s+1)}.\label{PPtan}
\end{multline}

On the other hand, the function $\fZ^l_{mn}(\cos\theta^e)$ can be
expressed via the hypergeometric function. Using hypergeometric-type
formulae for $P^l_{mn}$ \cite{Vil65,Var06}, we have at $m\geq n$
\begin{multline}
\fZ^l_{mn}(\cos\theta^e)=\bi^{m-n}\sqrt{\frac{\Gamma(l+m+1)\Gamma(l-n+1)}{\Gamma(l-m+1)\Gamma(l+n+1)}}
\cos^{2l}\frac{\theta}{2}\cos^{2l}\frac{\phi}{2}\times\\
\sum^l_{k=-l}\tg^{m-k}\frac{\theta}{2}\tan^{k-n}\frac{\phi}{2}\times\\
\times\hypergeom{2}{1}{m-l,-k-l}{m-k+1}{-\tg^2\frac{\theta}{2}}
\hypergeom{2}{1}{k-l,-n-l}{k-n+1}{-\tan^2\frac{\phi}{2}},\quad m\geq
k,\;k\geq n; \label{PBFtan1}
\end{multline}
\begin{multline}
\fZ^l_{mn}(\cos\theta^e)=\sqrt{\frac{\Gamma(l+m+1)\Gamma(l-n+1)}{\Gamma(l-m+1)\Gamma(l+n+1)}}
\cos^{2l}\frac{\theta}{2}\cos^{2l}\frac{\phi}{2}\times\\
\sum^l_{k=-l}\bi^{m+n-2k}\tg^{m-k}\frac{\theta}{2}\tan^{n-k}\frac{\phi}{2}\times\\
\times\hypergeom{2}{1}{m-l,-k-l}{m-k+1}{-\tg^2\frac{\theta}{2}}
\hypergeom{2}{1}{n-l,-k-l}{n-k+1}{-\tan^2\frac{\phi}{2}},\quad m\geq
k,\;n\geq k; \label{PBFtan2}
\end{multline}
and at $n\geq m$
\begin{multline}
\fZ^l_{mn}(\cos\theta^e)=\bi^{n-m}\sqrt{\frac{\Gamma(l-m+1)\Gamma(l+n+1)}{\Gamma(l+m+1)\Gamma(l-n+1)}}
\cos^{2l}\frac{\theta}{2}\cos^{2l}\frac{\phi}{2}\times\\
\sum^l_{k=-l}\tg^{k-m}\frac{\theta}{2}\tan^{n-k}\frac{\phi}{2}\times\\
\times\hypergeom{2}{1}{k-l,-m-l}{k-m+1}{-\tg^2\frac{\theta}{2}}
\hypergeom{2}{1}{n-l,-k-l}{n-k+1}{-\tan^2\frac{\phi}{2}},\quad k\geq
m,\;n\geq k; \label{PBFtan3}
\end{multline}
\begin{multline}
\fZ^l_{mn}(\cos\theta^e)=\sqrt{\frac{\Gamma(l-m+1)\Gamma(l+n+1)}{\Gamma(l+m+1)\Gamma(l-n+1)}}
\cos^{2l}\frac{\theta}{2}\cos^{2l}\frac{\phi}{2}\times\\
\sum^l_{k=-l}\bi^{2k-m-n}\tg^{k-m}\frac{\theta}{2}\tan^{k-n}\frac{\phi}{2}\times\\
\times\hypergeom{2}{1}{k-l,-m-l}{k-m+1}{-\tg^2\frac{\theta}{2}}
\hypergeom{2}{1}{k-l,-n-l}{k-n+1}{-\tan^2\frac{\phi}{2}},\quad k\geq
m,\;k\geq n. \label{PBFtan4}
\end{multline}
By way of example let us calculate matrix elements
$\fM^l_{mn}(g)=e^{-\bi m\varphi^e}\fZ^l_{mn}(\cos\theta^e)e^{-\bi
n\psi^e}$ at $l=0,\,1/2,\,1$, where $\fZ^l_{mn}(\cos\theta^e)$ is
defined via (\ref{PPtan}) or (\ref{PBFtan1})--(\ref{PBFtan4}). The
representation matrices at $l=0,\,\frac{1}{2},\,1$ have the
following form:
\begin{gather}
T_0(\varphi^e,\theta^e,\psi^e)=1,\label{T0}\\[0.3cm]
T_{\frac{1}{2}}(\varphi^e,\theta^e,\psi^e)=\ar\begin{pmatrix}
\fM^{\frac{1}{2}}_{-\frac{1}{2}-\frac{1}{2}} &
\fM^{\frac{1}{2}}_{\frac{1}{2}-\frac{1}{2}}\\
\fM^{\frac{1}{2}}_{-\frac{1}{2}\frac{1}{2}} &
\fM^{\frac{1}{2}}_{\frac{1}{2}\frac{1}{2}}
\end{pmatrix}=\ar\begin{pmatrix}
e^{\frac{\bi}{2}\varphi^e}\fZ^{\frac{1}{2}}_{-\frac{1}{2}-\frac{1}{2}}e^{\frac{\bi}{2}\psi^e}
&
e^{\frac{\bi}{2}\varphi^e}\fZ^{\frac{1}{2}}_{-\frac{1}{2}\frac{1}{2}}e^{-\frac{\bi}{2}\psi^e}\\
e^{-\frac{\bi}{2}\varphi^e}\fZ^{\frac{1}{2}}_{\frac{1}{2}-\frac{1}{2}}e^{\frac{\bi}{2}\psi^e}
&
e^{-\frac{\bi}{2}\varphi^e}\fZ^{\frac{1}{2}}_{\frac{1}{2}\frac{1}{2}}e^{-\frac{\bi}{2}\psi^e}
\end{pmatrix}=\nonumber\\[0.3cm]
=\ar\begin{pmatrix}
e^{\frac{\bi}{2}\varphi^e}\cos\frac{\theta^e}{2}e^{\frac{\bi}{2}\psi^e}
&
\bi e^{\frac{\bi}{2}\varphi^e}\sin\frac{\theta^e}{2}e^{-\frac{\bi}{2}\psi^e}\\
\bi
e^{-\frac{\bi}{2}\varphi^e}\sin\frac{\theta^e}{2}e^{\frac{\bi}{2}\psi^e}
&
e^{-\frac{\bi}{2}\varphi^e}\cos\frac{\theta^e}{2}e^{-\frac{\bi}{2}\psi^e}
\end{pmatrix}=\nonumber\\[0.3cm]
{\renewcommand{\arraystretch}{1.3} =\begin{pmatrix}
\left[\cos\frac{\theta}{2}\cos\frac{\phi}{2}-
\sin\frac{\theta}{2}\sin\frac{\phi}{2}\right]
e^{\frac{\bi(\varphi+\varsigma+\psi+\chi))}{2}} &
\bi\left[\cos\frac{\theta}{2}\sin\frac{\phi}{2}+
\sin\frac{\theta}{2}\cos\frac{\phi}{2}\right]
e^{\frac{\bi(\varphi+\varsigma-\psi-\chi)}{2}} \\
\bi\left[\cos\frac{\theta}{2}\sin\frac{\phi}{2}+
\sin\frac{\theta}{2}\cos\frac{\phi}{2}\right]
e^{\frac{\bi(-\varphi-\varsigma+\psi+\chi)}{2}} &
\left[\cos\frac{\theta}{2}\cos\frac{\phi}{2}-
\sin\frac{\theta}{2}\sin\frac{\phi}{2}\right]
e^{\frac{-\bi(\varphi+\varsigma+\psi+\chi)}{2}}
\end{pmatrix}},\label{T1}
\end{gather}
\begin{gather}
T_1(\varphi^e,\theta^e,\psi^e)=\ar\begin{pmatrix}
\fM^1_{-1-1} & \fM^1_{-10} & \fM^1_{-11}\\
\fM^1_{0-1} & \fM^1_{00} & \fM^1_{01}\\
\fM^1_{1-1} & \fM^1_{10} & \fM^1_{11}
\end{pmatrix}=\ar
\begin{pmatrix}
e^{\bi\varphi^e}\fZ^1_{-1-1}e^{\bi\psi^e} &
e^{\bi\varphi^e}\fZ^1_{-10} & e^{\bi\varphi^e}
\fZ^1_{-11}e^{-\bi\psi^e}\\
\fZ^1_{0-1}e^{\bi\psi^e} & \fZ^1_{00} & \fZ^1_{01}e^{-\bi\psi^e}\\
e^{-\bi\varphi^e}\fZ^1_{1-1}e^{\bi\psi^e} & e^{-\bi\psi^e}\fZ^1_{10}
& e^{-\bi\varphi^e}\fZ^1_{11}e^{-\bi\psi^e}
\end{pmatrix}=\nonumber\\[0.3cm]
=\ar\begin{pmatrix}
e^{\bi\varphi^e}\cos^2\frac{\theta^e}{2}e^{\bi\psi^e} &
\frac{\bi}{\sqrt{2}}e^{\bi\varphi^e}\sin\theta^e & -e^{\bi\varphi^e}
\sin^2\frac{\theta^e}{2}e^{-\bi\psi^e}\\
\frac{\bi}{\sqrt{2}}\sin\theta^ee^{\bi\psi^e} & \cos\theta^e &
\frac{\bi}{\sqrt{2}}\sin\theta^ee^{-\bi\psi^e}\\
-e^{-\bi\varphi^e}\sin^2\frac{\theta^e}{2}e^{\bi\psi^e} &
\frac{\bi}{\sqrt{2}}e^{-\bi\varphi^e}\sin\theta^e &
e^{-\bi\varphi^e}\cos^2\frac{\theta^e}{2}e^{-\bi\psi^e}
\end{pmatrix}=\nonumber
\end{gather}
\begin{multline}
{\renewcommand{\arraystretch}{1.1}=\left(\begin{array}{cc}\scr
\left[\cos^2\frac{\theta}{2}\cos^2\frac{\phi}{2}-\frac{\sin\theta\sin\phi}{2}+
\sin^2\frac{\theta}{2}\sin^2\frac{\phi}{2}\right]
e^{\bi(\varphi+\varsigma+\psi+\chi)} &\scr
\left[\frac{\bi}{\sqrt{2}}(\cos\theta\sin\phi+\sin\theta\cos\phi)\right]
e^{\bi(\varphi+\varsigma)} \\
\scr\left[\frac{\bi}{\sqrt{2}}(\cos\theta\sin\phi+\sin\theta\cos\phi)\right]
e^{\bi(\psi+\chi)} &\scr
\cos\theta\cos\phi-\sin\theta\sin\phi \\
\scr-\left[\cos^2\frac{\theta}{2}\sin^2\frac{\phi}{2}+
\frac{\sin\theta\sin\phi}{2}+
\sin^2\frac{\theta}{2}\cos^2\frac{\phi}{2}\right]
e^{\bi(-\varphi-\varsigma+\psi+\chi)} &\scr
\left[\frac{\bi}{\sqrt{2}}(\cos\theta\sin\phi+\sin\theta\cos\phi)\right]
e^{-\bi(\varphi+\varsigma)}
\end{array}\right.}\\
{\renewcommand{\arraystretch}{1.1}\left.\begin{array}{c}\scr
-\left[\cos^2\frac{\theta}{2}\sin^2\frac{\phi}{2}+\frac{\sin\theta\sin\phi}{2}+
\sin^2\frac{\theta}{2}\cos^2\frac{\phi}{2}\right]
e^{\bi(\varphi+\varsigma-\psi-\chi)} \\
\scr\left[\frac{\bi}{\sqrt{2}}(\cos\theta\sin\phi+\sin\theta\cos\phi)\right]
e^{-\bi(\psi+\chi)} \\
\scr\left[\cos^2\frac{\theta}{2}\cos^2\frac{\phi}{2}-\frac{\sin\theta\sin\phi}{2}+
\sin^2\frac{\theta}{2}\sin^2\frac{\phi}{2}\right]
e^{-\bi(\varphi+\varsigma+\psi+\chi)}
\end{array}\right).}\label{T2}
\end{multline}

Spherical functions of the second type
$f(\varphi^e,\theta^e)=\fM^m_l(\varphi^e,\theta^e,0)= e^{-\bi
m\varphi^e}\fZ^m_l(\cos\theta^e)$, where
\[
\fZ^m_l(\cos\theta^e)=\sum^l_{k=-l}P^l_{mk}(\cos\theta)P^k_l(\cos\phi)
\]
is {\it an associated hyperspherical function}, are defined on the
surface of the double 2-sphere (\ref{DBS}). The function
$\fZ^m_l(\cos\theta^e)$ is an eigenfunction of the Laplace-Beltrami
operator $\bigtriangleup_L(S^e_2)$ defined on the double 2-sphere,
\[
\bigtriangleup_L(S^e_2)=\frac{\partial^2}{\partial\theta^e{}^2}+
\ctg\theta^e\frac{\partial}{\partial\theta^e}+\frac{1}{\sin^2\theta^e}
\frac{\partial^2}{\partial\varphi^e{}^2}.
\]
Hypergeometric-type formulae for $\fZ^m_l(\cos\theta^e)$ are
\begin{multline}
\fZ_l^{m}(\cos\theta^e)=\bi^{m}\sqrt{\frac{\Gamma(l+m+1)}{\Gamma(l-m+1)}}
\cos^{2l}\frac{\theta}{2}\cos^{2l}\frac{\phi}{2}
\sum^l_{k=-l}\tg^{m-k}\frac{\theta}{2}\tan^{k}\frac{\phi}{2}\times\\
\times\hypergeom{2}{1}{m-l,-k-l}{m-k+1}{-\tg^2\frac{\theta}{2}}
\hypergeom{2}{1}{k-l,-l}{k+1}{-\tan^2\frac{\phi}{2}},\quad m\geq k;
\nonumber
\end{multline}
\begin{multline}
\fZ_l^{m}(\cos\theta^e)=\sqrt{\frac{\Gamma(l-m+1)}{\Gamma(l+m+1)}}
\cos^{2l}\frac{\theta}{2}\cos^{2l}\frac{\phi}{2}
\sum^l_{k=-l}\bi^{2k-m}\tg^{k-m}\frac{\theta}{2}\tan^{k}\frac{\phi}{2}\times\\
\times\hypergeom{2}{1}{k-l,-m-l}{k-m+1}{-\tg^2\frac{\theta}{2}}
\hypergeom{2}{1}{k-l,-l}{k+1}{-\tan^2\frac{\phi}{2}},\quad k\geq m.
\nonumber
\end{multline}
We obtain an important particular case from the previous formulae at
$m=n=0$. The function $\fZ_l(\cos\theta^e)\equiv
\fZ^l_{00}(\cos\theta^e)$ is called {\it a zonal hyperspherical
function}. The hypergeometric-type formula for $\fZ_l(\cos\theta^e)$
is
\begin{multline}
\fZ_l(\cos\theta^e)=
\cos^{2l}\frac{\theta}{2}\cos^{2l}\frac{\phi}{2}
\sum^l_{k=-l}\bi^{2k}\tg^{k}\frac{\theta}{2}\tan^{k}\frac{\phi}{2}\times\\
\times\hypergeom{2}{1}{k-l,-l}{k+1}{-\tg^2\frac{\theta}{2}}
\hypergeom{2}{1}{k-l,-l}{k+1}{-\tan^2\frac{\phi}{2}}. \nonumber
\end{multline}
In its turn, the function
$f(\dot{\varphi}^e,\dot{\theta}^e)=e^{\bi\dot{m}\dot{\varphi}^e}
\fZ^{\dot{m}}_{\dot{l}}(\cos\dot{\theta}^e)$ (or $f(\dot{\theta}^e)=
\fZ_{\dot{l}}(\cos\dot{\theta}^e)$) are defined on the surface of
dual sphere (\ref{DDS}). Explicit expressions and
hypergeometric-type formulae for $f(\dot{\varphi}^e,\dot{\theta}^e)$
are analogous to the previous expressions for
$f(\varphi^e,\theta^e)$.
\begin{sloppypar}
Spherical functions of the third type
$f(\varphi,\theta,\psi)=e^{-\bi m\varphi}P^l_{mn}(\cos\theta)e^{-\bi
n\psi}$ (or $f(\varsigma,\phi,\chi)=e^{-\bi
m\varsigma}P^l_{mn}(\cos\phi) e^{-\bi n\chi}$) are defined on the
surface of the real 3-sphere $S^3=\SO(4)/\SO(3)$. These functions
are general matrix elements of representations of the group
$\SO(3)$. Therefore, we have here representations of $\SO(4)$
restricted to the subgroup $\SO(3)$. Namely,\end{sloppypar}
\begin{equation}\label{Rest}
\hat{T}^l\downarrow^{\SO(4)}_{\SO(3)}=\sum^l_{m=0}\oplus Q^m,
\end{equation}
where spherical functions $f(\varphi,\theta,\psi)$ of the
representations $Q^m$ of $\SO(3)$ form an orthogonal basis in the
Hilbert space $L^2(S^3)$. Various expressions and
hypergeometric-type formulae for $f(\varphi,\theta,\psi)$ are given
in \cite{Vil65,Var06}.

Finally, spherical functions of the fourth type
$f(\varphi,\theta)=e^{-\bi m\varphi}P^m_{l}(\cos\theta)\sim
Y^m_l(\varphi,\theta)$ (or $f(\varsigma,\phi)=e^{-\bi
m\varsigma}P^m_{l}(\cos\phi)\sim Y^m_l(\varsigma,\phi)$) are defined
on the surface of the real 2-sphere. We have here representations
$\hat{T}^l\downarrow^{\SO(4)}_{\SO(3)}$ of the type (\ref{Rest}),
where associated spherical functions $f(\varphi,\theta)\sim
Y^m_l(\varphi,\theta)$ of $Q^m$ form an orthogonal basis in
$L^2(S^3)$. These representations are the most degenerate for the
group $\SO(4)$.

\section{Spherical functions of finite-dimensional representations
of $\SO_0(1,4)$} Let us come back to the de Sitter group
$\SO_0(1,4)$. It has been shown in the section 1 that spherical
functions of the first type $f(\fq)=\fM^\sigma_{mn}(\fq)=e^{-\bi
m\varphi^q}\fZ^\sigma_{mn}(\cos\theta^q)e^{-\bi n\psi^q}$ are
defined on the group manifold $\fS_{10}$ of $\SO_0(1,4)$. With the
aim to find an explicit form of hyperspherical function
$\fZ^\sigma_{mn}(\cos\theta^q)$ of the group $\SO_0(1,4)$, we will
use the addition theorem defined by the formulae
(\ref{Add1})--(\ref{Add4}). Let
$\cos(\theta+\phi-\bi\tau)=\cos(\theta^e-\bi\tau)=\cos\theta^q$ and
$\varphi_2=0$, then the formulae (\ref{Add2})--(\ref{Add4}) take the
form
\begin{eqnarray}
\cos\theta^q&=&\cos\theta^e\ch\tau+\bi\sin\theta^e\sh\tau,\nonumber\\
e^{\bi\varphi}&=&\frac{\sin\theta^e\ch\tau-\bi\cos\theta^e\sh\tau}{\sin\theta^q}=1,\nonumber\\
e^{\frac{\bi(\varphi+\psi)}{2}}&=&\frac{\cos\frac{\theta^e}{2}\ch\frac{\tau}{2}+
\bi\sin\frac{\theta^e}{2}\sh\frac{\tau}{2}}{\cos\frac{\theta^q}{2}}=1.
\nonumber
\end{eqnarray}
Hence it follows that $\varphi=\psi=0$ and formula (\ref{Add1}) can
be written as
\begin{equation}\label{HFSO14}
\fZ^\sigma_{mn}(\cos\theta^q)=\sum^\sigma_{k=-\sigma}\fZ^\sigma_{mk}(\cos\theta^e)
\fP^\sigma_{kn}(\ch\tau),
\end{equation}
where $\fZ^\sigma_{mn}(\cos\theta^e)$ is the hyperspherical function
of the compact subgroup $\SO(4)$ (see the formula (\ref{HFSO4})):
\[
\fZ^\sigma_{mk}(\cos\theta^e)=\sum^\sigma_{t=-\sigma}
P^\sigma_{mt}(\cos\theta)P^\sigma_{tk}(\cos\phi).
\]
It is easy to verify that if we take
$\cos(\theta+\phi-\bi\tau)=\cos(\phi+\theta^c)=\cos\theta^q$ and
$\varphi_2=0$ in the formulae (\ref{Add2})--(\ref{Add4}), then we
arrive at the function
\[
\fZ^\sigma_{mn}(\cos\theta^q)=\sum\limits^\sigma_{k=-\sigma}
P^\sigma_{mk}(\cos\phi)\fZ^\sigma_{kn}(\cos\theta^c),
\]
where
\[
\fZ^\sigma_{kn}(\cos\theta^c)=\sum\limits^\sigma_{t=-\sigma}
P^\sigma_{kt}(\cos\theta)\fP^\sigma_{tn}(\cosh\tau)
\]
is the hyperspherical function of the subgroup $\SO_0(1,3)$. In such
a way, the hyperspherical function $\fZ^\sigma_{mn}(\cos\theta^q)$
can be factorized with respect to the subgroups $\SO(4)$ and
$\SO_0(1,3)$.

Further, taking into account the expression for
$\fZ^\sigma_{mk}(\cos\theta^e)$, we can rewrite (\ref{HFSO14}) in
the following form:
\begin{equation}\label{HFSO14b}
\fZ^\sigma_{mn}(\cos\theta^q)=\sum^\sigma_{k=-\sigma}
\sum^\sigma_{t=-\sigma}P^\sigma_{mt}(\cos\theta)
P^\sigma_{tk}(\cos\phi)\fP^\sigma_{kn}(\cosh\tau).
\end{equation}
Analogously, for the factorization of
$\fZ^\sigma_{mn}(\cos\theta^q)$ with respect to the Lorentz subgroup
$\SO_0(1,3)$ we have
\[
\fZ^\sigma_{mn}(\cos\theta^q)=\sum^\sigma_{k=-\sigma}\sum^\sigma_{t=-\sigma}
P^\sigma_{mk}(\cos\phi)P^\sigma_{kt}(\cos\theta)\fP^\sigma_{tn}(\ch\tau).
\]
We consider here only the factorization of
$\fZ^\sigma_{mn}(\cos\theta^q)$ with respect to the maximal compact
subgroup $\SO(4)$. Thus, the formulae (\ref{HFSO14}) and
(\ref{HFSO14b}) define {\it a hyperspherical function of the de
Sitter group} $\SO_0(1,4)$ with respect to $\SO(4)$. Further, using
(\ref{PPtan}), we obtain an explicit expression for
$\fZ^\sigma_{mn}(\cos\theta^q)$,
\begin{multline}
\fZ^\sigma_{mn}(\cos\theta^q)=
\sum^\sigma_{k=-\sigma}\sum^\sigma_{t=-\sigma}\bi^{m+k-2t}
\sqrt{\Gamma(\sigma-m+1)\Gamma(\sigma+m+1)\Gamma(\sigma-t+1)\Gamma(\sigma+t+1)}\times\\
\cos^{2\sigma}\frac{\theta}{2}\tg^{m-t}\frac{\theta}{2}\times\\[0.2cm]
\sum^{\min(\sigma-m,l+t)}_{j=\max(0,t-m)}
\frac{\bi^{2j}\tg^{2j}\dfrac{\theta}{2}}
{\Gamma(j+1)\Gamma(\sigma-m-j+1)\Gamma(\sigma+t-j+1)\Gamma(m-t+j+1)}\times\\[0.2cm]
\sqrt{\Gamma(\sigma-k+1)\Gamma(\sigma+k+1)\Gamma(\sigma-t+1)\Gamma(\sigma+t+1)}
\cos^{2\sigma}\frac{\phi}{2}\tan^{k-t}\frac{\phi}{2}\times\\[0.2cm]
\sum^{\min(\sigma-k,\sigma+t)}_{s=\max(0,t-k)}
\frac{\bi^{2s}\tan^{2s}\dfrac{\phi}{2}}
{\Gamma(s+1)\Gamma(\sigma-k-s+1)\Gamma(\sigma+t-s+1)\Gamma(k-t+s+1)}\times\\[0.2cm]
\sqrt{\Gamma(\sigma-n+1)\Gamma(\sigma+n+1)\Gamma(\sigma-k+1)\Gamma(\sigma+k+1)}
\ch^{2\sigma}\frac{\tau}{2}\tnh^{n-k}\frac{\tau}{2}\times\\[0.2cm]
\sum^{\min(\sigma-n,\sigma+k)}_{p=\max(0,k-n)}
\frac{\tnh^{2p}\dfrac{\tau}{2}}
{\Gamma(p+1)\Gamma(\sigma-n-p+1)\Gamma(\sigma+k-p+1)\Gamma(n-k+p+1)}.\label{PPBtan}
\end{multline}
It is obvious that the functions $\fZ^\sigma_{mn}(\cos\theta^q)$ can
also be reduced to hypergeometric functions. Namely, these functions
are expressed via the following multiple hypergeometric
series\footnote{The hyperspherical functions
$\fZ^\sigma_{mn}(\cos\theta^q)$ of $\SO_0(1,4)$,
$\fZ^l_{mn}(\cos\theta^e)$ of $\SO(4)$ and
$\fZ^l_{mn}(\cos\theta^c)$ of $\SO_0(1,3)$ can be written in the
form of hypergeometric functions of many variables
\cite{AK26,Ext76}. So, the functions $\fZ^l_{mn}(\cos\theta^e)$ and
$\fZ^l_{mn}(\cos\theta^c)$ can be expressed via the Appell
functions, $\fZ^l_{mn}(\cos\theta^e)\sim
\Appell{4}{a_1,a_2}{a_3,a_4}{x_1;x_2}$ and
$\fZ^l_{mn}(\cos\theta^c)\sim\Appell{4}{a_1,a_2}{a_3,a_4}{x_1;y_1}$,
where $x_1=\tan^2\theta/2$, $x_2=\tan^2\phi/2$, $y_1=\tanh^2\tau/2$.
In its turn, the function $\fZ^\sigma_{mn}(\cos\theta^q)$ is reduced
to the Lauricella function, $\fZ^\sigma_{mn}(\cos\theta^q)\sim
\Lauricella{3}{a_1,a_2,a_3}{a_4,a_5}{x_1;x_2;y_1}$. From the
relations $\spin(4)\in\cl^+_{4,0}\simeq\cl_{0,3}$, where $\cl_{0,3}$
is the algebra of double biquaternions with a double quaternionic
division ring $\K\simeq\BH\oplus\BH$;
$\spin_+(1,3)\in\cl^+_{1,3}\simeq\cl_{3,0}$, where $\cl_{3,0}$ is
the algebra of complex biquaternions with a complex division ring
$\K\simeq\C$; $\spin_+(1,4)\in\cl^+_{1,4}\simeq\cl_{1,3}$, where
$\cl_{1,3}$ is the space-time algebra with a quaternionic division
ring $\K\simeq\BH$, we see that there is a close relationship
between hypercomplex angles of the group $\spin_+(p,q)$, division
rings of $\cl^+_{p,q}$ from the one hand and hypergeometric
functions of many variables from the other hand. A detailed
consideration of this relationship comes beyond the framework of
this paper and will be given in a separate work.} :
\begin{multline}
\fZ^\sigma_{mn}(\cos\theta^q)=\sqrt{\frac{\Gamma(\sigma+m+1)\Gamma(\sigma-n+1)}
{\Gamma(\sigma-m+1)\Gamma(\sigma+n+1)}}
\cos^{2\sigma}\frac{\theta}{2}\cos^{2\sigma}\frac{\phi}{2}\ch^{2\sigma}\frac{\tau}{2}\times\\
\sum^\sigma_{k=-\sigma}\sum^\sigma_{t=-\sigma}\bi^{m-k}\tg^{m-t}\frac{\theta}{2}\tan^{t-k}\frac{\phi}{2}
\tanh^{k-n}\frac{\tau}{2}\times\\
\hypergeom{2}{1}{m-\sigma,-t-\sigma}{m-t+1}{-\tg^2\frac{\theta}{2}}
\hypergeom{2}{1}{t-\sigma,-k-\sigma}{t-k+1}{-\tan^2\frac{\phi}{2}}
\hypergeom{2}{1}{k-\sigma,-n-\sigma}{k-n+1}{\tanh^2\frac{\tau}{2}},\\
\quad m\geq t,\;t\geq k,\;k\geq n; \label{PPBFtan1}
\end{multline}
\begin{multline}
\fZ^\sigma_{mn}(\cos\theta^q)=\sqrt{\frac{\Gamma(\sigma+m+1)\Gamma(\sigma-n+1)}
{\Gamma(\sigma-m+1)\Gamma(\sigma+n+1)}}
\cos^{2\sigma}\frac{\theta}{2}\cos^{2\sigma}\frac{\phi}{2}\ch^{2\sigma}\frac{\tau}{2}\times\\
\sum^\sigma_{k=-\sigma}\sum^\sigma_{t=-\sigma}\bi^{m+k-2t}\tg^{m-t}\frac{\theta}{2}\tan^{k-t}\frac{\phi}{2}
\tanh^{k-n}\frac{\tau}{2}\times\\
\hypergeom{2}{1}{m-\sigma,-t-\sigma}{m-t+1}{-\tg^2\frac{\theta}{2}}
\hypergeom{2}{1}{k-\sigma,-t-\sigma}{k-t+1}{-\tan^2\frac{\phi}{2}}
\hypergeom{2}{1}{k-\sigma,-n-\sigma}{k-n+1}{\tanh^2\frac{\tau}{2}},\\
\quad m\geq t,\;k\geq t,\;k\geq n; \label{PPBFtan2}
\end{multline}
\begin{multline}
\fZ^\sigma_{mn}(\cos\theta^q)=\sqrt{\frac{\Gamma(\sigma-m+1)\Gamma(\sigma+n+1)}
{\Gamma(\sigma+m+1)\Gamma(\sigma-n+1)}}
\cos^{2\sigma}\frac{\theta}{2}\cos^{2\sigma}\frac{\phi}{2}\ch^{2\sigma}\frac{\tau}{2}\times\\
\sum^\sigma_{k=-\sigma}\sum^\sigma_{t=-\sigma}\bi^{k-m}\tg^{t-m}\frac{\theta}{2}\tan^{k-t}\frac{\phi}{2}
\tanh^{n-k}\frac{\tau}{2}\times\\
\hypergeom{2}{1}{t-\sigma,-m-\sigma}{t-m+1}{-\tg^2\frac{\theta}{2}}
\hypergeom{2}{1}{k-\sigma,-t-\sigma}{k-t+1}{-\tan^2\frac{\phi}{2}}
\hypergeom{2}{1}{n-\sigma,-k-\sigma}{n-k+1}{\tanh^2\frac{\tau}{2}},\\
\quad t\geq m,\;k\geq t,\;n\geq k; \label{PPBFtan3}
\end{multline}
\begin{multline}
\fZ^\sigma_{mn}(\cos\theta^q)=\sqrt{\frac{\Gamma(\sigma-m+1)\Gamma(\sigma+n+1)}
{\Gamma(\sigma+m+1)\Gamma(\sigma-n+1)}}
\cos^{2\sigma}\frac{\theta}{2}\cos^{2\sigma}\frac{\phi}{2}\ch^{2\sigma}\frac{\tau}{2}\times\\
\sum^\sigma_{k=-\sigma}\sum^\sigma_{t=-\sigma}\bi^{2t-m-k}\tg^{t-m}\frac{\theta}{2}\tan^{t-k}\frac{\phi}{2}
\tanh^{n-k}\frac{\tau}{2}\times\\
\hypergeom{2}{1}{t-\sigma,-m-\sigma}{t-m+1}{-\tg^2\frac{\theta}{2}}
\hypergeom{2}{1}{t-\sigma,-k-\sigma}{t-k+1}{-\tan^2\frac{\phi}{2}}
\hypergeom{2}{1}{n-\sigma,-k-\sigma}{n-k+1}{\tanh^2\frac{\tau}{2}},\\
\quad t\geq m,\;t\geq k,\;n\geq k; \label{PPBFtan4}
\end{multline}
\begin{multline}
\fZ^\sigma_{mn}(\cos\theta^q)=\sqrt{\frac{\Gamma(\sigma-m+1)\Gamma(\sigma-n+1)}
{\Gamma(\sigma+m+1)\Gamma(\sigma+n+1)}}
\cos^{2\sigma}\frac{\theta}{2}\cos^{2\sigma}\frac{\phi}{2}\ch^{2\sigma}\frac{\tau}{2}\times\\
\sum^\sigma_{k=-\sigma}\sum^\sigma_{t=-\sigma}\bi^{k-m}
\frac{\Gamma(\sigma+k+1)}{\Gamma(\sigma-k+1)}
\tg^{t-m}\frac{\theta}{2}\tan^{k-t}\frac{\phi}{2}
\tanh^{k-n}\frac{\tau}{2}\times\\
\hypergeom{2}{1}{t-\sigma,-m-\sigma}{t-m+1}{-\tg^2\frac{\theta}{2}}
\hypergeom{2}{1}{k-\sigma,-t-\sigma}{k-t+1}{-\tan^2\frac{\phi}{2}}
\hypergeom{2}{1}{k-\sigma,-n-\sigma}{k-n+1}{\tanh^2\frac{\tau}{2}},\\
\quad t\geq m,\;k\geq t,\;k\geq n; \label{PPBFtan5}
\end{multline}
\begin{multline}
\fZ^\sigma_{mn}(\cos\theta^q)=\sqrt{\frac{\Gamma(\sigma-m+1)\Gamma(\sigma-n+1)}
{\Gamma(\sigma+m+1)\Gamma(\sigma+n+1)}}
\cos^{2\sigma}\frac{\theta}{2}\cos^{2\sigma}\frac{\phi}{2}\ch^{2\sigma}\frac{\tau}{2}\times\\
\sum^\sigma_{k=-\sigma}\sum^\sigma_{t=-\sigma}\bi^{2t-m-k}
\frac{\Gamma(\sigma+k+1)}{\Gamma(\sigma-k+1)}
\tg^{t-m}\frac{\theta}{2}\tan^{t-k}\frac{\phi}{2}
\tanh^{k-n}\frac{\tau}{2}\times\\
\hypergeom{2}{1}{t-\sigma,-m-\sigma}{t-m+1}{-\tg^2\frac{\theta}{2}}
\hypergeom{2}{1}{t-\sigma,-k-\sigma}{t-k+1}{-\tan^2\frac{\phi}{2}}
\hypergeom{2}{1}{k-\sigma,-n-\sigma}{k-n+1}{\tanh^2\frac{\tau}{2}},\\
\quad t\geq m,\;t\geq k,\;k\geq n; \label{PPBFtan6}
\end{multline}
\begin{multline}
\fZ^\sigma_{mn}(\cos\theta^q)=\sqrt{\frac{\Gamma(\sigma+m+1)\Gamma(\sigma+n+1)}
{\Gamma(\sigma-m+1)\Gamma(\sigma-n+1)}}
\cos^{2\sigma}\frac{\theta}{2}\cos^{2\sigma}\frac{\phi}{2}\ch^{2\sigma}\frac{\tau}{2}\times\\
\sum^\sigma_{k=-\sigma}\sum^\sigma_{t=-\sigma}\bi^{m-k}
\frac{\Gamma(\sigma-k+1)}{\Gamma(\sigma+k+1)}
\tg^{m-t}\frac{\theta}{2}\tan^{t-k}\frac{\phi}{2}
\tanh^{n-k}\frac{\tau}{2}\times\\
\hypergeom{2}{1}{m-\sigma,-t-\sigma}{m-t+1}{-\tg^2\frac{\theta}{2}}
\hypergeom{2}{1}{t-\sigma,-k-\sigma}{t-k+1}{-\tan^2\frac{\phi}{2}}
\hypergeom{2}{1}{n-\sigma,-k-\sigma}{n-k+1}{\tanh^2\frac{\tau}{2}},\\
\quad m\geq t,\;t\geq k,\;n\geq k; \label{PPBFtan7}
\end{multline}
\begin{multline}
\fZ^\sigma_{mn}(\cos\theta^q)=\sqrt{\frac{\Gamma(\sigma+m+1)\Gamma(\sigma+n+1)}
{\Gamma(\sigma-m+1)\Gamma(\sigma-n+1)}}
\cos^{2\sigma}\frac{\theta}{2}\cos^{2\sigma}\frac{\phi}{2}\ch^{2\sigma}\frac{\tau}{2}\times\\
\sum^\sigma_{k=-\sigma}\sum^\sigma_{t=-\sigma}\bi^{m+k-2t}
\frac{\Gamma(\sigma-k+1)}{\Gamma(\sigma+k+1)}
\tg^{m-t}\frac{\theta}{2}\tan^{k-t}\frac{\phi}{2}
\tanh^{n-k}\frac{\tau}{2}\times\\
\hypergeom{2}{1}{m-\sigma,-t-\sigma}{m-t+1}{-\tg^2\frac{\theta}{2}}
\hypergeom{2}{1}{k-\sigma,-t-\sigma}{k-t+1}{-\tan^2\frac{\phi}{2}}
\hypergeom{2}{1}{n-\sigma,-k-\sigma}{n-k+1}{\tanh^2\frac{\tau}{2}},\\
\quad m\geq t,\;k\geq t,\;n\geq k. \label{PPBFtan8}
\end{multline}

As is known, matrix elements of finite-dimensional representations
of $\SO_0(1,4)$ are expressed via the functions
$f(\fq)=\fM^\sigma_{mn}(\fq)=e^{-\bi
m\varphi^q}\fZ^\sigma_{mn}(\cos\theta^q)e^{-\bi n\psi^q}$, where
$\fZ^\sigma_{mn}(\cos\theta^q)$ is defined by (\ref{PPBtan}) or
(\ref{PPBFtan1})--(\ref{PPBFtan8})\footnote{The functions
$f(\fq)=\fM^\sigma_{mn}(\fq)$ are eigenfunctions of the
Laplace-Beltrami operator $\bigtriangleup_L(\fS_{10})=-F$ defined on
the group manifold $\fS_{10}$ of $\SO_0(1,4)$. An explicit
expression for $\bigtriangleup_L(\fS_{10})=-F$ is given by the
formula (\ref{FKO}).}. For example, let us calculate matrices of
finite-dimensional representations at $\sigma=0,\,\frac{1}{2},\,1$:
\begin{gather}
T_0(\varphi^q,\theta^q,\psi^q)=1,\label{T3}\\[0.3cm]
T_{\frac{1}{2}}(\varphi^q,\theta^q,\psi^q)=\ar\begin{pmatrix}
\fM^{\frac{1}{2}}_{-\frac{1}{2}-\frac{1}{2}} &
\fM^{\frac{1}{2}}_{\frac{1}{2}-\frac{1}{2}}\\
\fM^{\frac{1}{2}}_{-\frac{1}{2}\frac{1}{2}} &
\fM^{\frac{1}{2}}_{\frac{1}{2}\frac{1}{2}}
\end{pmatrix}=\ar\begin{pmatrix}
e^{\frac{\bi}{2}\varphi^q}\fZ^{\frac{1}{2}}_{-\frac{1}{2}-\frac{1}{2}}e^{\frac{\bi}{2}\psi^q}
&
e^{\frac{\bi}{2}\varphi^q}\fZ^{\frac{1}{2}}_{-\frac{1}{2}\frac{1}{2}}e^{-\frac{\bi}{2}\psi^q}\\
e^{-\frac{\bi}{2}\varphi^q}\fZ^{\frac{1}{2}}_{\frac{1}{2}-\frac{1}{2}}e^{\frac{\bi}{2}\psi^q}
&
e^{-\frac{\bi}{2}\varphi^q}\fZ^{\frac{1}{2}}_{\frac{1}{2}\frac{1}{2}}e^{-\frac{\bi}{2}\psi^q}
\end{pmatrix}=\nonumber\\[0.3cm]
=\ar\begin{pmatrix}
e^{\frac{\bi}{2}\varphi^q}\cos\frac{\theta^q}{2}e^{\frac{\bi}{2}\psi^q}
&
\bi e^{\frac{\bi}{2}\varphi^q}\sin\frac{\theta^q}{2}e^{-\frac{\bi}{2}\psi^q}\\
\bi
e^{-\frac{\bi}{2}\varphi^q}\sin\frac{\theta^q}{2}e^{\frac{\bi}{2}\psi^q}
&
e^{-\frac{\bi}{2}\varphi^q}\cos\frac{\theta^q}{2}e^{-\frac{\bi}{2}\psi^q}
\end{pmatrix}=\nonumber\\[0.3cm]
\end{gather}
\begin{multline}
{\renewcommand{\arraystretch}{1.3} =\left(\begin{array}{c}\scr
\left[\cos\frac{\theta}{2}\ch\frac{\tau}{2}\cos\frac{\phi}{2}-
\sin\frac{\theta}{2}\ch\frac{\tau}{2}\sin\frac{\phi}{2}+
\bi\sin\frac{\theta}{2}\sh\frac{\tau}{2}\cos\frac{\phi}{2}+
\bi\cos\frac{\theta}{2}\sh\frac{\tau}{2}\sin\frac{\phi}{2}\right]
e^{\frac{1}{2}(\epsilon+\varepsilon+\omega+\bi\varphi+\bi\psi-\bj\chi+\bk\varsigma)} \\
\scr\left[\cos\frac{\theta}{2}\sh\frac{\tau}{2}\cos\frac{\phi}{2}-
\sin\frac{\theta}{2}\sh\frac{\tau}{2}\sin\frac{\phi}{2}+
\bi\sin\frac{\theta}{2}\ch\frac{\tau}{2}\cos\frac{\phi}{2}+
\bi\cos\frac{\theta}{2}\ch\frac{\tau}{2}\sin\frac{\phi}{2}\right]
e^{\frac{1}{2}(\varepsilon+\omega-\epsilon+\bi\psi-\bi\varphi-\bj\chi-\bk\varsigma)}
\end{array}
\right.}\\
{\renewcommand{\arraystretch}{1.1} \left.\begin{array}{c}
\scr\left[\cos\frac{\theta}{2}\sh\frac{\tau}{2}\cos\frac{\phi}{2}-
\sin\frac{\theta}{2}\sh\frac{\tau}{2}\sin\frac{\phi}{2}+
\bi\cos\frac{\theta}{2}\ch\frac{\tau}{2}\sin\frac{\phi}{2}+
\bi\sin\frac{\theta}{2}\ch\frac{\tau}{2}\cos\frac{\phi}{2}\right]
e^{\frac{1}{2}(\epsilon-\varepsilon-\omega+\bi\varphi-\bi\psi+\bj\chi+\bk\varsigma)} \\
\scr\left[\cos\frac{\theta}{2}\ch\frac{\tau}{2}\cos\frac{\phi}{2}-
\sin\frac{\theta}{2}\ch\frac{\tau}{2}\sin\frac{\phi}{2}+
\bi\cos\frac{\theta}{2}\sh\frac{\tau}{2}\sin\frac{\phi}{2}+
\bi\sin\frac{\theta}{2}\sh\frac{\tau}{2}\cos\frac{\phi}{2}\right]
e^{\frac{1}{2}(-\epsilon-\varepsilon-\omega-\bi\varphi-\bi\psi+\bj\chi-\bk\varsigma)}
\end{array}\right)},\label{T4}
\end{multline}
\begin{gather}
T_1(\varphi^q,\theta^q,\psi^q)=\ar\begin{pmatrix}
\fM^1_{-1-1} & \fM^1_{-10} & \fM^1_{-11}\\
\fM^1_{0-1} & \fM^1_{00} & \fM^1_{01}\\
\fM^1_{1-1} & \fM^1_{10} & \fM^1_{11}
\end{pmatrix}=\ar
\begin{pmatrix}
e^{\bi\varphi^q}\fZ^1_{-1-1}e^{\bi\psi^q} &
e^{\bi\varphi^q}\fZ^1_{-10} & e^{\bi\varphi^q}
\fZ^1_{-11}e^{-\bi\psi^q}\\
\fZ^1_{0-1}e^{\bi\psi^q} & \fZ^1_{00} & \fZ^1_{01}e^{-\bi\psi^q}\\
e^{-\bi\varphi^q}\fZ^1_{1-1}e^{\bi\psi^q} &
e^{-\bi\varphi^q}\fZ^1_{10} &
e^{-\bi\varphi^q}\fZ^1_{11}e^{-\bi\psi^q}
\end{pmatrix}=\nonumber\\[0.3cm]
=\ar\begin{pmatrix}
e^{\bi\varphi^q}\cos^2\frac{\theta^q}{2}e^{\bi\psi^q} &
\frac{\bi}{\sqrt{2}}e^{\bi\varphi^q}\sin\theta^q & -e^{\bi\varphi^q}
\sin^2\frac{\theta^q}{2}e^{-\bi\psi^q}\\
\frac{\bi}{\sqrt{2}}\sin\theta^qe^{\bi\psi^q} & \cos\theta^q &
\frac{\bi}{\sqrt{2}}\sin\theta^qe^{-\bi\psi^q}\\
-e^{-\bi\varphi^q}\sin^2\frac{\theta^q}{2}e^{\bi\psi^q} &
\frac{\bi}{\sqrt{2}}e^{-\bi\varphi^q}\sin\theta^q &
e^{-\bi\varphi^q}\cos^2\frac{\theta^q}{2}e^{-\bi\psi^q}
\end{pmatrix},\label{T5}
\end{gather}
where
\[
\sin\theta^q=\sin\theta\cos\phi\ch\tau+\cos\theta\sin\phi\ch\tau-
\bi\cos\theta\cos\phi\sh\tau+\bi\sin\theta\sin\phi\sh\tau,
\]
\[
\cos\theta^q=\cos\theta\cos\phi\ch\tau-\sin\theta\sin\phi\ch\tau+
\bi\cos\theta\sin\phi\sh\tau+\bi\sin\theta\cos\phi\sh\tau,
\]
\begin{multline}
\sin^2\frac{\theta^q}{2}=\sin^2\frac{\theta}{2}\cos^2\frac{\phi}{2}
\ch^2\frac{\tau}{2}+\frac{1}{2}\sin\theta\sin\phi\ch\tau+
\cos^2\frac{\theta}{2}\sin^2\frac{\phi}{2}\ch^2\frac{\tau}{2}-\\
-\frac{\bi}{2}\left(\sin\frac{\theta}{2}\cos\frac{\phi}{2}+
\cos\frac{\theta}{2}\sin\frac{\phi}{2}\right)\sh\tau-
\cos^2\frac{\theta}{2}\cos^2\frac{\phi}{2}\sh^2\frac{\tau}{2}-
\sin^2\frac{\theta}{2}\sin^2\frac{\phi}{2}\sh^2\frac{\tau}{2},\nonumber
\end{multline}
\begin{multline}
\cos^2\frac{\theta^q}{2}=\cos^2\frac{\theta}{2}\cos^2\frac{\phi}{2}
\ch^2\frac{\tau}{2}-\frac{1}{2}\sin\theta\sin\phi\ch\tau+
\sin^2\frac{\theta}{2}\sin^2\frac{\phi}{2}\ch^2\frac{\tau}{2}+\\
+\frac{\bi}{2}\left(\sin\frac{\theta}{2}\cos\frac{\phi}{2}+
\cos\frac{\theta}{2}\sin\frac{\phi}{2}\right)\sh\tau-
\sin^2\frac{\theta}{2}\cos^2\frac{\phi}{2}\sh^2\frac{\tau}{2}-
\cos^2\frac{\theta}{2}\sin^2\frac{\phi}{2}\sh^2\frac{\tau}{2}.\nonumber
\end{multline}
It is easy to see that $T_{\frac{1}{2}}(\varphi^q,\theta^q,\psi^q)$
is the fundamental representation (\ref{Elem4}) of $\Sp(1,1)$.

Spherical functions of the second type
$f(\varphi^q,\theta^q)=\fM^m_\sigma(\varphi^q,\theta^q,0)=e^{-im\varphi^q}
\fZ^m_\sigma(\cos\theta^q)$, where
\begin{equation}
\fZ^m_\sigma(\cos\theta^q)=\sum^\sigma_{k=-\sigma}\sum^\sigma_{t=-\sigma}
P^\sigma_{mt}(\cos\theta)
P^\sigma_{tk}(\cos\phi)\fP^k_\sigma(\cosh\tau),\nonumber
\end{equation}
is {\it an associated hyperspherical function} of $\SO_0(1,4)$, are
defined on the surface of the quaternion 2-sphere $S^2_q$.
$\fZ^m_\sigma(\cos\theta^q)$ are eigenfunctions of the
Laplace-Beltrami operator $\bigtriangleup_L(S^2_q)=-F$,
\[
\bigtriangleup_L(S^2_q)=\frac{\partial^2}{\partial{\theta^q}^2}+
\cot\theta^q\frac{\partial}{\partial\theta^q}+
\frac{1}{\sin^2\theta^q}\frac{\partial^2}{\partial{\varphi^q}^2}.
\]
Hypergeometric-type formulae for $\fZ^m_\sigma(\cos\theta^q)$ are
\begin{multline}
\fZ^m_\sigma(\cos\theta^q)=\sqrt{\frac{\Gamma(\sigma+m+1)}{\Gamma(\sigma-m+1)}}
\cos^{2\sigma}\frac{\theta}{2}\cos^{2\sigma}\frac{\phi}{2}\ch^{2\sigma}\frac{\tau}{2}\times\\
\sum^\sigma_{k=-\sigma}\sum^\sigma_{t=-\sigma}\bi^{m-k}\tg^{m-t}\frac{\theta}{2}\tan^{t-k}\frac{\phi}{2}
\tanh^{k}\frac{\tau}{2}\times\\
\hypergeom{2}{1}{m-\sigma,-t-\sigma}{m-t+1}{-\tg^2\frac{\theta}{2}}
\hypergeom{2}{1}{t-\sigma,-k-\sigma}{t-k+1}{-\tan^2\frac{\phi}{2}}
\hypergeom{2}{1}{k-\sigma,-\sigma}{k+1}{\tanh^2\frac{\tau}{2}},\\
\quad m\geq t,\;t\geq k; \nonumber
\end{multline}
\begin{multline}
\fZ^m_\sigma(\cos\theta^q)=\sqrt{\frac{\Gamma(\sigma+m+1)}{\Gamma(\sigma-m+1)}}
\cos^{2\sigma}\frac{\theta}{2}\cos^{2\sigma}\frac{\phi}{2}\ch^{2\sigma}\frac{\tau}{2}\times\\
\sum^\sigma_{k=-\sigma}\sum^\sigma_{t=-\sigma}\bi^{m+k-2t}\tg^{m-t}\frac{\theta}{2}\tan^{k-t}\frac{\phi}{2}
\tanh^{k}\frac{\tau}{2}\times\\
\hypergeom{2}{1}{m-\sigma,-t-\sigma}{m-t+1}{-\tg^2\frac{\theta}{2}}
\hypergeom{2}{1}{k-\sigma,-t-\sigma}{k-t+1}{-\tan^2\frac{\phi}{2}}
\hypergeom{2}{1}{k-\sigma,-\sigma}{k+1}{\tanh^2\frac{\tau}{2}},\\
\quad m\geq t,\;k\geq t; \nonumber
\end{multline}
\begin{multline}
\fZ^m_\sigma(\cos\theta^q)=\sqrt{\frac{\Gamma(\sigma-m+1)}{\Gamma(\sigma+m+1)}}
\cos^{2\sigma}\frac{\theta}{2}\cos^{2\sigma}\frac{\phi}{2}\ch^{2\sigma}\frac{\tau}{2}\times\\
\sum^\sigma_{k=-\sigma}\sum^\sigma_{t=-\sigma}\bi^{k-m}
\frac{\Gamma(\sigma+k+1)}{\Gamma(\sigma-k+1)}
\tg^{t-m}\frac{\theta}{2}\tan^{k-t}\frac{\phi}{2}
\tanh^{k}\frac{\tau}{2}\times\\
\hypergeom{2}{1}{t-\sigma,-m-\sigma}{t-m+1}{-\tg^2\frac{\theta}{2}}
\hypergeom{2}{1}{k-\sigma,-t-\sigma}{k-t+1}{-\tan^2\frac{\phi}{2}}
\hypergeom{2}{1}{k-\sigma,-\sigma}{k+1}{\tanh^2\frac{\tau}{2}},\\
\quad t\geq m,\;k\geq t; \nonumber
\end{multline}
\begin{multline}
\fZ^m_\sigma(\cos\theta^q)=\sqrt{\frac{\Gamma(\sigma-m+1)}{\Gamma(\sigma+m+1)}}
\cos^{2\sigma}\frac{\theta}{2}\cos^{2\sigma}\frac{\phi}{2}\ch^{2\sigma}\frac{\tau}{2}\times\\
\sum^\sigma_{k=-\sigma}\sum^\sigma_{t=-\sigma}\bi^{2t-m-k}
\frac{\Gamma(\sigma+k+1)}{\Gamma(\sigma-k+1)}
\tg^{t-m}\frac{\theta}{2}\tan^{t-k}\frac{\phi}{2}
\tanh^{k}\frac{\tau}{2}\times\\
\hypergeom{2}{1}{t-\sigma,-m-\sigma}{t-m+1}{-\tg^2\frac{\theta}{2}}
\hypergeom{2}{1}{t-\sigma,-k-\sigma}{t-k+1}{-\tan^2\frac{\phi}{2}}
\hypergeom{2}{1}{k-\sigma,-\sigma}{k+1}{\tanh^2\frac{\tau}{2}},\\
\quad t\geq m,\;t\geq k. \nonumber
\end{multline}
The latter formulae hold at any $k$ when $\sigma$ is an half-integer
number. When $\sigma$ is an integer number, these formulae hold at
$k=0,1,\ldots,\sigma-1,\sigma$. At $k=-\sigma,-\sigma+1,\ldots,0$ we
must replace the function
$\hypergeom{2}{1}{k-\sigma,-\sigma}{k+1}{\tanh^2\frac{\tau}{2}}$ via
$\hypergeom{2}{1}{-\sigma,-k-\sigma}{-k+1}{\tanh^2\frac{\tau}{2}}$
and $\tanh^k\frac{\tau}{2}$ via $\tanh^{-k}\frac{\tau}{2}$.

At $m=n=0$ we obtain {\it a zonal hyperspherical function}
$\fZ_\sigma(\cos\theta^q)\equiv \fZ^\sigma_{00}(\cos\theta^q)$ of
the group $\SO_0(1,4)$. Namely,
\begin{multline}
\fZ_\sigma(\cos\theta^q)=
\cos^{2\sigma}\frac{\theta}{2}\cos^{2\sigma}\frac{\phi}{2}\ch^{2\sigma}\frac{\tau}{2}\times\\
\sum^\sigma_{k=-\sigma}\sum^\sigma_{t=-\sigma}\bi^{k}
\frac{\Gamma(\sigma+k+1)}{\Gamma(\sigma-k+1)}
\tg^{t}\frac{\theta}{2}\tan^{k-t}\frac{\phi}{2}
\tanh^{k}\frac{\tau}{2}\times\\
\hypergeom{2}{1}{t-\sigma,-\sigma}{t+1}{-\tg^2\frac{\theta}{2}}
\hypergeom{2}{1}{k-\sigma,-t-\sigma}{k-t+1}{-\tan^2\frac{\phi}{2}}
\hypergeom{2}{1}{k-\sigma,-\sigma}{k+1}{\tanh^2\frac{\tau}{2}},
\quad k\geq t; \nonumber
\end{multline}
\begin{multline}
\fZ_\sigma(\cos\theta^q)=
\cos^{2\sigma}\frac{\theta}{2}\cos^{2\sigma}\frac{\phi}{2}\ch^{2\sigma}\frac{\tau}{2}\times\\
\sum^\sigma_{k=-\sigma}\sum^\sigma_{t=-\sigma}\bi^{-k}
\frac{\Gamma(\sigma-k+1)}{\Gamma(\sigma+k+1)}
\tg^{-t}\frac{\theta}{2}\tan^{t-k}\frac{\phi}{2}
\tanh^{-k}\frac{\tau}{2}\times\\
\hypergeom{2}{1}{-\sigma,-t-\sigma}{-t+1}{-\tg^2\frac{\theta}{2}}
\hypergeom{2}{1}{t-\sigma,-k-\sigma}{t-k+1}{-\tan^2\frac{\phi}{2}}
\hypergeom{2}{1}{-\sigma,-k-\sigma}{-k+1}{\tanh^2\frac{\tau}{2}},
\quad t\geq k. \nonumber
\end{multline}
In its turn, the functions $f(\dot{\varphi}^q,\dot{\theta}^q)=e^{\bi
m\dot{\varphi}^q} \fZ^{\dot{m}}_{\dot{\sigma}}(\cos\dot{\theta}^q)$
(or $f(\dot{\theta}^q)=\fZ_{\dot{\sigma}}(\cos\dot{\theta}^q)$) are
defined on the surface of the dual quaternion sphere $\dot{S}^2_q$.
Explicit expressions and hypergeometric-type formulae for
$f(\dot{\varphi}^q,\dot{\theta}^q)$ are analogous to the previous
expressions for $f(\varphi^q,\theta^q)$.

Spherical functions of the fourth type
$f(\varphi,\theta,\psi)=\fM^\sigma_{mn}(\varphi,\theta,\psi)=
e^{-\bi m\varphi}P^\sigma_{mn}(\cos\theta)e^{-\bi n\psi}$ (or
$f(\varsigma,\phi,\chi)=\fM^\sigma_{mn}(\varsigma,\phi,\chi)=
e^{-\bi m\varsigma}P^\sigma_{mn}(\cos\phi) e^{-\bi n\chi}$) are
defined on the surface of the real 3-sphere $S^3=\SO(4)/\SO(3)$. Let
$L^2(S^3)$ be a Hilbert space of the functions defined on the sphere
$S^3$ in the space $\R^4$. Since $S^3\sim\SO_0(1,4)/P\sim K/M$, then
the representations of the principal nonunitary (spherical) series
$T_{\omega\sigma}$ are defined by the complex number $\sigma$ and an
irreducible unitary representation $\omega$ of the subgroup
$M=\SO(3)$. Thus, representations of the group $\SO_0(1,4)$, which
have a class 1 with respect to $K=\SO(4)$, are realized in the space
$L^2(S^3)$. At this point, spherical functions of the
representations $Q^m$ of $\SO(4)$ form an orthogonal basis in
$L^2(S^3)$. Therefore, we have here representations of $\SO_0(1,4)$
restricted to the subgroup $\SO(4)$:
\[
\hat{T}^\sigma\downarrow^{\SO_0(1,4)}_{\SO(4)}=\sum^l_{m=0}\oplus
Q^m.
\]

\subsection{Spherical functions on the hyperboloid and their
applications to hydrogen atom problem} In 1935, using a
stereographic projection of the momentum space onto a
four-dimensional sphere, Fock showed \cite{Fock} that
Schr\"{o}dinger equation for hydrogen atom is transformed into an
integral equation for hyperspherical functions defined on the
surface of the four-dimensional sphere. This discovery elucidates an
intrinsic nature of an additional degeneration of the energy levels
in hydrogen atom, and also it allows one to write important
relations for wavefunctions (for example, Fock wrote simple
expressions for the density matrix of the system of wavefunctions
for energy levels with an arbitrary quantum number $n$). In 1968,
authors of the work \cite{BBC68} showed that Fock integral equation
can be written in the form of a Klein-Gordon-type equation for
spherical functions defined on the surface of the four-dimensional
hyperboloid. ``Square roots" of the Klein-Gordon-type equation are
Dirac-like equations (in the paper \cite{BBC68} these equations are
called Majorana-type equations), or more general
Gel'fand-Yaglom-type equations \cite{GY48}. Equations of this type
were first considered by Dirac in 1935 \cite{Dir35}. Here there is
an analogy with the usual formulation of the Dirac equation for a
hydrogen atom in the Minkowski space-time, but the main difference
lies in the fact that Dirac-like equations are defined on the
four-dimensional hyperboloid\footnote{As is known, this hyperboloid
can be understood as the four-dimensional Minkowski space-time
endowed globally with a constant negative curvature.} immersed into
a five-dimensional de Sitter space.

So, spherical functions of the third type
$f(\epsilon,\tau,\varepsilon,\omega)=
\fM^\sigma_{mn}(\epsilon,\tau,\varepsilon,\omega)=e^{
-m\epsilon}\fP^\sigma_{mn}(\ch\tau)e^{-n(\varepsilon+\omega)}$ are
defined on the upper sheet $H^4_+$ of the four-dimensional
hyperboloid $\ld\bx,\bx\rd=1$, where $\fP^\sigma_{mn}(\cosh\tau)$ is
a Jacobi function\footnote{Representations of the group
$\SU(1,1)\simeq\SL(2,\R)$, known also as a three-dimensional Lorentz
group, are expressed via the functions
$\fP^\sigma_{mn}(\cosh\tau)$.} considered in details by Vilenkin
\cite{Vil65}. The functions
$\fM^\sigma_{mn}(\epsilon,\tau,\varepsilon,\omega)$ are
eigenfunctions of the Laplace-Beltrami operator
$\bigtriangleup_L(H^4_+)=-F$ defined on $H^4_+$:
\[
\left[\bigtriangleup_L(H^4_+)-\sigma(\sigma+3)\right]
\fM^\sigma_{mn}(\epsilon,\tau,\varepsilon,\omega)=0,
\]
where
\[
\bigtriangleup_L(H^4_+)=-\frac{\partial^2}{\partial\tau^2}-\coth\tau
\frac{\partial}{\partial\tau}-\frac{1}{\sinh^2\tau}\left[
\frac{\partial^2}{\partial\epsilon^2}-2\ch\tau\frac{\partial^2}
{\partial\epsilon\partial(\varepsilon+\omega)}+\frac{\partial^2}
{\partial(\varepsilon+\omega)^2}\right].
\]
Or,
\[
\left[-\frac{d^2}{d\tau^2}-\coth\tau\frac{d}{d\tau}+\frac{m^2+n^2-2mn\ch\tau}
{\sh^2\tau}-\sigma(\sigma+3)\right]\fP^\sigma_{mn}(\ch\tau)=0.
\]
After substitution $y=\ch\tau$ this equation can be rewritten as
\[
\left[(y^2-1)\frac{d^2}{dy^2}+2y\frac{d}{dy}-\frac{m^2+n^2-2mny}{y^2-1}+
\sigma(\sigma+3)\right]\fP^\sigma_{mn}(y)=0.
\]

Let us construct a quasiregular representation of the group
$\SO_0(1,4)$ on the functions $f(x)$ from $H^4_+$, where
$x=(\epsilon,\tau,\varepsilon,\omega)$. Let $L^2(H^4_+)$ be a
Hilbert space of the functions on the hyperboloid $H^4_+$ with a
scalar product
\[
\langle
f_1,f_2\rangle=\int\limits_{H^4_+}\overline{f_1(x)}f_2(x)d\mu(x)=
\int\limits^\infty_0\int\limits^\infty_0\int\limits^\infty_0
\int\limits^\infty_0\overline{\fP^{\sigma_1}_{mn}(\cosh\tau)}
\fP^{\sigma_2}_{mn}(\cosh\tau)e^{-2m\epsilon-2n(\omega+\varepsilon)}
\sinh\tau d\tau d\epsilon d\varepsilon d\omega,
\]
where $d\mu(x)$ is an invariant measure on $H^4_+$ with respect to
$\SO_0(1,4)$. This measure is defined by an equality
$d\mu(x)=\sinh\tau d\tau d\epsilon d\varepsilon d\omega$. In
accordance with (\ref{QEA2}) the range of variables $\epsilon$,
$\tau$, $\varepsilon$, $\omega$ is $(-\infty,+\infty)$, but we
consider here the upper sheet of the hyperboloid; therefore, the
range of these variables is $(0,\infty)$. {\it A quasiregular
representation} $T$ in the space $L^2(H^4_+)$ is defined by the
formula
\[
T(\fq)f(x)=f(\fq^{-1}x),\quad x\in H^4_+.
\]
It is easy to show that this representation is unitary. However, $T$
is reducible, and in accordance with Gel'fand-Graev theorem
\cite{GGV62} is decomposed into a direct integral of irreducible
representations $T^\sigma$ of the principal unitary series
($\sigma=-3/2+i\rho$, $0<\rho<\infty$).

Analogously, a quasiregular representation of the group $\SO_0(1,4)$
in a Hilbert space $L^2(C^4_+)$ of the functions on the upper sheet
$C^4_+$ of the cone $C^4$
($C^4_+:\;x^2_0-x^2_1-x^2_2-x^2_3-x^2_4=0,\;x_0>0$) has the
following form:
\[
T(g)f(x)=f(g^{-1}x),\quad x\in C^4_+.
\]
This representation is unitary with respect to a scalar product
\[
\langle
f_1,f_2\rangle=\int\limits_{C^4_+}\overline{f_1(x)}f_2(x)d\mu(x)
\]
defined on $L^2(C^4_+)$. Here $d\mu(x)$ is an invariant measure on
$C^4_+$ with respect to $\SO_0(1,4)$. This representation is
reducible.  Irreducible unitary representations of the group
$\SO_0(1,4)$ can be constructed in a Hilbert space of homogeneous
functions on the cone \cite{Vil65}.

Let us consider applications of the spherical functions
$f(\epsilon,\tau,\varepsilon,\omega)$ to hydrogen and antihydrogen
atom problems (about antihydrogen atom see \cite{EH99}). As it has
been shown in the work \cite{BBC68} when the internal motion can be
described by algebraic methods, as in the case of hydrogen atom, the
proposed equation for the motion of the system as a whole (motion of
the c.m.) is equivalent to a Majorana-type equation, free from the
well-known difficulties such as a spacelike solution. As is known,
the Bethe-Salpeter equation for two spinors of masses $m_1$ and
$m_2$,
\[
(\hat{p}_1-m_1)(\hat{p}_2-m_2)\psi(p_1,p_2)=\frac{\bi}{2\pi}
\int\int G(p_1,p_2;p^\prime_1,p^\prime_2)\psi(p^\prime_1,p^\prime_2)
dp^\prime_1dp^\prime_2,
\]
in the ladder approximation can be written as follows:
\[
(c_1\hat{P}^{(1)}+\hat{p}^{(1)}-m_1)(c_2\hat{P}^{(2)}-\hat{p}^{(2)}-m_2)
\psi_P(p)=\frac{\bi}{2\pi}\int G(q)\psi_P(p+q)dq,
\]
where
\begin{gather}
P=p_1+p_2,\quad p=c_2p_1-c_1p_2,\nonumber\\
c_1=m_1/(m_1+m_2),\quad c_2=m_2/(m_1+m_2),\nonumber
\end{gather}
the metric is $g_{\mu\nu}=+1,\,-1,\,-1,\,-1$, and the superscripts
on $\hat{P}^{(i)}$ and $\hat{p}^{(i)}$ refer to the $\gamma$
matrices. In this case, projection operators can be defined as
\begin{equation}\label{Proj}
\Lambda_\pm^{(i)}=[\cE_i(p)\pm\cK_i]/2\cE_i(p),
\end{equation}
where
\begin{gather}
\cE_i=\left[P^2(m^2_i-p^2)+(p\cdot P)^2\right]^{1/2},\nonumber\\
\cK_1=\left[m_1\hat{P}^{(1)}-\bi
P^\mu\sigma_{\mu\nu}^{(1)}p^\nu\right],\quad\cK_2=\left[m_1\hat{P}^{(2)}+\bi
P^\mu\sigma_{\mu\nu}^{(2)}p^\nu\right]\nonumber
\end{gather}
with
\[
\sigma^{(i)}_{\mu\nu}=(1/2\bi)\left[\gamma^{(i)}_\mu,\gamma^{(i)}_\nu\right].
\]
Further, using the operators (\ref{Proj}), we obtain
\begin{equation}\label{Equat}
(P^2-\cK^2_1-\cK^2_2)\varphi(p^T)=-(\Lambda^{(1)}_+\Lambda^{(2)}_+
-\Lambda^{(1)}_-\Lambda^{(2)}_-)\hat{P}^{(1)}\hat{P}^{(2)}\int
G(p^T-l)\varphi(l)\delta(l\cdot p)dl,
\end{equation}
where $p^T_\mu=p_\mu-p^Lu_\mu$ is the transverse relative momenta,
and $p^L=p\cdot u$, $u^\mu=P^\mu/|P|$,
$\phi(l)=\int^{+\infty}_{-\infty}\psi(l,q^L)dq^L$. The approximation
\[
\Lambda^{(1)}_+\Lambda^{(2)}_+ -\Lambda^{(1)}_-\Lambda^{(2)}_-=+1
\]
means that we take only positive-energy states for the constituents.
On the other hand, the choice
\[
\Lambda^{(1)}_+\Lambda^{(2)}_+ -\Lambda^{(1)}_-\Lambda^{(2)}_-=-1
\]
would have meant taking only negative-energy states for the system
and would correspond to charge conjugation for the c.m. motion.

Since $\Lambda^i_+=1$ is equivalent to
\[
\cK_i=\cE_i=(m^2_i-(p^T)^2)^{1/2}|P|,
\]
then the equation (\ref{Equat}) can be written as
\[
\left[P^2-|P|(m_1+m_2-(p^T)^2/2\mu)\right]\varphi(p^T)=P^2\int
G(p^T-l)\varphi(l)\delta(l\cdot P)dl,
\]
where
\[
\mu=\frac{m_1m_2}{m_1+m_2}.
\]
In the case of hydrogen atom this equation has the form
\begin{equation}\label{Hyd}
\left[|P|-(m_1+m_2-(p^T)^2/2\mu)\right]\varphi(p^T)=|P|\frac{e^2}{2\pi}
\int\frac{1}{(p^T-l)^2}\delta(l\cdot P)\varphi(l)dl.
\end{equation}
Using the Fock stereographic projection \cite{Fock,BI66}
\[
\xi_\mu=2ap_\mu(a^2-p^2),\quad\xi_4=(a^2+p^2)/(a^2-p^2),\quad
\mu=0,\ldots, 3,
\]
where $p^2=p_\mu p^\mu$ and $a$ is an arbitrary constant, we will
project stereographically the four-dimensional $p$-space on a
five-dimensional hyperboloid. This projection allows us to rewrite
the equation (\ref{Hyd}) in the form of a Klein-Gordon-type equation
\begin{equation}\label{KG}
(P^2-\cK^2)\Psi_P=0
\end{equation}
with
\[
\cK=m_1+m_2-\mu e^4/2N,
\]
and $N^2$ is the operator $D^T+1$, where $D^T$ is the angular part
of the four-dimensional Laplace operator. $\Psi_P(\xi_a)$ form a
basis for a representation of the de Sitter group $\SO_0(1,4)$. A
``square root" of the Klein-Gordon-type equation (\ref{KG}) is a
Majorana-type equation
\begin{equation}\label{MTE1}
\left[\Gamma\cdot P-(m_1+m_2)N+e^4\mu/2N\right]\Psi_P=0
\end{equation}
or,
\begin{equation}\label{MTE2}
\left[\Gamma\cdot P+(m_1+m_2)N-e^4\mu/2N\right]\dot{\Psi}_P=0
\end{equation}
where $\Gamma$-matrices behave like components of a five-vector in
$\SO_0(1,4)$. Equations (\ref{MTE1}) and (\ref{MTE2}) describe
hydrogen and antihydrogen atoms, respectively.

In the equations (\ref{KG})--(\ref{MTE2}) the functions $\Psi_P$ are
eigenfunctions of the Laplace-Beltrami operator defined on the
surface of the five-dimensional hyperboloid (more precisely
speaking, on the upper sheet $H^4_+$ of this hyperboloid for the
equation (\ref{MTE1}) and on the lower sheet $H^4_-$ for
(\ref{MTE2})). As it has been shown previously, this hyperboloid is
a homogeneous space of the de Sitter group $\SO_0(1,4)$. On the
other hand, spherical functions $\Psi_p$ are solutions of the
equations (\ref{KG})--(\ref{MTE2}), that is, they are wavefunctions,
and for that reason $\Psi_P$ play a crucial role in the hydrogen
(antihydrogen) atom problem.

Let us consider in brief solutions (wavefunctions) of the
Majorana-type equations (\ref{MTE1}) and (\ref{MTE2}). With this end
in view we must introduce an {\it inhomogeneous de Sitter group}
$\ISO_0=\SO_0(1,4)\odot T_5$, which is a semidirect product of the
subgroup $\SO_0(1,4)$ (connected component) of five-dimensional
rotations and a subgroup $T_5$ of five-dimensional translations of
the de Sitter space $\R^{1,4}$. The subgroup $T_5$ is a direct
product of five one-dimensional translation groups $T_1$,
$T_5=T_1\otimes T_1\otimes T_1\otimes T_1\otimes T_1$. At this
point, each group $T_1$ is isomorphic to the group $\R^+$ of all
positively defined real numbers. At the restriction to $H^4_+$, the
maximal homogeneous space $\cM_{15}=\R^{1,4}\times\fS_{10}$ of
$\ISO_0(1,4)$ is reduced to $\cM_9=\R^{1,4}\times H^4_+$. Let
$F(x,\epsilon,\tau,\varepsilon,\omega)$ be a square integrable
function on $\cM_9$, that is,
\[
\int\limits_{H^4_+}\int\limits_{T_5}|F|^2d^5xd^4g<+\infty,
\]
then in the case of finite-dimensional representations of
$\SO_0(1,4)$ there is an expansion of
$F(x,\epsilon,\tau,\varepsilon,\omega)$ in a Fourier-type integral
\begin{equation}\label{FTI}
F(x,\epsilon,\tau,\varepsilon,\omega)=\sum^\infty_{\sigma=0}
\sum^\sigma_{m,n=-\sigma}\int\limits_{T_5}\boldsymbol{\alpha}^\sigma_{mn}
e^{\bi
px}e^{-m\epsilon-n(\varepsilon+\omega)}\fP^\sigma_{mn}(\cosh\tau)d^5x,
\end{equation}
where
\[
\boldsymbol{\alpha}^\sigma_{mn}=\frac{(-1)^{m-n}(2\sigma+3)}{16\pi^2}
\int\limits_{H^4_+}\int\limits_{T_5}Fe^{-\bi
px}\fP^\sigma_{mn}(\cosh\tau)e^{-m\epsilon-n(\varepsilon+\omega)}d^5xd^4g,
\]
and $d^4g=\sinh\tau d\tau d\epsilon d\varepsilon d\omega$ is a Haar
measure on the hyperboloid $H^4_+$.

Further, let $T$ be an unbounded region in $\R^{1,4}$ and let
$\Sigma$ be a surface of the hyperboloid $H^4_+$ (correspondingly,
$\dot{\Sigma}$, for the sheet $H^4_-$), then it needs to find a
function
$\boldsymbol{\psi}(g)=(\psi^m_P(g),\dot{\psi}^{\dot{m}}_P(g))^T$ in
the all region $T$. $\boldsymbol{\psi}(g)$ is a continuous function
(everywhere in $T$), including the surfaces $\Sigma$ and
$\dot{\Sigma}$. At this point,
$\left.\phantom{\frac{x}{x}}\psi^m_{P}(g)\right|_\Sigma= F_{m}(g)$,
$\left.\phantom{\frac{x}{x}}\dot{\psi}^{\dot{m}}_P(g)\right|_{\dot{\Sigma}}=
\dot{F}_{\dot{m}}(g)$, where $F_{m}(g)$ and $\dot{F}_{\dot{m}}(g)$
are square integrable functions (boundary conditions) defined on the
surfaces $\Sigma$ and $\dot{\Sigma}$, respectively.

Following the method proposed in \cite{Var03,Var03c,Var03d,Var05},
we can find solutions of the boundary value problem in the form of
Fourier type series
\begin{equation}\label{FT1}
\psi^m_P=\sum^\infty_{\sigma=0}\sum_k\boldsymbol{f}_{\sigma mk}(r)
\sum^\sigma_{n=-\sigma}\boldsymbol{\alpha}^m_{\sigma n}
\fM^\sigma_{mn}(\epsilon,\tau,\varepsilon,\omega),
\end{equation}
\begin{equation}\label{FT2}
\dot{\psi}^{\dot{m}}_P(g)=\sum^\infty_{\dot{\sigma}=0}\sum_{\dot{k}}
\overset{\ast}{\boldsymbol{f}}_{\dot{\sigma}\dot{m}\dot{k}}(r^\ast)
\sum^{\dot{\sigma}}_{\dot{n}=-\dot{\sigma}}\boldsymbol{\alpha}^{\dot{m}}
_{\dot{\sigma}\dot{n}}\overset{\ast}{\fM}{}^{\dot{\sigma}}_{\dot{m}\dot{n}}
(\epsilon,\tau,\varepsilon,\omega),
\end{equation}
where
\[
\boldsymbol{\alpha}^m_{\sigma n}=\frac{(-1)^n(2\sigma+3)}{16\pi^2}
\int\limits_{H^4_+}F_m\fM^\sigma_{mn}(\epsilon,\tau,\varepsilon,\omega)
\sinh\tau d\tau d\epsilon d\varepsilon d\omega,
\]
\[
\boldsymbol{\alpha}^{\dot{m}}_{\dot{\sigma}
\dot{n}}=\frac{(-1)^{\dot{n}}(2\dot{\sigma}+3)}{16\pi^2}
\int\limits_{H^4_-}F_{\dot{m}}\overset{\ast}{\fM}{}^{\dot{\sigma}}
_{\dot{m}\dot{n}}(\epsilon,\tau,\varepsilon,\omega) \sinh\tau d\tau
d\epsilon d\varepsilon d\omega.
\]
The indices $k$ and $\dot{k}$ numerate equivalent representations.
$\fM^\sigma_{mn}(\epsilon,\tau,\varepsilon,\omega)$
($\overset{\ast}{\fM}{}^{\dot{\sigma}}
_{\dot{m}\dot{n}}(\epsilon,\tau,\varepsilon,\omega)$) are
hyperspherical functions defined on the surface $\Sigma$
($\dot{\Sigma}$) of the four-dimensional hyperboloid $H^4$ of the
radius $r$ ($r^\ast$) ($H^4$ can be understood as a four-dimensional
sphere with an imaginary radius $r$), $\boldsymbol{f}_{\sigma
mk}(r)$ and
$\overset{\ast}{\boldsymbol{f}}_{\dot{\sigma}\dot{m}\dot{k}}(r^\ast)$
are radial functions. Taking into account the subgroup $T_5$, we can
rewrite the wavefunctions (\ref{FT1}) and (\ref{FT2}) in terms of
Fourier-type integrals (\ref{FTI}) (field operators).

\section{Spherical functions of unitary representations of $\SO_0(1,4)$}
Spherical functions $\fM^l_{mn}(\varphi^q,\theta^q,\psi^q)$,
considered in the section 4, define matrix elements of non-unitary
finite-dimensional representations of the group $\SO_0(1,4)$.
Following the analogue between $\spin_+(1,3)\simeq\SL(2,\C)$ and
$\spin_+(1,4)\simeq\Sp(1,1)$, we can define finite-dimensional
(spinor) representations of $\SO_0(1,4)$ in the space of symmetric
polynomials $\Sym_{(k,r)}$ as follows\footnote{As is known, any
proper Lorentz transformation $\fg$ corresponds to a fractional
linear transformation of the complex plane with the matrix
$\begin{pmatrix}\alpha & \beta\\ \gamma &
\delta\end{pmatrix}\in\SL(2,\C)$ \cite{GMS}. In its turn, any proper
de Sitter transformation $\fq$ can be identified with a fractional
linear transformation $w=(az+b)(cz+d)^{-1}$ of the anti-quaternion
plane with the matrix $\begin{bmatrix}a & b\\ c &
d\end{bmatrix}\in\Sp(1,1)$ (about quaternion and anti-quaternion
planes and their fractional linear transformations see
\cite{Roz55}).}:
\begin{equation}\label{TenRep}
T_{\fq}q(z,\overline{z})=(cz+d)^{l_0+l_1-1}
\overline{(cz+d)}^{l_0-l_1+1}q\left(\frac{az+b}{cz+d};
\frac{\overline{az+b}}{\overline{cz+d}}\right),
\end{equation}
where $a,b,c,d\in\BH$, $k=l_0+l_1-1$, $r=l_0-l_1+1$, and the pair
$(l_0,l_1)$ defines an irreducible representation of $\SO_0(1,4)$ in
the Diximier-Str\"{o}m basis \cite{Dix61,Str69}:
\[
M_3\mid j^\prime,m^\prime,q,q;l_0,m,l_1\rangle=m^\prime\mid
j^\prime,m^\prime,q,q;l_0,m,l_1\rangle,
\]
\[
M_\pm\mid j^\prime,m^\prime,q,q;l_0,m,l_1\rangle=\left[(j^\prime\mp
m^\prime)(j^\prime\pm m^\prime+1)\right]^{\frac{1}{2}}\mid
j^\prime,m^\prime+1,q,q;l_0,m,l_1\rangle,
\]
\begin{multline}
P_3\mid
j^\prime,m^\prime,q,q;l_0,m,l_1\rangle=-\alpha(j^\prime+1;q,q)
\left[(j^\prime+1)^2-m^2\right]^{\frac{1}{2}}\mid
j^\prime+1,m^\prime,q,q;l_0,m,l_1\rangle+\\
+\frac{m^\prime(q+1)q}{j^\prime(j^\prime+1)}\mid
j^\prime,m^\prime,q,q;l_0,m,l_1\rangle-\alpha(j^\prime;q,q)
\left[{j^\prime}^2-m^2\right]^{\frac{1}{2}}\mid
j^\prime-1,m^\prime,q,q;l_0,m,l_1\rangle,\nonumber
\end{multline}
\begin{multline}
P_\pm\mid
j^\prime,m^\prime,q,q;l_0,m,l_1\rangle=\\
=\pm\alpha(j^\prime+1;q,q) \left[(j^\prime\pm
m^\prime+1)(j^\prime\pm m^\prime+2)\right]^{\frac{1}{2}}\mid
j^\prime+1,m^\prime\pm
1,q,q;l_0,m,l_1\rangle+\\
+\frac{(q+1)q}{j^\prime(j^\prime+1)}\left[(j^\prime\mp m^\prime)
(j^\prime\pm m^\prime+1)\right]^{\frac{1}{2}}\mid
j^\prime,m^\prime\pm
1,q,q;l_0,m,l_1\rangle\mp\\
\mp\alpha(j^\prime;q,q)\left[(j^\prime\mp m^\prime)(j^\prime\mp
m^\prime-1)\right]^{\frac{1}{2}}\mid j^\prime-1,m^\prime\pm
1,q,q;l_0,m,l_1\rangle,\nonumber
\end{multline}
\begin{multline}
P_0\mid
j^\prime,m^\prime,q,q;l_0,m,l_1\rangle=a(q,q;l_0,l_1)\left[(q+j^\prime+2)
(q-j^\prime+1)\right]^{\frac{1}{2}}\mid
j^\prime,m^\prime,q+1,q;l_0,m,l_1\rangle+\\
+a(q-1,q;l_0,l_1)\left[(q+j^\prime+1)(q-j^\prime)\right]^{\frac{1}{2}}\mid
j^\prime,m^\prime,q-1,q;l_0,m,l_1\rangle+\\
+b(q,q;l_0,l_1)\left[(j^\prime-q)(j^\prime+q+1)\right]^{\frac{1}{2}}\mid
j^\prime,m^\prime,q,q+1;l_0,m,l_1\rangle+\\
+b(q,q-1;l_0,l_1)\left[(j^\prime+q)(j^\prime-q+1)\right]^{\frac{1}{2}}\mid
j^\prime,m^\prime,q,q-1;l_0,m,l_1\rangle,\nonumber
\end{multline}
where $M_\pm=M_1\pm\bi M_2$, $P_\pm=P_1\pm\bi P_2$ and
\[
\alpha(j^\prime;q,q)=\frac{1}{j^\prime}\left[\frac{({j^\prime}^2-q^2)
((q+1)^2-{j^\prime}^2)}{(2j^\prime+1)(2j^\prime-1)}\right]^{\frac{1}{2}},
\]
\[
a(q,q;l_0,l_1)=\left[\frac{(q-l_0+1)(q+l_0+2)((q+\frac{3}{2})^2+l^2_1)}
{4(2q+1)(q+1)}\right]^{\frac{1}{2}},
\]
\[
b(q,q;l_0,l_1)=\left[\frac{(l_0-q)(l_0+q+1)((q+\frac{1}{2})^2+l^2_1)}
{4(2q+1)(q+1)}\right]^{\frac{1}{2}}.
\]

The relations between the numbers $l_0$, $l_1$ and $\sigma$,
$\dot{\sigma}$ are given by the following formulae:
\[
(l_0,l_1)=(\sigma,\sigma+1),\quad(l_0,l_1)=(-\dot{\sigma},\dot{\sigma}+1),
\]
whence it immediately follows that
\begin{equation}\label{RelLL1}
\sigma=\frac{l_0+l_1-1}{2},\quad\dot{\sigma}=\frac{l_0-l_1+1}{2}.
\end{equation}

In the case of principal series representations of $\SO_0(1,4)$ we
have\footnote{This relation is a particular case of the most general
formula $l_1=-\frac{1}{2}(n-1)+\bi\rho$ for the principal series
representations of $\SO_0(1,n)$ \cite{Boy71}.}
$l_1=-\frac{3}{2}+\bi\rho$, $\rho\in\R$. Using formulae
(\ref{HFSO14b}), (\ref{PPBtan}) and (\ref{RelLL1}), we find that
matrix elements of the principal series representations of the group
$\SO_0(1,4)$ have the form
\begin{multline}
\fM^{-\frac{3}{2}+\bi\rho,l_0}_{mn}(\fq)=
e^{-m(\epsilon+\bi\varphi+\bk\varsigma)-n(\varepsilon+\omega+\bi\psi-\bj\chi)}
\times\\
\sum^{l_0}_{k=-l_0}\sum^{l_0}_{t=-l_0}\bi^{m+k-2t}
\sqrt{\Gamma(l_0-m+1)\Gamma(l_0+m+1)\Gamma(l_0-t+1)\Gamma(l_0+t+1)}\times\\
\cos^{2l_0}\frac{\theta}{2}\tg^{m-t}\frac{\theta}{2}\times\\[0.2cm]
\sum^{\min(l_0-m,l_0+t)}_{j=\max(0,t-m)}
\frac{\bi^{2j}\tg^{2j}\dfrac{\theta}{2}}
{\Gamma(j+1)\Gamma(l_0-m-j+1)\Gamma(l_0+t-j+1)\Gamma(m-t+j+1)}\times\\[0.2cm]
\sqrt{\Gamma(l_0-k+1)\Gamma(l_0+k+1)\Gamma(l_0-t+1)\Gamma(l_0+t+1)}
\cos^{2l_0}\frac{\phi}{2}\tan^{k-t}\frac{\phi}{2}\times\\[0.2cm]
\sum^{\min(l_0-k,l_0+t)}_{s=\max(0,t-k)}
\frac{\bi^{2s}\tan^{2s}\dfrac{\phi}{2}}
{\Gamma(s+1)\Gamma(l_0-k-s+1)\Gamma(l_0+t-s+1)\Gamma(k-t+s+1)}\times\\[0.2cm]
\sqrt{\Gamma(-\tfrac{1}{2}+\bi\rho-n)\Gamma(-\tfrac{1}{2}+\bi\rho+n)
\Gamma(-\tfrac{1}{2}+\bi\rho-k)\Gamma(-\tfrac{1}{2}+\bi\rho+k)}
\ch^{-3+2\bi\rho}\frac{\tau}{2}\tnh^{n-k}\frac{\tau}{2}\times\\[0.2cm]
\sum^{\infty}_{p=\max(0,k-n)} \frac{\tnh^{2p}\dfrac{\tau}{2}}
{\Gamma(p+1)\Gamma(-\tfrac{1}{2}+\bi\rho-n-p)
\Gamma(-\tfrac{1}{2}+\bi\rho+k-p)\Gamma(n-k+p+1)}.\label{PPBtanP}
\end{multline}
From the latter expression it follows that spherical function
$f(\fq)$ of the principal series can be defined by means of the
function
\[
\fM^{-\frac{3}{2}+\bi\rho,l_0}_{mn}(\fq)=
e^{-m(\epsilon+\bi\varphi+\bk\varsigma)}\fZ^{-\frac{3}{2}+\bi\rho,l_0}_{mn}
(\cos\theta^q)e^{-n(\varepsilon+\omega+\bi\psi-\bj\chi)},
\]
where
\[
\fZ^{-\frac{3}{2}+\bi\rho,l_0}_{mn}(\cos\theta^q)=
\sum^{l_0}_{k=-l_0}\sum^{l_0}_{t=-l_0}P^{l_0}_{mt}(\cos\theta)
P^{l_0}_{tk}(\cos\phi)\fP^{-\frac{3}{2}+\bi\rho}_{kn}(\cosh\tau).
\]

Let us now express the spherical function
$\fM^{-\frac{3}{2}+\bi\rho,l_0}_{mn}(\fq)$ of the principal series
representations of $\SO_0(1,4)$ via multiple hypergeometric series.
Using formulae (\ref{PPBtanP}) and
(\ref{PPBFtan1})--(\ref{PPBFtan2}), we find
\begin{multline}
\fM^{-\frac{3}{2}+\bi\rho,l_0}_{mn}(\fq)=
e^{-m(\epsilon+\bi\varphi+\bk\varsigma)-n(\varepsilon+\omega+\bi\psi-\bj\chi)}
\times\\
\sqrt{\frac{\Gamma(l_0+m+1)\Gamma(-\frac{1}{2}+\bi\rho-n)}{\Gamma(l_0-m+1)
\Gamma(-\frac{1}{2}+\bi\rho+n)}}
\cos^{2l_0}\frac{\theta}{2}\cos^{2l_0}\frac{\phi}{2}\ch^{-3+2\bi\rho}\frac{\tau}{2}
\times\\
\sum^{l_0}_{k=-l_0}\sum^{l_0}_{t=-l_0}\bi^{m-k}
\sqrt{\frac{\Gamma(l_0-k+1)\Gamma(-\frac{1}{2}+\bi\rho+k)}
{\Gamma(l_0+k+1)\Gamma(-\frac{1}{2}+\bi\rho-k)}}
\tg^{m-t}\frac{\theta}{2}\tan^{t-k}\frac{\phi}{2}
\tanh^{k-n}\frac{\tau}{2}\times\\
\hypergeom{2}{1}{m-l_0,-t-l_0}{m-t+1}{-\tg^2\frac{\theta}{2}}
\hypergeom{2}{1}{t-l_0,-k-l_0}{t-k+1}{-\tan^2\frac{\phi}{2}}\times\\
\hypergeom{2}{1}{k+\frac{3}{2}-\bi\rho,-n+\frac{3}{2}-\bi\rho}{k-n+1}{\tanh^2\frac{\tau}{2}},
\quad m\geq t,\;t\geq k,\;k\geq n; \label{PPBFtanP1}
\end{multline}
\begin{multline}
\fM^{-\frac{3}{2}+\bi\rho,l_0}_{mn}(\fq)=
e^{-m(\epsilon+\bi\varphi+\bk\varsigma)-n(\varepsilon+\omega+\bi\psi-\bj\chi)}
\times\\
\sqrt{\frac{\Gamma(l_0+m+1)\Gamma(-\frac{1}{2}+\bi\rho-n)}{\Gamma(l-m_0+1)
\Gamma(-\frac{1}{2}+\bi\rho+n)}}
\cos^{2l_0}\frac{\theta}{2}\cos^{2l_0}\frac{\phi}{2}
\ch^{-3+2\bi\rho}\frac{\tau}{2}\times\\
\sum^{l_0}_{k=-l_0}\sum^{l_0}_{t=-l_0}\bi^{m+k-2t}
\sqrt{\frac{\Gamma(l_0-k+1)\Gamma(-\frac{1}{2}+\bi\rho+k)}
{\Gamma(l_0+k+1)\Gamma(-\frac{1}{2}+\bi\rho-k)}}
\tg^{m-t}\frac{\theta}{2} \tan^{k-t}\frac{\phi}{2}
\tanh^{k-n}\frac{\tau}{2}\times\\
\hypergeom{2}{1}{m-l_0,-t-l_0}{m-t+1}{-\tg^2\frac{\theta}{2}}
\hypergeom{2}{1}{k-l_0,-t-l_0}{k-t+1}{-\tan^2\frac{\phi}{2}}\times\\
\hypergeom{2}{1}{k+\frac{3}{2}-\bi\rho,-n+\frac{3}{2}-\bi\rho}{k-n+1}
{\tanh^2\frac{\tau}{2}}, \quad m\geq t,\;k\geq t,\;k\geq n;
\label{PPBFtanP2}
\end{multline}
\begin{multline}
\fM^{-\frac{3}{2}+\bi\rho,l_0}_{mn}(\fq)=
e^{-m(\epsilon+\bi\varphi+\bk\varsigma)-n(\varepsilon+\omega+\bi\psi-\bj\chi)}
\times\\
\sqrt{\frac{\Gamma(l_0-m+1)\Gamma(-\frac{1}{2}+\bi\rho+n)}{\Gamma(l_0+m+1)
\Gamma(-\frac{1}{2}+\bi\rho-n)}}
\cos^{2l_0}\frac{\theta}{2}\cos^{2l_0}\frac{\phi}{2}
\ch^{-3+3\bi\rho}\frac{\tau}{2}\times\\
\sum^{l_0}_{k=-l_0}\sum^{l_0}_{t=-l_0}\bi^{k-m}
\sqrt{\frac{\Gamma(l_0+k+1)\Gamma(-\frac{1}{2}+\bi\rho-k)}
{\Gamma(l_0-k+1)\Gamma(-\frac{1}{2}+\bi\rho+k)}}
\tg^{t-m}\frac{\theta}{2}\tan^{k-t}\frac{\phi}{2}
\tanh^{n-k}\frac{\tau}{2}\times\\
\hypergeom{2}{1}{t-l_0,-m-l_0}{t-m+1}{-\tg^2\frac{\theta}{2}}
\hypergeom{2}{1}{k-l_0,-t-l_0}{k-t+1}{-\tan^2\frac{\phi}{2}}\times\\
\hypergeom{2}{1}{n+\frac{3}{2}-\bi\rho,-k+\frac{3}{2}-\bi\rho}{n-k+1}
{\tanh^2\frac{\tau}{2}}, \quad t\geq m,\;k\geq t,\;n\geq k;
\label{PPBFtanP3}
\end{multline}
\begin{multline}
\fM^{-\frac{3}{2}+\bi\rho,l_0}_{mn}(\fq)=
e^{-m(\epsilon+\bi\varphi+\bk\varsigma)-n(\varepsilon+\omega+\bi\psi-\bj\chi)}
\times\\
\sqrt{\frac{\Gamma(l_0-m+1)\Gamma(-\frac{1}{2}+\bi\rho+n)}{\Gamma(l_0+m+1)
\Gamma(-\frac{1}{2}+\bi\rho-n)}}
\cos^{2l_0}\frac{\theta}{2}\cos^{2l_0}\frac{\phi}{2}
\ch^{-3+2\bi\rho}\frac{\tau}{2}\times\\
\sum^{l_0}_{k=-l_0}\sum^{l_0}_{t=-l_0}\bi^{2t-m-k}
\sqrt{\frac{\Gamma(l_0+k+1)\Gamma(-\frac{1}{2}+\bi\rho-k)}
{\Gamma(l_0-k+1)\Gamma(-\frac{1}{2}+\bi\rho+k)}}
\tg^{t-m}\frac{\theta}{2}\tan^{t-k}\frac{\phi}{2}
\tanh^{n-k}\frac{\tau}{2}\times\\
\hypergeom{2}{1}{t-l_0,-m-l_0}{t-m+1}{-\tg^2\frac{\theta}{2}}
\hypergeom{2}{1}{t-l_0,-k-l_0}{t-k+1}{-\tan^2\frac{\phi}{2}}\times\\
\hypergeom{2}{1}{n+\frac{3}{2}-\bi\rho,-k+\frac{3}{2}-\bi\rho}{n-k+1}
{\tanh^2\frac{\tau}{2}}, \quad t\geq m,\;t\geq k,\;n\geq k;
\label{PPBFtanP4}
\end{multline}
\begin{multline}
\fM^{-\frac{3}{2}+\bi\rho,l_0}_{mn}(\fq)=
e^{-m(\epsilon+\bi\varphi+\bk\varsigma)-n(\varepsilon+\omega+\bi\psi-\bj\chi)}
\times\\
\sqrt{\frac{\Gamma(l_0-m+1)\Gamma(-\frac{1}{2}+\bi\rho-n)}{\Gamma(l_0+m+1)
\Gamma(-\frac{1}{2}+\bi\rho+n)}}
\cos^{2l_0}\frac{\theta}{2}\cos^{2l_0}\frac{\phi}{2}
\ch^{-3+2\bi\rho}\frac{\tau}{2}\times\\
\sum^{l_0}_{k=-l_0}\sum^{l_0}_{t=-l_0}\bi^{k-m}
\sqrt{\frac{\Gamma(l_0+k+1)\Gamma(-\frac{1}{2}+\bi\rho+k)}
{\Gamma(l_0-k+1)\Gamma(-\frac{1}{2}+\bi\rho-k)}}
\tg^{t-m}\frac{\theta}{2}\tan^{k-t}\frac{\phi}{2}
\tanh^{k-n}\frac{\tau}{2}\times\\
\hypergeom{2}{1}{t-l_0,-m-l_0}{t-m+1}{-\tg^2\frac{\theta}{2}}
\hypergeom{2}{1}{k-l_0,-t-l_0}{k-t+1}{-\tan^2\frac{\phi}{2}}\times\\
\hypergeom{2}{1}{k+\frac{3}{2}-\bi\rho,-n+\frac{3}{2}-\bi\rho}{k-n+1}
{\tanh^2\frac{\tau}{2}},
\quad t\geq m,\;k\geq t,\;k\geq n; \label{PPBFtanP5}
\end{multline}
\begin{multline}
\fM^{-\frac{3}{2}+\bi\rho,l_0}_{mn}(\fq)=
e^{-m(\epsilon+\bi\varphi+\bk\varsigma)-n(\varepsilon+\omega+\bi\psi-\bj\chi)}
\times\\
\sqrt{\frac{\Gamma(l_0-m+1)\Gamma(-\frac{1}{2}+\bi\rho-n)}{\Gamma(l_0+m+1)
\Gamma(-\frac{1}{2}+\bi\rho+n)}}
\cos^{2l_0}\frac{\theta}{2}\cos^{2l_0}\frac{\phi}{2}
\ch^{-3+2\bi\rho}\frac{\tau}{2}\times\\
\sum^{l_0}_{k=-l_0}\sum^{l_0}_{t=-l_0}\bi^{2t-m-k}
\sqrt{\frac{\Gamma(l_0+k+1)\Gamma(-\frac{1}{2}+\bi\rho+k)}
{\Gamma(l_0-k+1)\Gamma(-\frac{1}{2}+\bi\rho-k)}}
\tg^{t-m}\frac{\theta}{2}\tan^{t-k}\frac{\phi}{2}
\tanh^{k-n}\frac{\tau}{2}\times\\
\hypergeom{2}{1}{t-l_0,-m-l_0}{t-m+1}{-\tg^2\frac{\theta}{2}}
\hypergeom{2}{1}{t-l_0,-k-l_0}{t-k+1}{-\tan^2\frac{\phi}{2}}\times\\
\hypergeom{2}{1}{k+\frac{3}{2}-\bi\rho,-n+\frac{3}{2}-\bi\rho}{k-n+1}
{\tanh^2\frac{\tau}{2}},
\quad t\geq m,\;t\geq k,\;k\geq n; \label{PPBFtanP6}
\end{multline}
\begin{multline}
\fM^{-\frac{3}{2}+\bi\rho,l_0}_{mn}(\fq)=
e^{-m(\epsilon+\bi\varphi+\bk\varsigma)-n(\varepsilon+\omega+\bi\psi-\bj\chi)}
\times\\
\sqrt{\frac{\Gamma(l_0+m+1)\Gamma(-\frac{1}{2}+\bi\rho+n)}{\Gamma(l_0-m+1)
\Gamma(-\frac{1}{2}+\bi\rho-n)}}
\cos^{2l_0}\frac{\theta}{2}\cos^{2l_0}\frac{\phi}{2}
\ch^{-3+2\bi\rho}\frac{\tau}{2}\times\\
\sum^{l_0}_{k=-l_0}\sum^{l_0}_{t=-l_0}\bi^{m-k}
\sqrt{\frac{\Gamma(l_0-k+1)\Gamma(-\frac{1}{2}+\bi\rho-k)}
{\Gamma(l_0+k+1)\Gamma(-\frac{1}{2}+\bi\rho+k)}}
\tg^{m-t}\frac{\theta}{2}\tan^{t-k}\frac{\phi}{2}
\tanh^{n-k}\frac{\tau}{2}\times\\
\hypergeom{2}{1}{m-l_0,-t-l_0}{m-t+1}{-\tg^2\frac{\theta}{2}}
\hypergeom{2}{1}{t-l_0,-k-l_0}{t-k+1}{-\tan^2\frac{\phi}{2}}\times\\
\hypergeom{2}{1}{n+\frac{3}{2}-\bi\rho,-k+\frac{3}{2}-\bi\rho}{n-k+1}
{\tanh^2\frac{\tau}{2}}, \quad m\geq t,\;t\geq k,\;n\geq k;
\label{PPBFtanP7}
\end{multline}
\begin{multline}
\fM^{-\frac{3}{2}+\bi\rho,l_0}_{mn}(\fq)=
e^{-m(\epsilon+\bi\varphi+\bk\varsigma)-n(\varepsilon+\omega+\bi\psi-\bj\chi)}
\times\\
\sqrt{\frac{\Gamma(l_0+m+1)\Gamma(-\frac{1}{2}+\bi\rho+n)}{\Gamma(l_0-m+1)
\Gamma(-\frac{1}{2}+\bi\rho-n)}}
\cos^{2l_0}\frac{\theta}{2}\cos^{2l_0}\frac{\phi}{2}
\ch^{-3+2\bi\rho}\frac{\tau}{2}\times\\
\sum^{l_0}_{k=-l_0}\sum^{l_0}_{t=-l_0}\bi^{m+k-2t}
\sqrt{\frac{\Gamma(l_0-k+1)\Gamma(-\frac{1}{2}+\bi\rho-k)}
{\Gamma(l_0+k+1)\Gamma(-\frac{1}{2}+\bi\rho+k)}}
\tg^{m-t}\frac{\theta}{2}\tan^{k-t}\frac{\phi}{2}
\tanh^{n-k}\frac{\tau}{2}\times\\
\hypergeom{2}{1}{m-l_0,-t-l_0}{m-t+1}{-\tg^2\frac{\theta}{2}}
\hypergeom{2}{1}{k-l_0,-t-l_0}{k-t+1}{-\tan^2\frac{\phi}{2}}\times\\
\hypergeom{2}{1}{n+\frac{3}{2}-\bi\rho,-k+\frac{3}{2}-\bi\rho}{n-k+1}
{\tanh^2\frac{\tau}{2}}, \quad m\geq t,\;k\geq t,\;n\geq k.
\label{PPBFtanP8}
\end{multline}

Spherical functions of the second type
$f(\varphi^q,\theta^q)=\fM^m_{-\frac{3}{2}+\bi\rho,l_0}(\varphi^q,\theta^q,0)$
of the principal series are defined as
\[
\fM^m_{-\frac{3}{2}+\bi\rho,l_0}(\varphi^q,\theta^q,0)=e^{-\bi
m\varphi^q}\fZ^m_{-\frac{3}{2}+\bi\rho,l_0}(\cos\theta^q),
\]
where
\[
\fZ_{-\frac{3}{2}+\bi\rho,l_0}^{m}(\cos\theta^q)=
\sum^{l_0}_{k=-l_0}\sum^{l_0}_{t=-l_0}P^{l_0}_{mt}(\cos\theta)
P^{l_0}_{tk}(\cos\phi)\fP_{-\frac{3}{2}+\bi\rho}^{k}(\cosh\tau).
\]
Hypergeometric-type formulae for the functions
$f(\varphi^q,\theta^q)$ follow directly from (\ref{PPBFtanP1}) to
(\ref{PPBFtanP8}) at $n=0$.

Spherical functions of the third type
$f(\epsilon,\tau,\varepsilon,\omega)=
\fM^{-\frac{3}{2}+\bi\rho}_{mn}(\epsilon,\tau,\varepsilon,\omega)$
for the principal series representations have the form
\[
\fM^{-\frac{3}{2}+\bi\rho}_{mn}(\epsilon,\tau,\varepsilon,\omega)=
e^{-m\epsilon}\fP^{-\frac{3}{2}+\bi\rho}_{mn}(\ch\tau)e^{-n(\varepsilon+\omega)}.
\]
The hypergeometric-type formulae are
\begin{multline}
\fM^{-\frac{3}{2}+\bi\rho}_{mn}(\epsilon,\tau,\varepsilon,\omega)=
e^{-m\epsilon-n(\varepsilon+\omega)}\sqrt{\frac{\Gamma(\bi\rho+m-\frac{1}{2})
\Gamma(\bi\rho-n-\frac{1}{2})}{\Gamma(\bi\rho-m-\frac{1}{2})
\Gamma(\bi\rho+n-\frac{1}{2})}}\times\\
\ch^{-3+2\bi\rho}\frac{\tau}{2}\tanh^{m-n}\frac{\tau}{2}
\hypergeom{2}{1}{m-\bi\rho-\frac{1}{2},-n-\bi\rho-\frac{1}{2}}{m-n+1}
{\tanh^2\frac{\tau}{2}},\quad m\ge n;\nonumber
\end{multline}
\begin{multline}
\fM^{-\frac{3}{2}+\bi\rho}_{mn}(\epsilon,\tau,\varepsilon,\omega)=
e^{-m\epsilon-n(\varepsilon+\omega)}\sqrt{\frac{\Gamma(\bi\rho+n-\frac{1}{2})
\Gamma(\bi\rho-m-\frac{1}{2})}{\Gamma(\bi\rho-n-\frac{1}{2})
\Gamma(\bi\rho+m-\frac{1}{2})}}\times\\
\ch^{-3+2\bi\rho}\frac{\tau}{2}\tanh^{n-m}\frac{\tau}{2}
\hypergeom{2}{1}{n-\bi\rho-\frac{1}{2},-m-\bi\rho-\frac{1}{2}}{n-m+1}
{\tanh^2\frac{\tau}{2}},\quad n\ge m.\nonumber
\end{multline}

In like manner we can define conjugated spherical functions
$f(\fq)=\fM^{-\frac{3}{2}-\bi\rho,l_0}_{\dot{m}\dot{n}}(\fq)$,
$f(\dot{\varphi}^q,\dot{\theta}^q)=\fM^{\dot{m}}_{-\frac{3}{2}-\bi\rho,l_0}
(\dot{\varphi}^q,\dot{\theta}^q,0)$ and
$\dot{f}(\epsilon,\tau,\varepsilon,\omega)=\fM^{-\frac{3}{2}-\bi\rho}_{\dot{m}
\dot{n}}(\epsilon,\tau,\varepsilon,\omega)$, since a conjugated
representation of $\SO_0(1,4)$ is defined by the pair
$\pm(l_0,-l_1)$.


\begin{thebibliography}{00}
\bibitem{AAV69} A. K. Agamaliev, N. M. Atakishiev, I. A. Verdiev,
``Invariant Expansion of Solutions of Relativistic Equations," Yad.
Fiz. {\bf 9}, 201--211 (1969).
\bibitem{All85} B. Allen, ``Vacuum states in de Sitter space," Phys.
Rev. D{\bf 32}, 3136 (1985).
\bibitem{AK26} P. Appell, M. J. Kamp\'{e} de F\'{e}riet, {\em Fonctions
hyp\'{e}rgeom\'{e}triques et hypersph\'{e}ques. Polynomes d'Hermite}
(Gauthier--Villars, Paris, 1926).
\bibitem{BI66} M. Bander, C. Itzykson, ``Group Theory and the
Hydrogen Atom," Rev. Mod. Phys. {\bf 38}, 330--358 (1966).
\bibitem{BBC68} G. Bisiacchi, P. Budini, and G. Calucci, ``Majorana
Equations for Composite Systems," Phys. Rev. {\bf 172}, 1508--1515
(1968).
\bibitem{Boy71} C. P. Boyer, ``Matrix Elements for the Most
Degenerate Continuous Principal Series of Representations of
$SO(p,1)$," J. Math. Phys. {\bf 12}, 1599--1603 (1971).
\bibitem{BM96} J. Bros, U. Moschella, ``Two-point Functions and Quantum Fields in
de Sitter Universe,"  Rev. Math. Phys. {\bf 8} 327-392 (1996).
\bibitem{Dir35} P. A. M. Dirac, ``The electron wave equation in de Sitter
space," Annals of Math. {\bf 36}, 657 (1935).
\bibitem{Dix61} J. Dixmier, ``Repr\'{e}sentations int\'{e}grables du
groupe de De Sitter," Bull. Soc. math. France, {\bf 89}, 9--41
(1961).
\bibitem{EH99} J. Eades, F. J. Hartmann, ``Forty years of
antiprotons," Rev. Mod. Phys. {\bf 71}, 373--419 (1999).
\bibitem{Ext76} H. Exton, {\em Multiple Hypergeometric Functions and Applications}
(Ellis Horwood, Chicester, 1976).
\bibitem{Fock} V. A. Fock, ``Zur Theorie des Wasserstoffatoms,"
Zs. f. Phys. {\bf 98}, 145--154 (1935).
\bibitem{GY48} I. M. Gel'fand, A. M. Yaglom, ``General
relativistic--invariant equations and infinite--dimensional
representations of the Lorentz group," Zh. Ehksp. Teor. Fiz. {\bf
18}, 703--733 (1948).
\bibitem{GGV62} I. M. Gel'fand, M. I. Graev, N. Ya. Vilenkin, {\em Integral
geometry and related questions of group representations}
(Fizmatgiz, Moskow, 1962) [in Russian].
\bibitem{GMS} I. M. Gel'fand, R. A. Minlos, Z. Ya. Shapiro, {\em Representations
of the Rotation and Lorentz Groups and their Applications} (Pergamon
Press, Oxford, 1963).
\bibitem{Hel78} S. Helgason, {\em Differential geometry, Lie groups and symmetric
spaces} (Academic Press, New York, 1978).
\bibitem{Kar78} M. Karoubi, {\em K-Theory. An Introduction}
(Springer-Verlag, Berlin, 1979).
\bibitem{KK85} A. U. Klimyk, I. I. Kachurik, {\em Calculating Methods in
Group Representations Theory} (Visha Shkola, Kiev, 1986) [in
Russian].
\bibitem{MRT05} S. Moradi, S. Rouhani, M. V. Takook, ``Discrete
Symmetries for Spinor Field in de Sitter Space," Phys. Lett. B {\bf
613}, 74--82 (2005).
\bibitem{Nai58} M. A. Naimark,
{\em Linear Representations of the Lorentz Group} (Pergamon, London,
1964).
\bibitem{Roz55} B. A. Rozenfel'd, {\em Non-Euclidean Geometries} (Gostehizdat,
Moscow, 1955)
[in Russian].
\bibitem{Str69} S. Str\"{o}m, ``On the decomposition of a unitary representation
of (1+4) de Sitter group with respect to representations of the
Lorentz group," Arkiv f\"{o}r Fysik {\bf 40}, 1--33 (1969).
\bibitem{Tak63} R. Takahashi, ``Sur les repr\'{e}sentations
unitaries des groupes de Lorentz g\'{e}n\'{e}ralis\'{e}s," Bull.
Soc. math. France, {\bf 91}, 289--433 (1963).
\bibitem{Var03} V. V. Varlamov, ``General Solutions of Relativistic
Wave Equations," Int. J. Theor. Phys. {\bf 42}, No. 3, 583--633
(2003).
\bibitem{Var03c} V. V. Varlamov, `` Relativistic wavefunctions on the
Poincar\'{e} group," J. Phys. A: Math. Gen. {\bf 37}, 5467--5476
(2004).
\bibitem{Var03d} V. V. Varlamov, ``Maxwell field on the Poincar\'{e}
group," Int. J. Mod. Phys. {\bf A20}, 4095--4112 (2005).
\bibitem{Var04} V. V. Varlamov, ``Universal Coverings of Orthogonal
Groups," Adv. Appl. Clifford Algebras {\bf 14}, 81--168 (2004).
\bibitem{Var05} V. V. Varlamov, ``General Solutions of Relativistic
Wave Equations II: Arbitrary Spin Chains," preprint math-ph/0503058
(2005), to appear in Int. J. Theor. Phys.
\bibitem{Var05a} V. V. Varlamov, ``$CPT$ groups for spinor field in
de Sitter space," Phys. Lett. B {\bf 631}, 187-191 (2005).
\bibitem{Var06} V. V. Varlamov, ``Relativistic spherical functions
on the Lorentz group," J. Phys. A: Math. Gen. {\bf 39}, 805--822
(2006).
\bibitem{Vil65} N. Ya. Vilenkin, {\em Special Functions and the Theory of Group
Representations} (AMS, Providence, 1968).
\bibitem{VK90} N. Ya. Vilenkin, A. U. Klimyk,
{\em Representations of Lie Groups and Special Functions}, vols.
1--3. (Dordrecht: Kluwer Acad. Publ., 1991--1993).
\end{thebibliography}
\end{document}